\documentclass[11pt,letterpaper]{article}
\usepackage{geometry}
\usepackage{amsfonts}
\usepackage{amsmath}
\usepackage{amssymb}
\usepackage{mathrsfs}
\usepackage{mathpartir}

\usepackage{amsthm}
\newtheorem{thm}{Theorem}
\newtheorem{lemma}[thm]{Lemma}
\newtheorem{definition}{Definition}

\usepackage{color}
\usepackage{comment}
\usepackage{xstring}

\usepackage{hyperref}

\usepackage{graphicx}

\usepackage{multicol}

\usepackage{array}
\usepackage{longtable}
\usepackage{needspace}

\usepackage{listings}
\title{A Modular Approach to Metatheoretic \\ Reasoning for Extensible Languages}
\author{Dawn Michaelson \and Gopalan Nadathur \and Eric Van Wyk}
\date{}

\begin{document}

\maketitle

\renewcommand{\bar}[1]{\ensuremath{\overline{#1}}}

\newcommand{\imp}{\ensuremath{\supset}}
\renewcommand{\emptyset}{\ensuremath{\varnothing}}
\newcommand{\forallx}[2]{\forall #1.\, #2}
\newcommand{\existsx}[2]{\exists #1.\, #2}

\newcommand{\ntset}[1]{\ensuremath{\mathscr{C}^{#1}}}
\newcommand{\constrset}[1]{\ensuremath{\mathbb{C}^{#1}}}
\newcommand{\relset}[1]{\ensuremath{\mathscr{R}^{#1}}}
\newcommand{\ruleset}[1]{\ensuremath{\mathbb{R}^{#1}}}
\newcommand{\transrelset}[1]{\ensuremath{\mathscr{T}^{#1}}}
\newcommand{\transruleset}[1]{\ensuremath{\mathbb{T}^{#1}}}
\newcommand{\trset}[1]{\ensuremath{\mathbb{S}^{#1}}}

\newcommand{\lemmaset}{\ensuremath{\mathscr{L}}}

\newcommand{\sequent}[3]{\ensuremath{#1:#2 \longrightarrow #3}}
\newcommand{\sequentctx}[1]{\ensuremath{#1:}}
\newcommand{\sequentsansctx}[2]{\ensuremath{#1 \longrightarrow #2}}

\newcommand{\setize}[1]{\ensuremath{\{ #1 \}}}

\newcommand{\oneTo}[2]{\setize{#1_1,\ldots, #1_{#2}}}
\newcommand{\Eall}{\ensuremath{E, E_1, \ldots, E_n}}

\newcommand{\plus}[3]{\ensuremath{\mathit{plus}(#1, #2, #3)}}

\newcommand{\typeexpr}[3]{\ensuremath{#1 \vdash #2 : #3}}
\newcommand{\typestmt}[3]{\ensuremath{#1 \vdash #2, #3}} 
\newcommand{\evalexpr}[3]{\ensuremath{#1 \vdash #2 \Downarrow #3}}
\newcommand{\evalstmtRel}{\Downarrow}
\newcommand{\evalstmt}[3]{\ensuremath{(#1,~#2) \evalstmtRel #3}}
\newcommand{\evalstmtT}[3]{\ensuremath{(#1,~#2)\ \transRel{\evalstmtRel}\ #3}}
\newcommand{\evalstmtES}[4]{\ensuremath{(#1,~#2) \Downarrow_{ES} (#3, #4)}}
\newcommand{\vars}[2]{\ensuremath{\mathit{vars}(\mathit{#1}, \mathit{#2})}}
\newcommand{\val}[1]{\ensuremath{\mathit{value}(\mathit{#1})}}
\newcommand{\lookuptype}[3]{\ensuremath{\mathit{lkp\!Ty}(#1, #2, #3)}}
\newcommand{\notpresenttype}[2]{\ensuremath{\mathit{notBoundTy}(#1, #2)}}
\newcommand{\lookupval}[3]{\ensuremath{\mathit{lkp\!Val}(#1, #2, #3)}}
\newcommand{\updateCtx}[4]{\ensuremath{\mathit{update}(#1, #2, #3, #4)}}
\newcommand{\removeCtx}[3]{\ensuremath{\mathit{remove}(#1, #2, #3)}}
\newcommand{\consval}[3]{\ensuremath{#1:#2;~#3}}

\newcommand{\joinlevel}[3]{\ensuremath{\mathit{join}(#1, #2, #3)}}
\newcommand{\lookupsec}[3]{\ensuremath{\mathit{lkpSec}(#1, #2, #3)}}
\newcommand{\notpresentsec}[2]{\ensuremath{\mathit{notBoundSec}(#1, #2)}}
\newcommand{\exprlevel}[3]{\ensuremath{#1 \vdash \mathit{level}(#2,#3)}}
\newcommand{\secure}[4]{\ensuremath{#1\ #2 \vdash \mathit{secure}(#3,#4)}}
\newcommand{\eqpublicvals}[3]{\ensuremath{\mathit{eqpublicvals}(#1, #2, #3)}}

\newcommand{\proj}[1]{\ensuremath{\mathit{proj}_{#1}}}
\newcommand{\projectexpr}[2]{\ensuremath{\proj{e}(#1, #2)}}
\newcommand{\projectstmt}[2]{\ensuremath{\proj{s}(#1, #2)}}
\newcommand{\projectty}[2]{\ensuremath{\proj{ty}(#1, #2)}}

\newcommand{\frag}{\ensuremath{\pi}}
\newcommand{\hproven}[5]{\ensuremath{\mathit{pp}_{#3}(#1,~#2,~#4,~#5)}}
\newcommand{\proven}[4]{\ensuremath{\mathit{proven}(#1,~#2,~#3,~#4)}}
\newcommand{\unknown}[1][]{\ensuremath{\iota_{#1}}}
\newcommand{\gen}[2]{\ensuremath{\mathit{GenExt}(#1,#2)}}
\newcommand{\seqRelSans}[1]{\sim_{#1}}
\newcommand{\seqRel}[3]{\ensuremath{#1 \seqRelSans{#3} #2}}
\newcommand{\transRel}[1]{\ensuremath{{#1}_{P}}}
\newcommand{\dropT}[1]{\ensuremath{\mathit{dropP}(#1)}}
\newcommand{\addT}[1]{\ensuremath{\mathit{addP}(#1)}}
\newcommand{\extSize}[1]{\ensuremath{{#1}_{ES}}}
\newcommand{\seqTRelnewsans}[2][]{\ensuremath{\sqsubseteq^{#1}_{#2}}}
\newcommand{\seqTRelnew}[4][]{\ensuremath{#3 \seqTRelnewsans[#1]{#2} #4}}
\newcommand{\subst}[3]{#3[\![#1/#2]\!]}
\newcommand{\appsubst}[2]{#2[#1]}
\newcommand{\substcomp}[2]{\ensuremath{#2 \circ #1}}
\newcommand{\mgu}{mgu}

\newcommand{\langComp}[2]{\ensuremath{#1 \lhd #2}}
\newcommand{\proofComp}[1]{\ensuremath{\mathit{compose}(#1)}}

\newcommand{\fig}[1]{Figure~\IfBeginWith{#1}{fig:}{\ref{#1}}{\ref{fig:#1}}}
\renewcommand{\sec}[1]{Section~\IfBeginWith{#1}{sec:}{\ref{#1}}{\IfBeginWith{#1}{subsec:}{\ref{#1}}{\ref{sec:#1}}}}
\newcommand{\subsec}[1]{Section~\IfBeginWith{#1}{sec:}{\ref{#1}}{\IfBeginWith{#1}{subsec:}{\ref{#1}}{\ref{subsec:#1}}}}
\newcommand{\property}[1]{Property~\IfBeginWith{#1}{eq:}{\ref{#1}}{\ref{eq:#1}}}
\newcommand{\projConstr}[1]{Projection Constraint~\IfBeginWith{#1}{eq:}{\ref{#1}}{\ref{eq:#1}}}
\newcommand{\refequation}[1]{Equation~\IfBeginWith{#1}{eq:}{\ref{#1}}{\ref{eq:#1}}}
\newcommand{\reflemma}[1]{Lemma~\IfBeginWith{#1}{thm:}{\ref{#1}}{\ref{thm:#1}}}
\newcommand{\refthm}[1]{Theorem~\IfBeginWith{#1}{thm:}{\ref{#1}}{\ref{thm:#1}}}
\newcommand{\refdef}[1]{Definition~\IfBeginWith{#1}{def:}{\ref{#1}}{\ref{def:#1}}}
\newcommand{\refapp}[1]{Appendix~\IfBeginWith{#1}{app:}{\ref{#1}}{\ref{app:#1}}}

\newcommand{\eg}{{\it e.g.}}
\newcommand{\ie}{{\it i.e.}}

\newcommand{\id}{{\sl id}}
\newcommand{\cut}{{\sl cut}}
\newcommand{\contL}{{\sl c}$\mathcal{L}$}
\newcommand{\botL}{$\bot\mathcal{L}$}
\newcommand{\topR}{$\top\mathcal{R}$}
\newcommand{\orL}{$\lor\mathcal{L}$}
\newcommand{\orR}{$\lor\mathcal{R}$}
\newcommand{\andL}{$\land\mathcal{L}$}
\newcommand{\andR}{$\land\mathcal{R}$}
\newcommand{\impL}{$\supset\!\!\mathcal{L}$}
\newcommand{\impR}{$\supset\!\!\mathcal{R}$}
\newcommand{\forallL}{$\forall\mathcal{L}$}
\newcommand{\forallR}{$\forall\mathcal{R}$}
\newcommand{\existsL}{$\exists\mathcal{L}$}
\newcommand{\existsR}{$\forall\mathcal{R}$}
\newcommand{\defL}{{\sl def}$\mathcal{L}$}
\newcommand{\defR}{{\sl def}$\mathcal{R}$}
\newcommand{\ind}[2]{\ensuremath{\mbox{\sl ind}^{#1}_{#2}}} 
 
\newcommand{\inv}[1]{#1^c_{\mbox{\scriptsize{\sl inv}}}}

\begin{abstract}

\noindent This paper concerns the development of metatheory for
extensible languages. 
  It uses as its starting point a view that programming languages
  tailored to specific application domains are to be
  constructed by composing components from an open library of
  independently-developed extensions to a host language.
  In the elaboration of this perspective, static analyses (such as
  typing) and dynamic semantics (such as evaluation) are 
  described via relations whose specifications are distributed across
  the host language and extensions and are given in a rule-based
  fashion.
  Metatheoretic properties, which ensure that static analyses
  accurately gauge runtime behavior, are represented in this context
  by formulas over such relations.
  These properties may be fundamental to the language, introduced by
  the host language, or they may pertain to analyses introduced by
  individual extensions.
  We expose the problem of \emph{modular metatheory}, \ie, the notion
  that proofs of relevant properties can be constructed by reasoning
  independently within each component in the library.
  To solve this problem, we propose the twin ideas of decomposing proofs
  around language fragments and of reasoning generically about
  extensions based on broad, a priori constraints imposed on their
  behavior.
  We establish the soundness of these styles of reasoning by showing
  how complete proofs of the properties can be automatically
  constructed for any language obtained by composing the independent
  parts.
  Mathematical precision is given to our discussions by framing them
  within a logic that encodes inductive rule-based specifications via
  least fixed-point definitions.
  We also sketch the structure of a practical system for metatheoretic
  reasoning for extensible languages based on the ideas developed.
\end{abstract}

\section{Introduction}
\label{sec:introduction}

This paper concerns the \emph{simultaneous treatment} of the notions of
metatheory and language extensibility, two concepts that have been
noted to be important to modern programming languages. 
Metatheory pertains to the association of properties with a
programming language that apply to \emph{all} programs written in the
language. 
A conventional application of metatheory is in ensuring that relevant
static analyses of programs can be used to draw sound conclusions about
their runtime behavior.
Language extensibility, on the other hand, refers to the provision of
a framework for smoothly adding new features to a language to satisfy
particular needs of users.
The ability to provide such a framework has taken on a special
significance with the modern-day permeation of computing into 
varied domains and the corresponding need to tailor languages to
particular groups of users.
While mechanisms for establishing properties of programming languages and for
building in language extensibility have been investigated
independently in the past, the focus on treating the two issues
\emph{simultaneously} raises interesting new questions. 
This paper articulates some of these questions and provides answers
to them.

The discussions in this paper are based on 
a particular approach to supporting extensibility that has an
established versatility~\cite{ekman2007oopsla,erdweg11oopsla,kaminski2017oopsla}.
This approach assumes a common core---identified as the \emph{host
language}---to every language that is of interest. 
It then assumes an \emph{open library of independently-developed
extensions} around the host language.
In this context, a language for a particular computational task is
obtained by combining a selected set of extensions with the host
language. 
Coherence in this model requires attempts to create such
combinations will always succeed, \ie{} there will not be unwanted
interference between any of the participating components.
More specifically, extensions typically contribute new syntactic
constructs and may also add to the semantic attributes and analyses
based on such attributes.
Well-definedness in this context requires such additions
interact constructively; for instance, compositions should not lead to
syntactic ambiguity, should not preclude deterministic parsing, and
semantic attributes should be well-defined for the resulting language.
Moreover, the ability to interact
should be determinable
\emph{locally} for each component, independently of knowledge about
the specific components in a particular composite language.
These desiderata have been considered in past work (\eg,
see~\cite{schwerdfeger2009pldi,kaminski2012sle}), 
and we assume that means for ensuring such coherence have been
employed as the starting point for our work.

Metatheoretic properties for a language are usually stated with
respect to the syntactic constructs it offers and are based on their
associated semantic attributes.
The traditional approach to developing proofs for such properties
relies fundamentally on a \emph{closed world} assumption: the
constructs provided by the language and the semantic attributes
associated with them are known completely at the time a proof is
constructed. 
This assumption, which is reasonable when the language is provided as
a monolithic whole, is not an acceptable one relative to an
extensibility framework. 
In this context, the complete language is known only when different
pieces are composed together, a point at which it is too late to try
to establish its metatheoretic properties. 
A related observation is that the process of reasoning about a
specific fragment of a language is best carried out by the designer(s)
of that fragment rather than someone who is using it in a
composition. 
Thus, to serve the extensionality context properly, it seems necessary
to \emph{modularize} the reasoning process. 
More specifically, a means must be provided for each component to
develop what will ultimately constitute a part of the proof of a
property for a composite language while being oblivious, in a sense,
to what other components may constitute the overall language. 
Moreover, devices should exist for \emph{automatically} stitching
together these separately-developed proof fragments to yield a proof
of the property for the full language at the time the
composition is determined. 

Our objective in this paper is to realize the vision described above. 
In developing our ideas, we distinguish between two kinds of
properties. 
The first sort of property is that which is intrinsic to the character
of the language and hence should be known to the host language as well
as to each extension as it is conceived.
A canonical example of such a property is \emph{type preservation},
\ie, the proposition that the types of program fragments are preserved
under evaluation steps.
We refer to such properties as \emph{foundational properties}.
For a property of this kind, it is possible to realize an obvious form
of modularization: each component contributes a ``partial proof'' that
pertains to the constructs it introduces and these different pieces
are combined on demand into a complete proof.
Note, however, that care must be exercised in constructing
the partial proofs in the absence of the closed world assumption.
One of the contributions of this paper is that it identifies 
automatically enforceable and practically acceptable constraints on
the manner in which semantic attributes are specified and proofs are
constructed that suffice to enforce such care.
In particular, it is shown, when these constraints are followed,
there is a sound means for combining the different proof
fragments to yield a complete proof for any combination of the
component parts. 

We also consider another variety of properties, ones that might be
introduced at a later stage of development, typically by an extension.
As an example, consider an extension that layers ``security''
capabilities on a pre-existing language structure.
Such an extension may also introduce a special form of static analysis
meant to ensure that programs are well-behaved from a security
perspective and a corresponding metatheoretic property verifying
the soundness of this analysis. 
We refer to such properties here as \emph{auxiliary properties}.
A modularization of the kind just discussed is not possible for 
the proof of such a property.
Instead, the proof must be provided in its entirety by the extension
introducing it.
However, to be effective in constructing such a proof, that component
must still get help from the other components with which it could be
interacting.  
One important question in this context is how constructs from a priori
unknown extensions are to be modeled in the reasoning process.
The extensibility framework allows extensions to describe
``projections'' into the host language for their constructs as a means
for other extensions to understand them at a distance.
We propose to use such projections as a vehicle for circumscribing
the behavior of extensions relative to the host language.
In this context, the host language posits \emph{projection
  constraints} as foundational properties to which extensions may assume
adherence in reasoning about auxiliary properties they introduce. 
Other extensions contribute to the proof by showing the behavior
of their constructs is compatible with the projection constraints.

We develop these ideas in the rest of the paper.
We begin by describing the framework for extensibility that provides
the basis for our work in \sec{framework}.
\sec{problem} elaborates on the issue of metatheory and challenges in
its modular realization relative to this extensibility framework.
Showing how these challenges can be met requires us to be more 
specific about the logical structure of properties and proofs.
Towards this end, we base our discussions on a specific logic for 
reasoning about relational
specifications~\cite{baelde14jfr,gacek11ic}, which we describe  
in \sec{abella}.  
The next two sections take up a discussion of the two kinds of
metatheoretic properties we have described above, in the process
playing out the ideas to their modular treatment that we have
sketched.
The technical developments in Sections~\ref{sec:framework}
through~\ref{sec:auxiliaryProperties} are 
grounded by a running example comprising a simple host language for
imperative programming; an extension for building, inspecting, and
manipulating lists; and another extension for annotating variable
declarations with a security level (either \emph{public} or \emph{private}) 
and a static information-flow analysis is intended to ensure that
values in private variables do not leak into public ones. In this
context, we use type preservation and the soundness of the
information-flow analysis as vehicles for illustrating the modular
development of proofs for the two kinds of properties. 
\sec{implementation} discusses how the ideas developed in this paper
can be deployed in a framework for modular reasoning and, more
specifically, introduces the {\it Extensibella} and {\it Sterling}
systems~\cite{mmel.website} that embody some of this thinking. 
Finally, \sec{relatedWork} discusses related work and \sec{conclusion}
summarizes this paper and discusses directions for future work.

\section{A Framework for Extensibility}
\label{sec:framework}

Language specifications are typically based on describing the relevant
syntactic expressions and then describing relations over these
expressions; these relations identify statically-determined
attributes and constraints that must be satisfied by them for an
expression to be well-formed or dynamically-determined attributes that
formalize notions such as the evaluation of expressions.
In the framework that provides the basis for the work in this paper,
expression forms and the relations associated with them may be
introduced by the host language or by extensions. 
The formal specification in both cases is provided via a 7-tuple
written as
$\langle \ntset{}, \constrset{}, 
 \relset{}, \ruleset{},
 \transrelset{}, \transruleset{}, \trset{} \rangle$,
with specific constraints on how this form is to be used in the case of
a host language or an extension.
The first two elements of this 7-tuple identify the syntactic
constructs introduced by the language component and the next two
describe the relations relevant to the collection of expressions.
It is possible in our framework for an extension to introduce
relations applying to constructs from other extensions in the same
language library that it does
not know about directly.
The last three items in the specification provide a means for
realizing this possibility.

In the rest of this section, we explain the details of this scheme for
language specification and show how it moulds itself naturally to
providing for a framework for language extensibility.
The specification mechanisms are illustrated by a small host language
and a couple of simple extensions that will also lend themselves to
illustrations of the mechanisms for modular reasoning
developed later in the paper.

\subsection{The Specification of Syntax}

In descriptions meant for a language user, the focus is usually on
spelling out the concrete structure of expressions.
The view that is relevant to this paper is of a different kind.
Here we are concerned with abstract syntax that identifies the principle
constructs and the participating components of each expression.
Moreover, this internal view will have to pay attention not only to
expression categories that manifest themselves explicitly in language
constructs, but also those that play a role in describing attributes
that are to be associated with language expressions.

The formal description of these aspects is entrusted to the first two
components in the 7-tuple
presentation ($\ntset{}$ and $\constrset{}$).
The first step in this direction is to identify the set of syntactic
categories, a task accomplished through $\ntset{}$.
A set of constructors is then associated with each
expression category by $\constrset{}$.
These constructors take as arguments expressions of other known
syntactic categories.
We depict the association in a 
form that looks like a datatype declaration in a functional
programming language.
A presentation in this form should be understandable without further
comment for the host language.
Some of the sets of constructors for syntactic categories given by the
host language will be the complete set, while others may be extended
by extensions.
The syntactic categories for which
extensions may introduce new constructors are determined by the
$\transrelset{}$ element of the tuple, as discussed below.

Extensions are part of a library built around a given host language.
Thus, all the host language syntactic categories and expression
constructors are assumed to be known at the outset in an extension.
An extension can augment this collection by including new syntactic
categories in the $\ntset{}$ corresponding to it.
$\constrset{}$ may add to the expression forms for extensible host language
categories and also identify such forms for the categories it adds.
We will write $\mathit{category(c)}$ for a constructor $c \in
\constrset{}$ to indicate the syntactic category in $\ntset{}$ of the
phrase the constructor $c$ builds.

We illustrate the structure described above by considering the example
of a language library with a host language and two extensions.
The host language has a few statement forms and some simple expression
forms.  One extension adds lists and the other
augments variables with security annotations to support an
information-flow analysis.
The syntax specification pertinent to these components is shown in
Figure~\ref{fig:syntax}; we write $\ntset{H}$ and $\constrset{H}$ for
the specification for the host language, $\ntset{L}$ and
$\constrset{L}$ for that for the list extension, and $\ntset{S}$ and
$\constrset{S}$ for the extension that introduces security
annotations.
The host language has extensible categories for statements ($s$), expressions
($e$), and types ($ty$).  There are also (unspecified) categories for
the names of variables ($n$) and integer literals ($i$),
as well as categories for typing contexts ($\Gamma$) mapping names
to types and value contexts ($\gamma$) mapping names to values, a subset of
expressions.
Note the categories $\Gamma$ and $\gamma$ that correspond to
typing and value contexts do not usually have a manifestation in
the externally visible syntax for the language.
They are primarily needed in specifying typing and related properties
for expression forms as we shall see in the next subsection and are
thus part of the specification of the (internal) syntax.
The latter four categories are non-extensible, \ie, extensions may not
add constructors building expressions of these types.

The intended interpretation of most of the constructs introduced by the host
language should be self-explanatory, though some explanation may be
useful. 
The constructor $\mathit{decl}$ serves to represent the declaration of
a new variable $n$ of type $ty$ with its initial value given by expression $e$,
\emph{seq} is for representing the sequential
composition of statements, \emph{ifte} is for representing a conditional
statement, and \emph{while} is for representing a while loop.
Constructs representing variable references ($\mathit{var}$), integer
($\mathit{intlit}$) and Boolean literals, and some traditional binary
and unary operators, are provided.

The list extension introduces constructs representing the creation and inspection
of lists and their names should indicate their intended meaning. It also
introduces a new type for list types and a new statement form representing
simultaneous assignment to two variables the head and tail of a list
(\ie{} $\mathit{splitlist}(hd, tl, e)$ represents evaluating $e$ into a list and
assigning the first element to $hd$ and the rest of the list to $tl$),
but does not add any new
syntactic categories and thus $\ntset{L}$ is empty.
The security extension augments the host language category $s$ to
allow variables to be identified as ones whose contents must be
protected (\emph{private}) or ones whose contents may be viewed by
the world (\emph{public}) through a new construct for variable
declarations that includes this information. It also adds new
syntactic categories for the security level indicators as well as
\emph{security contexts} mapping names to security levels and thus
\ntset{S} is $\setize{\mathit{sl}, \Sigma}$.

\begin{figure}
\begin{minipage}[t]{3in}
$\ntset{H} = \{ s, e, n, i, ty, \Gamma, \gamma \}$

\bigskip\noindent
$\constrset{H}$:

\begin{tabular}{lcl}
$s$ & ::= & $\mathit{skip}$ \\
    & $|$ & $\mathit{decl}(n, ty, e)$ \\
    & $|$ & $\mathit{assign}(n, e)$ \\
    & $|$ & $\mathit{seq}(s, s)$ \\
    & $|$ & $\mathit{ifte}(e, s, s)$ \\
    & $|$ & $\mathit{while}(e, s)$ \\
\\
$e$ & ::= & $\mathit{var}(n)$ \\
    & $|$ & $\mathit{intlit}(i)$ \\
    & $|$ & $\mathit{true}$ \\
    & $|$ & $\mathit{false}$ \\
    & $|$ & $\mathit{add}(e, e)$ \\
    & $|$ & $\mathit{eq}(e, e)$ \\
    & $|$ & $\mathit{gt}(e, e)$ \\
    & $|$ & $\mathit{not}(e)$ \\
\\

$\mathit{ty}$ & ::= &  $\mathit{int}$ \\
   & $|$ & $\mathit{bool}$ \\
\\
$\Gamma$ & ::= & $\mathit{nilty}$ \\
         & $|$ & $\mathit{consty}(n, ty, \Gamma)$ \\
\\
$\gamma$ & ::= & $\mathit{nilval}$ \\
         & $|$ & $\consval{n}{e}{\gamma}$ \\
\end{tabular}
\end{minipage}
\begin{minipage}[t]{3in}
$\ntset{L} = \{ \}$

\bigskip\noindent
$\constrset{L}$:

\begin{tabular}{lcl}
$e$ & ::= & $\mathit{nil}$ \\
    & $|$ & $\mathit{cons}(e, e)$ \\
    & $|$ & $\mathit{null}(e)$ \\
    & $|$ & $\mathit{head}(e)$ \\
    & $|$ & $\mathit{tail}(e)$ \\
\\
$\mathit{ty}$ & ::= &  $\mathit{list}(ty)$ \\
\\
$\mathit{s}$ & ::= &  $\mathit{splitlist}(n, n, e)$
\end{tabular}

\bigskip\noindent
\hrule{}
\bigskip\noindent
\\
$\ntset{S} = \{ \mathit{sl}, \Sigma \}$

\bigskip\noindent
$\constrset{S}$:

\begin{tabular}{lcl}
$s$ & ::= & $\mathit{secdecl(n, ty, sl, e)}$ \\
\\
$\mathit{sl}$ & ::=  & $\mathit{public}$ \\
    & $|$ & $\mathit{private}$ \\
\\
$\Sigma$ & ::= & $\mathit{nilsec}$ \\
      & $|$ & $\mathit{conssec}(n, sl, \Sigma)$ 
\end{tabular}

\end{minipage}
\caption{The syntax of the example host language ($H$), the list
  extension ($L$), 
  and the security extension ($S$).  Note that we use the notation
  $\consval{n}{e}{\gamma}$ for evaluation contexts $\gamma$ rather
  than a named constructor for conciseness in writing rules.}
\label{fig:syntax}
\end{figure}

\subsection{Describing Semantic Relations}

Relations between expressions in different syntactic categories play
varied roles in language specifications.
In our framework, such relations will be represented by predicates
that are typed by syntactic categories.
The element $\relset{}$ in the specification $\langle \ntset{}, \constrset{}, 
 \relset{}, \ruleset{},
 \transrelset{}, \transruleset{}, \trset{} \rangle$ identifies these
 predicates and their types.
We shall assume also that these relations are defined in a
syntax-directed and rule-based fashion, a practice
common in language descriptions.
The element $\ruleset{}$ in the 7-tuple presents the definitions of
relations in this form.

As with the syntax description, relations identified by the host
language are assumed to be part of the vocabulary of an extension in
the same language library.
Thus the $\ruleset{}$ component of an extension may include rules for
these relations, in addition to ones for the relations it introduces.
There is, however, a constraint on rules of the former kind.
In the extensibility framework, we would like to hold as fixed in a
meaningful sense the definition of host-introduced relations on
host-introduced constructs.
Towards realizing this desiderata, we require the identification of
one of the arguments of each relation as its \emph{primary
component}.
The intuition here is that the relation is \emph{about} that argument;
for example, a typing relation is about the term it types.
In keeping with this intuition, we shall refer to the relation as one
\emph{for} the primary component category. 
In writing relations in specifications, we shall indicate their
primary component by annotating it with an asterisk.
The stipulation then is that an extension may include a rule for a
relation introduced by the host language only if it pertains to a constructor
added to the syntactic category by that extension.

To illustrate this aspect of our framework, we build on the syntax
specification of the example host language and extensions from the
previous subsection.  The full language specification may be found in
\refapp{language}, including all the rules for the host language and
both extensions.
\fig{host-rules} gives the relations introduced by the host language
and some rules the host language introduces.
It introduces relations for looking up types in typing
contexts ($\lookuptype{\Gamma^*}{n}{ty}$) and value contexts
($\lookupval{\gamma^*}{n}{e}$), as well as a relation for checking a
name is not bound in a typing context ($\notpresenttype{\Gamma^*}{n}$).
Additionally, the host language introduces a relation for updating
assignments in value contexts ($\updateCtx{\gamma^*}{n}{e}{\gamma}$)
using a relation to remove the old binding
($\removeCtx{\gamma*}{n}{\gamma}$).

The host language also introduces relations over expressions and
statements.
One of these, $\val{e^*}$, identifies expression forms which are
values.  In the host language, these are integer and Boolean
constants.  We will refer to expressions $e$ for which $\val{e}$ holds
as \emph{values}.
Another relation over expressions, $\vars{e^*}{ns}$, identifies the
names used in an expression.
Some rules for both of these relations are shown in \fig{host-rules}.

Our host language also introduces typing relations over expressions
and statements.  These are the respective primary components of the
two relations, as typing is \emph{about} the term being typed.
Expression typing, written $\typeexpr{\Gamma}{e^*}{ty}$, that an
expression $e$ has type $ty$ under the typing context $\Gamma$, has
the expected rules for the host language constructs.
Statement typing, written $\typestmt{\Gamma}{s^*}{\Gamma'}$, indicates
the statement $s$ is well-typed under the initial typing context
$\Gamma$, producing the typing context $\Gamma'$ with updated type
bindings for variables.  There are a couple of things to note with
this relation.  First, in rule \textsc{TS-decl} in \fig{host-rules},
we see that declaring a new variable requires the variable not already
be assigned a type in the context ($\notpresenttype{\Gamma}{n}$).
Then, in rule \textsc{TS-ifte}, we see we throw away any updates to
the typing context from the branches, similar to how new scopes are
treated in languages like C and Java.  These choices allow us to keep
the example simple, avoiding the complications of encoding explicit
scopes into the typing and evaluation contexts.

Finally, the host language also introduces relations and rules
specifying possibly-nonterminating big-step evaluation for expressions
and statements, some rules for which are shown in \fig{host-rules}.  As
with typing, evaluation is \emph{about} the expression or statement
being evaluated, and thus these are the primary components of the
relation. 
The relation $\evalexpr{\gamma}{e^*}{e}$ specifies how the
primary-component expression evaluates to a value in the context of
the value store $\gamma$, with rules as expected.
Similarly, $\evalstmt{\gamma}{s^*}{\gamma'}$ describes the execution
of a statement in the context of a value store $\gamma$, relating this
and the statement $s$ to a store that results from the execution
containing updated values.

\begin{figure}

\begin{tabular}{ll}
  \hspace*{-0.15in}
  \relset{H} = & \hspace*{-0.15in}
                 \{\lookuptype{\Gamma^*}{n}{ty}, \ \
                 \notpresenttype{\Gamma^*}{n}, \ \
                 \lookupval{\gamma^*}{n}{e}, \ \
                 \val{e^*}, \ \
                 \vars{e^*}{2^{n}}, \\
               & \typeexpr{\Gamma}{e^*}{ty}, \ \
                 \typestmt{\Gamma}{s^*}{\Gamma}, \ \
                 \updateCtx{\gamma^*}{n}{e}{\gamma}, \ \
                 \removeCtx{\gamma^*}{n}{\gamma}, \ \
                 \evalexpr{\gamma}{e^*}{e}, \ \
                 \evalstmt{\gamma}{s^*}{\gamma}\}
\end{tabular}

\bigskip
\noindent $\ruleset{H}$ includes

\begin{minipage}[t]{3in}
\begin{center}
\framebox{$\val{e^*}$}
\end{center}
\[\inferrule*[Right=V-Int]{$ $}{\val{intlit(i)}}
\]

\[\inferrule*[Right=V-True]{$ $}{\val{\mathit{true}}}
\]

\[\inferrule*[Right=V-False]{$ $}{\val{\mathit{false}}}
\]

\begin{center}
\framebox{$\typeexpr{\Gamma}{e^*}{ty}$}
\end{center}
\[
  \inferrule*[Right=T-eq]
    {\typeexpr{\Gamma}{e_1}{int} \\ 
     \typeexpr{\Gamma}{e_2}{int}}
    {\typeexpr{\Gamma}{\mathit{eq(e_1, e_2)}}{\mathit{bool}}}
\]

\[
  \inferrule*[Right=T-not]
    {\typeexpr{\Gamma}{e}{bool}}
    {\typeexpr{\Gamma}{\mathit{not(e)}}{\mathit{bool}}}
\]

\[
  \inferrule*[Right=T-add]
    {\typeexpr{\Gamma}{e_1}{\mathit{int}} \\ 
     \typeexpr{\Gamma}{e_2}{\mathit{int}}}
    {\typeexpr{\Gamma}{\mathit{add(e_1, e_2)}}{\mathit{int}}}
\]

\begin{center}
\framebox{$\typestmt{\Gamma}{s^*}{\Gamma}$}
\end{center}
\[\inferrule*[right=TS-decl]
    {\typeexpr{\Gamma}{e}{ty} \\
     \notpresenttype{\Gamma}{n}}
    {\typestmt{\Gamma}{\mathit{decl(n, ty, e)}}{\mathit{consty(n, ty, \Gamma)}}}
\]

\[\inferrule*[right=TS-ifte]
    {\typeexpr{\Gamma}{e}{\mathit{bool}} \\
     \typestmt{\Gamma}{s_1}{\Gamma'} \\
     \typestmt{\Gamma}{s_1}{\Gamma''}} 
    {\typestmt{\Gamma}{\mathit{ifte(e, s_1, s_2)}}{\Gamma}}
\]
\end{minipage}
\begin{minipage}[t]{3.5in}
\begin{center}
\framebox{$\vars{e^*}{ns}$}
\end{center}
\[\inferrule*[Right=VR-var]
    {$ $}
    {\vars{\mathit{var(n)}}{\{n\}}}
\]

\[\inferrule*[Right=VR-intlit]
    {$ $}
    {\vars{\mathit{intlit(i)}}{\emptyset}}
\]

\[\inferrule*[Right=VR-eq]
    {\vars{e_1}{vr_1} \\
     \vars{e_2}{vr_2}}
    {\vars{\mathit{eq(e_1, e_2)}}{(vr_1 \cup vr_2)}}
\]

\begin{center}
\framebox{$\evalexpr{\gamma}{e^*}{e}$}
\end{center}
\[\inferrule*[right=E-eq-True]
     {\evalexpr{\gamma}{e_1}{v_1} \\
      \evalexpr{\gamma}{e_2}{v_2} \\
      v_1 = v_2}
     {\evalexpr{\gamma}{\mathit{eq(e_1, e_2)}}{\mathit{true}}}
\]

\[\inferrule*[right=E-eq-False]
     {\evalexpr{\gamma}{e_1}{v_1} \\ 
      \evalexpr{\gamma}{e_2}{v_2} \\
      v_1 \not = v_2}
     {\evalexpr{\gamma}{\mathit{eq(e_1, e_2)}}{\mathit{false}}}
\]

\[\inferrule*[right=E-add]
     {\evalexpr{\gamma}{e_1}{\mathit{intlit}(i_1)} \\
      \evalexpr{\gamma}{e_2}{\mathit{intlit}(i_2)} \\
      \plus{i_1}{i_2}{i}}
     {\evalexpr{\gamma}{\mathit{add(e_1, e_2)}}{\mathit{intlit}(i)}}
\]

\begin{center}
\framebox{$\evalstmt{\gamma}{s^*}{\gamma}$}
\end{center}
\[\inferrule*[right=X-seq]
    {\evalstmt{\gamma}{s_1}{\gamma'} \\
     \evalstmt{\gamma'}{s_2}{\gamma''}} 
    {\evalstmt{\gamma}{\mathit{seq(s_1, s_2)}}{\gamma''}}
\]

\[\inferrule*[right=X-ifte-True]
    {\evalexpr{\gamma}{e}{\mathit{true}} \\
     \evalstmt{\gamma}{s_1}{\gamma'}} 
    {\evalstmt{\gamma}{\mathit{ifte(e, s_1, s_2)}}{\gamma'}}
\]
\end{minipage}

\caption{Relations introduced by the host language ($\relset{H}$) and
  selected rules defining them ($\ruleset{H}$).  The full set of rules
  may be found in Appendix~\ref{app:language}.}
\label{fig:host-rules}
\end{figure}

These relations must also be defined with respect to terms in the
relevant syntactic categories built using constructors introduced by
the list extension $L$ and security extension $S$, some rules for
which are shown in \fig{list-host-rules} and \fig{security-host-rules}
respectively. 
Here we see our extensions satisfy the constraint that they may only
add rules for the host language's relations (those in \relset{H}) such that
the primary component of the conclusion of the rule is a pattern matching
a construct introduced by that extension. This means extensions
may define host relations for their constructs only; they cannot change
the rules for constructs from the host language nor those introduced by other
extensions.
Note this also precludes extending relations over non-extensible
types, such as the relation for looking up types
$\lookuptype{\Gamma}{n}{ty}$ with its primary component being the
non-extensible type-context category $\Gamma$, as new constructors
cannot be introduced by the extension for which to define such
relations.
We see, for example, some rules for typing list constructs in
\fig{list-host-rules}, identifying
which list forms are values, what their variables are, and how they
are evaluated. The security extension introduces only one construct in
a host-language-introduced category ($s$) and thus provides rules in
\fig{security-host-rules} for typing and
evaluation for that one construct, the secure declaration form. These
rules encode behavior that mimic the ones for the constructs of the
host language. 

\begin{figure}
\noindent $\ruleset{L}$ includes

\begin{minipage}[t]{3in}
\begin{center}
\framebox{$\val{e^*}$}
\end{center}
\[\inferrule*[Right=V-nil]
    {$ $}
    {\val{\mathit{nil}}}
\]

\[\inferrule*[Right=V-cons]
    {\val{e_1} \\ 
     \val{e_2}}
    {\val{\mathit{cons(e_1, e_2)}}}
\]

\begin{center}
\framebox{$\typeexpr{\Gamma}{e^*}{ty}$}
\end{center}
\[\inferrule*[Right=T-null]
    {\typeexpr{\Gamma}{e}{\mathit{list(ty)}}}
    {\typeexpr{\Gamma}{\mathit{null(e)}}{\mathit{bool}}}
\]

\[\inferrule*[Right=T-tail]
    {\typeexpr{\Gamma}{e}{\mathit{list(ty)}}}
    {\typeexpr{\Gamma}{\mathit{tail(e)}}{\mathit{list(ty)}}}
\]
\end{minipage}
\begin{minipage}[t]{3in}
\begin{center}
\framebox{$\vars{e^*}{2^{n}}$}
\end{center}
\[\inferrule*[Right=VR-cons]
    {\vars{e_1}{vr_1} \\ 
     \vars{e_2}{vr_2}}
    {\vars{\mathit{cons(e_1, e_2)}}{(vr_1 \cup vr_2)}}
\]

\[\inferrule*[Right=VR-head]
    {\vars{e}{vr}}
    {\vars{\mathit{head(e)}}{vr}}
\]

\begin{center}
\framebox{$\evalexpr{\gamma}{e^*}{e}$}
\end{center}
\[\inferrule*[Right=E-cons]
    {\evalexpr{\gamma}{e_1}{v_1} \\ 
     \evalexpr{\gamma}{e_2}{v_2}}
    {\evalexpr{\gamma}{\mathit{cons(e_1, e_2)}}
                      {\mathit{cons(v_1, v_2)}}}
\]
\[\inferrule*[Right=E-head]
    {\evalexpr{\gamma}{e}{\mathit{cons(v_1, v_2)}}}
    {\evalexpr{\gamma}{\mathit{head(e)}}{v_1}}
\]
\end{minipage}

\begin{center}
\framebox{$\typestmt{\Gamma}{s^*}{\Gamma}$}
\end{center}
\[\inferrule*[Right=TS-splitlist]
             {\typeexpr{\Gamma}{e}{\mathit{list(ty)}} \\
               \lookuptype{\Gamma}{n_{hd}}{ty} \\
               \lookuptype{\Gamma}{n_{tl}}{\mathit{list(ty)}}}
             {\typestmt{\Gamma}{\mathit{splitlist}(n_{hd}, n_{tl}, e)}{\Gamma}}
\]

\begin{center}
\framebox{$\evalstmt{\gamma}{s^*}{\gamma}$}
\end{center}
\[\inferrule*[Right=X-splitlist]
             {\evalexpr{\gamma}{e}{\mathit{cons}(v_1, v_2)} \\
               n_{hd} \neq n_{tl}}
             {\evalstmt{\gamma}{\mathit{splitlist}(n_{hd}, n_{tl}, e)}
              {\consval{n_{hd}}{v_1}
                {\consval{n_{tl}}{v_2}{\gamma}}}}
\]

\caption{Selected rules given by the list extension ($\ruleset{L}$)
  for the relations introduced by the host language ($\relset{H}$).
  The full set may be found in
  Appendix~\ref{app:language}.} \label{fig:list-host-rules}
\end{figure}

\begin{figure}
$\ruleset{S}$ includes

\begin{center}
\framebox{$\typestmt{\Gamma}{s^*}{\Gamma}$}
\end{center}
\[\inferrule*[right=TS-secdecl]
    {\typeexpr{\gamma}{e}{ty} \\
     \notpresenttype{\Gamma}{n}}
    {\typestmt{\Gamma}{\mathit{secdecl(n, ty, sl, e)}}{\mathit{consty(n, ty, \Gamma)}}}
\]

\begin{center}
\framebox{$\evalstmt{\gamma}{s^*}{\gamma}$}
\end{center}
\[\inferrule*[Right=X-secdecl]
    {\evalexpr{\gamma}{e}{v}}
    {\evalstmt{\gamma}{\mathit{secdecl(n, ty, sl, e)}}{\consval{n}{v}{\gamma}}}
\]

\caption{Rules given by the security extension ($\ruleset{S}$) for the
  relations introduced by the host language
  ($\relset{H}$).} \label{fig:security-host-rules}
\end{figure}

\subsection{Viewing Extensions at a Distance via Projections} \label{subsec:viewByProj}

An extension may want to introduce new relations and to define them in
meaningful ways even in the presence of other extensions, even though
it does not know the details of these other extensions.
For example, the security extension may want to introduce new
relations as a vehicle for utilizing the 
functionality provided by the host language and other extensions in
the language library but with the additional provision that the
visibility of sensitive data can be monitored.
These relations would identify a static analysis similar to typing,
where the relations are expected to indicate that information
from private variables does not escape into public ones in a given
program. 

The first step in extensions introducing new relations is describing
how the relations are defined for constructs the extension knows,
those introduced by itself and the host language.
\fig{security-security-rules} shows relations the
security extension might introduce and define to build the kind of
static analysis described above. 
As for typing and values, there is a relation for looking up values in
a security context ($\lookupsec{\Sigma^*}{n}{sl}$) and one for
checking names are not assigned levels
($\notpresentsec{\Sigma^*}{n}$).
The security equivalent of typing an expression, written
$\exprlevel{\Sigma}{e^*}{sl}$, determines a security level $sl$,
either $\mathit{public}$ or $\mathit{private}$, for the expression
$e$ under the security context $\Sigma$.  This determines whether any
variables used in $e$ are assigned $\mathit{private}$ in $\Sigma$.
We can see, in rule \textsc{L-eq}, that this relation uses the
$\mathit{join}$ relation relating two security levels to the
more secure of the two.

The $\mathit{secure}$ relation, $\secure{\Sigma}{sl}{s^*}{\Sigma'}$,
determines that a statement $s$, its primary component, is ``secure''
with respect to information flow in a security context $\Sigma$ when
executed at a particular security level $sl$.  As with typing, this
relation also has a new security context $\Sigma'$ with the updated
security bindings from any declarations within the statement.
The intent of this analysis is to identify, conservatively, those
programs in which data from variables declared as $\mathit{private}$
(using the $\mathit{secdecl}$ construct) do not influence those that are
$\mathit{public}$.
We see the mapping of names to security levels in rules
\textsc{S-decl}, \textsc{S-secdecl-private}, and
\textsc{S-secdecl-public} in \fig{security-security-rules}, and how
the security level at which the statement will be executed affects the
assignments allowed in \textsc{S-assign-private}
and \textsc{S-assign-public}.

In determining if loops are secure, if a private variable is used in
the condition, no public variables may be assigned in the body as this
could leak information about the value of the private variable.  Thus
in rule \textsc{S-while} the security level at which the loop is
executed, $sl''$, is joined with the security level of the condition
($sl$) and the security level of the context of the loop itself
($sl'$) to determine the level at which the body of the loop will be
executed.  For example, if the condition contains a private variable
then the loop body must be secure in a $\mathit{private}$ context,
preventing any assignment to public variables.  A similar analysis is
done for conditionals, with rules in \refapp{language}.

\begin{figure}
\begin{tabular}{ll}
  \hspace*{-0.15in}
  \relset{S} = & \hspace*{-0.15in} \{\lookupsec{\Sigma^*}{n}{\mathit{sl}}, \ \
                   \notpresentsec{\Sigma^*}{n}, \ \
                   \joinlevel{\mathit{sl^*}}{\mathit{sl}}{\mathit{sl}}, \ \
                   \exprlevel{\Sigma}{e^*}{\mathit{sl}}, \\
               &   \secure{\Sigma}{\mathit{sl}}{s^*}{\Sigma}\}
\end{tabular}

\bigskip
\noindent $\ruleset{S}$ includes

\begin{minipage}[t]{3in}
\begin{center}
\framebox{$\exprlevel{\Sigma}{e^*}{\mathit{sl}}$}
\end{center}
\[\inferrule*[Right=L-int]
    { }
    {\exprlevel{\Sigma}{\mathit{intlit(i)}}{\mathit{public}}}
\]

\[\inferrule*[Right=L-var]
    {\lookupsec{\Sigma}{n}{\ell}}
    {\exprlevel{\Sigma}{\mathit{var(n)}}{\ell}}
\]

\[\inferrule*[Right=L-eq]
    {\exprlevel{\Sigma}{e_1}{\ell_1} \\
     \exprlevel{\Sigma}{e_2}{\ell_2} \\
     \joinlevel{\ell_1}{\ell_2}{\ell}}
    {\exprlevel{\Sigma}{eq(e_1, e_2)}{\ell}}
\]
\end{minipage}
\begin{minipage}[t]{3in}
\begin{center}
\framebox{$\joinlevel{\mathit{sl}*}{\mathit{sl}}{\mathit{sl}}$}
\end{center}
\[\inferrule*[Right=J-public]
    { }
    {\joinlevel{\mathit{public}}{\mathit{public}}{\mathit{public}}}
\]

\[\inferrule*[Right=J-private-l]
    { }
    {\joinlevel{\mathit{private}}{\ell}{\mathit{private}}}
\]

\[\inferrule*[Right=J-private-r]
    { }
    {\joinlevel{\ell}{\mathit{private}}{\mathit{private}}}
\]
\end{minipage}

\begin{center}
\framebox{$\secure{\Sigma}{\mathit{sl}}{s^*}{\Sigma}$}
\end{center}
\[\inferrule*[Right=S-seq]
    {\secure{\Sigma}{\ell}{s_1}{\Sigma'} \\
     \secure{\Sigma'}{\ell}{s_2}{\Sigma''} }
    {\secure{\Sigma}{\ell}{\mathit{seq}(s_1, s_2)}{\Sigma''}}
\]

\[\inferrule*[Right=S-decl]
    {\exprlevel{\Sigma}{e}{\mathit{public}} \\
     \notpresentsec{\Sigma}{n}}
    {\secure{\Sigma}{\mathit{public}}
      {\mathit{decl(n, ty, e)}}
      {\mathit{conssec(n, public, \Sigma)}}}
\]

\[\inferrule*[Right=S-assign-private]
    {\exprlevel{\Sigma}{e}{\ell} \\
     \lookupsec{\Sigma}{n}{\mathit{private}}}
    {\secure{\Sigma}{\ell'}
      {\mathit{assign(n, e)}}{\Sigma}}
\]

\[\inferrule*[Right=S-assign-public]
    {\exprlevel{\Sigma}{e}{\mathit{public}} \\
     \lookupsec{\Sigma}{n}{\mathit{public}}}
    {\secure{\Sigma}{\mathit{public}}
      {\mathit{assign(n, e)}}{\Sigma}}
\]

\[\inferrule*[Right=S-while]
    {\exprlevel{\Sigma}{e}{\ell} \\
     \joinlevel{\ell'}{\ell}{\ell''} \\
     \secure{\Sigma}{\ell''}{s}{\Sigma'}}
    {\secure{\Sigma}{\ell'}{\mathit{while(e, s)}}{\Sigma}}
\]

\[\inferrule*[Right=S-secdecl-private]
    {\exprlevel{\Sigma}{e}{\ell} \\
     \notpresentsec{\Sigma}{n}}
    {\secure{\Sigma}{\ell'}
      {\mathit{secdecl(n, ty, private, e)}}
      {\mathit{conssec(n, private, \Sigma)}}}
\]

\[\inferrule*[Right=S-secdecl-public]
    {\exprlevel{\Sigma}{e}{\mathit{public}} \\
     \notpresentsec{\Sigma}{n}}
    {\secure{\Sigma}{\mathit{public}}
      {\mathit{secdecl(n, ty, public, e)}}
      {\mathit{conssec(n, public, \Sigma)}}}
\]

\caption{Selected rules in $\ruleset{S}$ for security relations
  introduced by $S$ ($\relset{S}$).  The full set of rules can be
  found in Appendix~\ref{app:language}.} \label{fig:security-security-rules}
\end{figure}

Observe the security extension defines the \emph{level} and
\emph{secure} relations over statements and expressions; these are
their respective primary components and are syntactic
categories introduced by the host language.
Because only the extension introducing a relation is aware of its
existence, the extension must bear the burden of defining it
completely.  This includes introducing rules defining it for
constructs introduced by the host language, as we have seen with
rules like \textsc{L-var} and \textsc{S-while}.
However, having noted this, we immediately see a difficulty in
defining a new relation, such as $\mathit{secure}$, completely.
There may be several different extensions participating in a library
and modularity requires they must not use knowledge of each other
in their construction,
yet multiple extensions may participate in a composite language.
How then is a relation defined by one extension to be assessed with
respect to expression forms introduced by another extension?
With specific reference to the example at hand, how are we to perform
a security analysis of statement and expression forms
introduced by the list extension?

To address this issue, our framework includes a mechanism, inspired by
forwarding in attribute grammars~\cite{vanwyk2002cc},
for viewing extensions ``at a distance.''
One component of this mechanism is a \emph{projection} of extension
constructs that will then be the basis
for their view in other extensions.
Each extensible category in the host language has a projection relation relating
a term built by an extension-introduced construct to its projection in
the same category.  These projection relations are identified in the
fifth item ($\transrelset{}$) of its tuple $\langle \ntset{},
\constrset{}, \relset{}, \ruleset{}, \transrelset{}, \transruleset{},
\trset{} \rangle$.  The existence of a projection relation in
$\transrelset{}$ is, in
fact, what determines in our framework whether a category is
extensible:  those with projection relations are extensible, those
without are not.
Extensions do not introduce projection relations,
and thus this part of the specification is empty for them.
Of course, the projection relation must be defined to be useful.
The rules that do so constitute the sixth item
($\transruleset{}$) in the specification of an
extension; note this part of the language specification is
irrelevant to the host language.
The final piece to the mechanism is a description of how relations
defined by an extension should be defined for constructs from other
extensions.
This part, which is meaningful only to extensions that are introducing
new relations, is represented by the $\trset{}$ item in the language
specification.
Specifically, $\trset{}$ comprises a collection of
\emph{projection rules} that identify the definitions of the relevant
relations, each rule using the projection relation for the category of
the relation's primary component.
At a technical level, projections in such rules are permitted only on
the primary component of the relation and this component is required
to be a schematic variable in the conclusion of the rule.  Thus these
rules copy the definition of the relation from the projection.

We illustrate this aspect of the framework in
\fig{projection} by showing the projection relations introduced by the
host language and the rules defining them in the extensions, as well
as completing the
definitions of the relations described by the security extension.
The host language
provides the projection relations $\projectexpr{e}{e}$ for expressions,
$\projectstmt{s}{s}$ for statements, and $\projectty{ty}{ty}$ for types
in \transrelset{H} shown in
\fig{projection}.
Because the other syntactic categories introduced by the host language
are not intended to be extensible, we do not introduce projection
relations for them.
The projection relations in our example language take the most basic
form of projection relations and do not involve any additional
arguments, but this can be useful in some cases. For example, a host
language might permit the use of an expression's type in determining its
projection, and in that case the relation for expressions might
include a typing context argument to permit determining an
expression's type (this would be written as \projectexpr{\Gamma,e}{e}).

Consider the projection rule \textsc{P-splitlist} for relation
$\mathit{proj_s}$ for the list-splitting construct \emph{splitlist}
introduced by $L$ in \transruleset{L} in \fig{projection}. A
representation of a list
split $\mathit{splitlist(n_{hd}, n_{tl}, e)}$ is projected to an
encoding of a
sequence of assignments accomplishing the splitting of the list.
The $\mathit{secure}$ relation is then defined for the
$\mathit{splitlist}$ construct by taking its projection to the
constructs representing this sequence
of assignments, checking the security of that sequence, then using
it to define security for the $\mathit{splitlist}$ construct.
This is achieved
by the projection rule \textsc{P-secure} that indicates a statement
$s$ is secure if its projection $s'$ is secure.

There is a subtlety concerning projections illustrated by the
rules for list expression forms.
These rules project expression forms contributed by the extension to
seemingly unrelated ones in the host language: for example, 
$\mathit{null(e)}$ is projected to $e$ and $\mathit{cons}(e_1,e_2)$ is
projected to $\mathit{eq}(e_1,e_2)$.
The way to understand these projections is that their purpose is to
preserve enough information about the expressions built using the
contructors from the extension so analyses described by other
extensions through the projections are still meaningful.
Thus, consider the $\mathit{level}$ relation from the security
extension.
This is intended to assess the security status of an expression which,
in the case of $\mathit{cons}(e_1,e_2)$, would depend on the
subexpressions $e_1$ and $e_2$.
Thus the key to a good projection in this case is that these
subexpressions be preserved.
Of course, an extension does not have a direct means for assessing the
kinds of analyses other extensions might want to carry out.
We will see later how meta-theoretic properties can help
identify the desiderata for good projections. 

\begin{figure}
  \noindent
  Projection relations in $\transrelset{H}$:  $\{ \proj{e} : e,\ \proj{s} : s,\ \proj{ty} : ty \}$ \\
  \\
  \noindent
  Projection relation rules in $\transruleset{L}$: \\ \noindent
  \begin{minipage}[t]{0.28\textwidth}
  \[\inferrule*[Right=P-null]
      { }
      {\projectexpr{\mathit{null(e)}}{e}}
  \]

  \[\inferrule*[Right=P-head]
      { }
      {\projectexpr{\mathit{head(e)}}{e}}
  \]
  \end{minipage}
  \begin{minipage}[t]{0.28\textwidth}
  \[\inferrule*[Right=P-tail]
     { }
     {\projectexpr{\mathit{tail(e)}}{e}}
  \]

  \[\inferrule*[Right=P-nil]
      { }
      {\projectexpr{\mathit{nil}}{\mathit{true}}}
  \]
  \end{minipage}
  \begin{minipage}[t]{0.4\textwidth}
  \[\inferrule*[Right=P-cons]
      { }
      {\projectexpr{\mathit{cons}(e_1, e_2)}{\mathit{eq}(e_1, e_2)}}
  \]

  \[\inferrule*[Right=P-list]
      { }
      {\projectty{\mathit{list}(ty)}{ty}}
  \]
  \end{minipage}
  \\[10pt]
  \[\inferrule*[right=P-splitlist]
      {n_{hd} \neq n_{tl}}
      {\proj{s}(\mathit{splitlist}(n_{hd}, n_{tl}, e), \\\\
          seq(seq(assign(n_{hd}, e), assign(n_{tl},
             tail(var(n_{hd})))), assign(n_{hd}, head(var(n_{hd})))))}
      \]

  \vspace{0.4cm}

  \noindent
  Projection relation rule in $\transruleset{S}$:
  \[\inferrule*[Right=P-secdecl]{ }{\projectstmt{\mathit{secdecl}(n, ty,
    sl, e)}{\mathit{decl}(n, ty, e)}}\]

  \noindent
  Projection rules in $\trset{S}$: \\
  \begin{longtable}{b{0.45\textwidth}b{0.55\textwidth}}
    \vspace*{-0.5in}
  \[\inferrule*[right=P-level]
      {\projectexpr{e}{e'} \\
       \exprlevel{\Sigma}{e'}{\ell}}
      {\exprlevel{\Sigma}{e}{\ell}}
  \]
  &
  \vspace*{-0.5in}
  \[\inferrule*[right=P-secure]
      {\projectstmt{s}{s'} \\
       \secure{\Sigma}{sl}{s'}{\Sigma'} }
      {\secure{\Sigma}{sl}{s}{\Sigma'}}
  \]
  \end{longtable}
  \vspace*{-0.55in}

\caption{Projection relations and rules defining them and projection rules for host $H$
  and extensions $L$ and $S$.  Note $\transrelset{L}$,
  $\transrelset{S}$, $\transruleset{H}$, $\trset{H}$, and $\trset{L}$
  are all empty and thus not shown.} \label{fig:projection}
\end{figure}

\subsection{Well-Formedness and Language Composition}
\label{sec:framework-composing-languages}

We conclude this section by making precise the preceding informal
description of the extensibility framework.
The framework is exercised by describing a \emph{language library}
that comprises a single host language and a collection of 
extensions via tuples of the form
$\langle \ntset{}, \constrset{}, \relset{}, \ruleset{}, \transrelset{},
\transruleset{}, \trset{} \rangle$.
We assume the host language to be given by the tuple
$\langle \ntset{H}, \constrset{H},
\relset{H}, \ruleset{H}, \transrelset{H}, \transruleset{H}, \trset{H}
\rangle$.
We describe below the constraints that must be satisfied for
such a collection to be deemed well-formed and we then explain how a
standalone language description is to be constructed from the
composition of a collection of extensions with the host language.

\subsubsection{Well-Formedness for Host Language and Extension Descriptions}
\label{sec:framework-wellformedness}

In the very first instance, there must be a unique
identification for every constructor and relation that is identified
in the specification of the host language and the different
extensions.
Thus, the constructors identified by $\constrset{}$ in each
component must be distinct as also must be the relations identified by
$\relset{}$ and $\transrelset{}$.
The specification of constructors in the
$\constrset{}$ set of the components and of predicates in the $\relset{}$ and
$\transrelset{}$ sets impose typing constraints on the symbols identified.
Well-formedness requires that these symbols be used in a way
respecting their typing throughout the language library's
specifications.

There is one further requirement of the host language specification:
$\transrelset{H}$ identifies at most one projection relation for each
category in $\ntset{H}$.
These relations associate terms in a particular category with their
projections, but may do so in a manner that is parameterized by other
terms, which may be used in defining projection rules as we
explain below.
A syntactic category that does not have an associated projection
relation is considered to be non-extensible.

An extension specification for an extension $E$, on the other hand,
must satisfy
conditions beyond the basic ones mentioned at the outset to be deemed
well-formed.
For an extension $E$, its specification being
$\langle \ntset{E}, \constrset{E},
\relset{E}, \ruleset{E}, \transrelset{E}, \transruleset{E}, \trset{E}
\rangle$, these conditions are the following:
\begin{enumerate}
\item For each constructor $c$ in $\constrset{E}$ building a syntactic
  category introduced by the host language
  ($\mathit{category}(c)\in\ntset{H}$), the category is an extensible
  category (there is a projection relation in $\transrelset{H}$ for
  $\mathit{category}(c)$).

\item
  For a rule in $\ruleset{E}$ that has $R(t_1,\ldots, t_n)$ as its
  conclusion where $R \in \relset{H}$ and the $i^{\rm th}$ 
  argument of $R$ is its primary component, it must be the case that
  $t_i$ has a constructor in $\constrset{E}$ as its top-level symbol.
  This is a formal rendition of the requirement that an extension
  must not modify the definition of a relation introduced by the host
  language as it pertains to host language constructs. 

\item For each rule introduced by an extension defining a projection
  (rule in $\transruleset{E}$), it must be the case that the
  projecting argument has a constructor in $\constrset{E}$ as its
  top-level symbol.

\item For each $R \in \relset{E}$ that has a host-language-introduced
  syntactic category as its primary component, there must be a rule
  for $R$ in the set of projection rules $\trset{E}$. 

\item Each rule in $\trset{E}$ for a relation $R$ whose primary
  component is its $i^{\rm th}$ argument must have the form
  \[
    \inferrule{T(x, x') \\
               R(t_1,\ldots, t_{i-1}, x', t_{i+1},\ldots, t_n)}
              {R(t_1,\ldots, t_{i-1}, x,  t_{i+1},\ldots, t_n)}
  \]
  where $x$ and $x'$ are schematic variables and $T$ is to be read as
  a predicate that uses the relevant projection relation from
  $\transrelset{H}$ to indicate that $x$ projects to $x'$ under
  conditions determined by the additional parameters for the relation.
  For example, if $\proj{}$ is a three-place projection relation whose
  second and third arguments identify the projected term and its
  projection and the first argument is an auxilary parameter, then
  $\proj{}(s,x,x')$ may be the instantiation of $T(x,x')$, where $s$
  is a term whose structure identifies a dependency on the conclusion
  of the rule. 
\end{enumerate}
Note we do not require an extension to introduce rules defining the
projection for the constructors it introduces.  Choosing not to define
the projection relation for a new constructor limits the extensibility
of the language, as relations introduced by other extensions cannot
hold for the new syntax.  This is, however, a choice left to an
extension writer.

Well-formedness as defined above can be statically assessed and in
isolation for the host language specification and with knowledge of
the host language alone for an extension specification. 
This is an important feature: the well-formedness of a component in 
an open language library should be assessible as it is
developed, independently of other components.
We will assume the specifications in a library have been
determined to be well-formed before any meta-theoretic analysis
related to them is undertaken. 
This is similar to an approach to determining independently the
well-formedness of language components specified using attribute
grammars~\cite{kaminski17phd,kaminski2012sle}.

\subsubsection{Language Composition}
\label{sec:framework-composition}
The host language $H$ and a collection of extensions $E_1,\ldots, E_n$
in a library can be composed to form a \emph{complete} language by the
process we describe below.
The language that results from such a composition is denoted by
$\langComp{H}{\oneTo{E}{n}}$.
A complete language is represented by a tuple of the form $\langle
\ntset{}, \constrset{}, \relset{}, \ruleset{} \rangle$.
The components \transrelset{}, \transruleset{}, and \trset{} in
specifications impact the composition process but they have do not
have a role in a complete language.

The composition process will \emph{instantiate} projection rules from one
extension with new constructors from another to specify how a relation
introduced by the former is defined on the constructs introduced by the
latter.
These instantiated rules, denoted \ruleset{\trset{}}, will be part of
\ruleset{} and are defined as follows:
\begin{tabbing} 
xxxxxxxxx\=\kill
\> $\ruleset{\trset{}} := \{\ r[c(\overline{y})/x] \ | \ $ \=
        $i \in \{1..n\},  j \in \{1..n\}, i \neq j,
           r \in  \trset{E_i},~ c \in \constrset{E_j},
           \mathit{category}(c) \in \ntset{H}$ \\
\> \> $\mathit{category}(x) = \mathit{category}(c)$, $\overline{y}$ is
             a sequence of variables not\\           
\>\>  appearing in $r$ of length the arity of $c\}.$
\end{tabbing}
Here $r[c(\overline{y})/x]$ represents the rule that results from a
projection rule by replacing the schematic variable $x$ that appears
in the primary component position in the conclusion of the rule with
the term $c(\overline{y})$.
Observe this instantiation is done only for extension constructs
for host language categories, and only when doing so is syntactically
valid ($\mathit{category}(x) = \mathit{category}(c)$).

An example is perhaps useful in understanding the intent underlying
the instantiation of projection rules.
Recall the list \emph{head} constructor in
$\constrset{L}$ and the \emph{level} projection rule \textsc{P-level}
in $\trset{S}$ from \fig{projection}.
Both are over the host category $e$
for expressions. The instantiation process above will create a new rule
for \emph{level} over the \emph{head} construct in \ruleset{\trset{}}:
\[
  \inferrule*[Right=L-head]
     {\projectexpr{\mathit{head}(e)}{e'} \\
      \exprlevel{\Sigma}{e'}{\mathit{sl}}} 
     {\exprlevel{\Sigma}{\mathit{head}(e)}{\mathit{sl}}}
\]
This illustrates how we apply the projection rules to new syntax to complete
the definitions of the extension-introduced relations.
Note that in the language composition each instance of a projection rule
that is added has its primary component built by a specific
constructor introduced by a different extension.

Given \ruleset{\trset{}}, the definition of the complete language
$\langle \ntset{}, \constrset{}, \relset{}, \ruleset{} \rangle$
resulting from the composition $\langComp{H}{\oneTo{E}{n}}$ is
straightforward:
\begin{itemize}
\item $\ntset{} := \ntset{H} \cup \bigcup_{i=1}^{n} \ntset{E_i}$.  The
  set of syntax categories is the union of all the syntax categories
  in the host language and all the extensions.

\item $\constrset{} := \constrset{H} \cup \bigcup_{i=1}^{n} \constrset{E_i}$.
  The set of syntax constructors is the union of all the syntax
  constructors in the host language and all the extensions.

\item $\relset{} := \relset{H} 
  \cup \transrelset{H}
  \cup \bigcup_{i=1}^{n}\relset{E_i}$.
  The set of relations in the composed language is the union of all
  the semantic relations in the host language and all the extensions, both the regular
  relations and the projection relations. 
 
\item $\ruleset{} := \ruleset{*} \cup \transruleset{*} \cup
\ruleset{\trset{}}$ where
\begin{itemize}
\item $\ruleset{*} := \ruleset{H} \cup \bigcup_{i=1}^{n} \ruleset{E_i}$.
  This is all the semantic relation rules in the host language and all
  the extensions.

\item $\transruleset{*} := \bigcup_{i=1}^{n} \transruleset{E_i}$. This
  is the collection of all the projection rules in all the extensions.

\end{itemize}

The construction of the set of relation rules $\ruleset{}$ 
includes
the rules from three sources: the rules given by the host language and
extensions in
their respective rule sets ($\ruleset{*}$), the projection
relation rules from the extensions ($\transruleset{*}$), and the
instantiations of the projection rules for constructors from
distinct extensions in \ruleset{\trset{}} as defined above.

\end{itemize}

In the sequel, we will need to talk about the language identified by a
library component.
If that component is the host language $H$, this will mean the
language $\langComp{H}{\emptyset}$.
If it is an extension $E$ (and $H$ is the host language), then it will
mean the language $\langComp{H}{\setize{E}}$.

\section{Metatheoretic Properties and Modular Reasoning}
\label{sec:problem}

As we have seen in the previous section, language specifications begin
with a description of syntax and then extend into the presentation of
semantics for expressions.
One objective of the semantic component is to formalize
the execution behavior of programs and program fragments; this is the
role, for instance, of the evaluation relation on expressions and statements
in the example in \sec{framework}.
However, semantic relations have an important additional purpose:
they provide for a statically-determined
attribution of properties to expressions that are intended to
translate into assurances of good dynamic behavior.
For example, the property of well-typing for a program
is supposed to guarantee its execution will be free of
type-related errors.
There is, however, a distance between the results of static analyses
and what actually happens with a program at runtime.
Metatheoretic properties, which relate different semantic features
of programs, are a means for closing this gap.
For example, in translating well-typing of programs into the intended
guarantee of their error-free execution, it is important to
\emph{verify} that types of expressions do not change under
evaluation. 
The metatheoretic property of type preservation relating
the typing and evaluation relations for program fragments has
exactly this purpose.

We are interested in this paper in a broad exploitation of static
analyses in the style described above.
In this context, metatheoretic properties will relate relations
over syntactic expressions. 
We will specifically focus on proving such properties that are
expressed via formulas of the form
\begin{equation*}
  \label{eq:propForm}
  \forallx{\overline{x}}{R(\overline{t}) \imp F}
\end{equation*}
where $\overline{x}$ is a sequence of variables, $R(\overline{t})$ is
an atomic formula that has $R$ as its predicate or relation symbol and
the sequence of terms $\overline{t}$ as its arguments and is
constructed using the variables in $\overline{x}$ and a 
vocabulary identified by specifications of the kind discussed in the
previous section, and $F$ is an arbitrary formula constructed
over the same vocabulary and variables.
We shall refer to the relation $R(\overline{t})$ as the \emph{key
  relation} of such a property.
We restrict attention to formulas in this form for simplicity in
discussion.
Note, however, that such a formula is equivalent to one with
the more general structure
$\forallx{\overline{x_1}}
          {F_1 \imp \cdots \imp
            \forallx{\overline{x_m}}
                    {F_m
                        \supset F}}$
where $F_i$ is $R(\overline{t})$ for some $i$ between $1$ and
$m$. Thus, our discussions in reality extend to formulas with this
richer syntax.

Proofs of the kind of properties in which we are interested must depend
on the definitions of the relations that participate in them.
Of particular importance is the manner in which the definition of the
relation $R$ plays a role in the argument.
There are two important ingredients to the typical treatment of this
relation when it is defined in a rule-based and syntax-directed
fashion.
First, it is assumed the relation is completely specified by the
rules pertaining to it, \ie{} that a closed-world assumption
applies to these rules.
This assumption justifies a case analysis style of argument based on
the definition of the relation.
Second, for the argument to be effective, it is usually necessary to
invoke induction with respect to the relation.
This induction is based, again, on the definition of the relation: the
assumption generally is the property of interest holds for any
instance of the relation $R$ that appears in the definition of the
instance for which it is being proved.\footnote{\label{fn:categories}
  Inductive
  arguments are often based on the structure of expressions in the
  syntactic category corresponding to one of the variables in
  $\overline{x}$. This is a special case of what is described here:
  these situations can be   visualized as ones in which the structure
  of the syntactic category is made explicit via a predicate and the
  property to be proved is relativized to the validity of that
  predicate.}

In the traditional setting for such proofs, the definition of a
language is assumed to be provided in its entirety at the outset. 
The closed-world assumption has an obvious validity in this
situation.
However, this assumption breaks down in our language extensibility
framework.
In this context, the identification of syntactic 
constructs, as well as the relations over them, is typically distributed
across language components that come together at a much later point in the
overall design.
This is too late a stage to be contemplating metatheoretic
properties or proofs for them for at least two reasons.
First, metatheoretic properties often play an intrinsic role in the
design of the language and are not meant to be ``stuck on'' after the
fact. 
Second, delaying the verification task till the point of composition
means the original designers of the particular language component,
who likely understand its functionality the best, will not have a role
to play in it.

The main thrust of this work is to address the issue
identified above.
We first consider the properties we have identified as
foundational ones.
We assume these properties to be spelled out by the host
language, based on semantic relations that are identified when the
framework starts to be elucidated. 
For example, the host language specified in \sec{framework} may
introduce the following property using as
atomic formulas relations it has introduced:
\begin{multline}
  \label{eq:exprTypePreservation}
  \forallx
  {\Gamma,e,\mathit{ty}_e,\gamma,v_e}
  {\evalexpr{\gamma}{e}{v_e} \imp
   \typeexpr{\Gamma}{e}{\mathit{ty}_e} \imp \\
   (\forallx{x,\mathit{ty}_x,v_x}
    {\lookuptype{\Gamma}{x}{\mathit{ty}_x} \land 
     \lookupval{\gamma}{x}{v_x} \imp
     \typeexpr{\mathit{nilty}}{v_x}{\mathit{ty}_x}}) \imp
   \typeexpr{\mathit{nilty}}{v_e}{\mathit{ty}_e}}
\end{multline}
This is a type preservation property over expressions: it asserts that
if an expression $e$ evaluates to a value $v_e$, $e$ has 
type $\mathit{ty}_e$, and the evaluation context respects the typing
determined by the typing context, then $v_e$ must also have type
$\mathit{ty}_e$. 
As we have noted, the proof of a property of the form
$\forallx{\overline{x}}{R(\overline{t}) \imp F}$
will usually call for a case analysis over the definition of the relation
$R$.
We propose to modularize the proof effort by delegating the
responsibility for dealing with the case for each  particular
syntactic construct to the component, whether the host language or an
extension, that introduces it. 
Thus, in the example presented, the argument for the cases of the
evaluation relation for different expression forms will be constructed
independently within the host language and the list and security
extensions.
For this idea to work, there must of course be a mechanism for
automatically combining the individual proof fragments into an actual
proof of the property for a complete language.
Care is needed in the construction of the proof fragments for this to
be possible.
For instance, a component that develops a proof fragment must be
sensitive to the fact that it has only a partial view of the full
language. 
Thus, any further case analysis that is needed must use a properly
delimited version of the closed-world assumption.
Fortunately, it is possible to describe criteria that are
automatically enforceable and not unduly restrictive
towards ensuring this property, as we show in
Section~\ref{sec:foundationalProperties}. 

We next consider auxiliary properties, those introduced by
extensions.
As a concrete example of why such a capability may be useful, consider
the security extension described in \sec{framework} that provides
mechanisms for dealing with sensitive data while using the
programming capabilities afforded by the host language and other extensions.
That extension also introduces the \emph{secure} relation that encapsulates
a static analysis meant to ensure the secrecy of data in the
course of execution. 
The soundness of this analysis needs, of course, to be validated.
Letting \eqpublicvals{\Sigma}{\gamma_A}{\gamma_B} be a shorthand for
the formula
\begin{align*}
  \forallx{x}{\lookupsec{\Sigma}{x}{\mathit{public}} \imp{} &
    (\forallx{v}{\lookupval{\gamma_A}{x}{v} \imp \lookupval{\gamma_B}{x}{v}})} \wedge \\
  & (\forallx{v}{\lookupval{\gamma_B}{x}{v} \imp \lookupval{\gamma_A}{x}{v}})
\end{align*}
we may encapsulate this soundness in the following metatheoretic property:
\begin{multline} \label{eq:secure-correct}
  \forallx{s,\Sigma,sl,\gamma_A,\gamma_B,\Sigma',\gamma_A',\gamma_B'}
          {\evalstmt{\gamma_A}{s}{\gamma_A'}\  \imp \
               \evalstmt{\gamma_B}{s}{\gamma_B'}\  \imp \
                      \secure{\Sigma}{sl}{s}{\Sigma'}\  \imp} \\
  \eqpublicvals{\Sigma}{\gamma_A}{\gamma_B}\ \imp\ 
  \eqpublicvals{\Sigma'}{\gamma_A'}{\gamma_B'}
\end{multline}
This formula formalizes the fact that information from private
variables does not leak into public data during the evaluation of a
program that satisfies the \emph{secure} relation: the formula holds
if it is the case that, when a ``secure'' statement is executed
in two different states ($\gamma_A$ and $\gamma_B$) that assign the
same values to public variables
(\eqpublicvals{\Sigma}{\gamma_A}{\gamma_B}) but possibly differing
values to private variables, then the resulting states will also
assign the same values to the public variables
(\eqpublicvals{\Sigma}{\gamma'_A}{\gamma'_B}).
Now, the proof of such a property can obviously not be
distributed to the different components in a composite language. 
However, the extension introducing the property would also not quite
be able to construct a proof of it by itself because  it does not have
a priori knowledge of all the extensions with which it will be composed.
The approach we develop for overcoming this difficulty uses
projections to describe constraints on the behavior of extensions.
More specifically, the host language identifies formulas
relating the attributes associated with extension constructs to those of
the host language constructs to which they project.
The task of verifying these formulas hold is then delegated to
each extension that builds on the host language. 
Once such properties have been established, the extension responsible
for proving an auxiliary property gets to use them in the course of
reasoning about the behavior of extensions about which it otherwise knows
nothing.
We develop a structure for modular reasoning based on these
ideas and establish its soundness in Section~\ref{sec:auxiliaryProperties}. 

An important aspect of our soundness arguments for the reasoning
structures we develop, whether for foundational or for auxiliary
properties, is that they are constructive: in particular, they
yield a method for assembling proofs developed independently for
each component into proofs for the relevant properties for the
complete language once its parts have been identified.

\section{A Structure for Reasoning about Relational Specifications}
\label{sec:abella}

The technical developments in this paper 
require us to be more specific about the structure of proofs.
Towards this end, we will frame such reasoning within a logic called 
$\cal G$~\cite{gacek11ic} that provides the basis for the 
Abella proof assistant~\cite{baelde14jfr}.
We present this logic below in a manner suitable for its use in this 
paper, in the process also motivating its choice for the kind of
metatheoretic reasoning that is of interest.
We will utilize the consistency result for the logic~\cite{gacek11ic}
implicitly in our soundness arguments in later sections.
In particular, we will read soundness for the methodology that we
articulate as the ability to generate a proof in the logic for a
metatheorem for a complete language from the (partial) proofs that
are provided by each participating component.

While $\cal G$ is a logic that is based on the simply-typed lambda
calculus and includes logical devices for treating binding notions,
only a limited, first-order version of it is needed in this work and
we will therefore present it as such.
From this perspective, its collection of terms and atomic formulas are 
determined in the usual way for a typed first-order logic once a set
of sorts and a set of constant and predicate or relation symbols with  
associated types have been identified. 
There is a natural correspondence between the sorts in the logic and
the syntactic categories identified in the component $\ntset{}$ of a
language specification in the style of Section~\ref{sec:framework}.
Further, this correspondence extends to one between the constructors
in the collection $\constrset{}$ in the language specification and the
constants in the logic and, similarly, the relations in
$\relset{}$ and $\transrelset{}$ in the language specification and 
predicate symbols in the logic. 
Arbitrary formulas are constructed from atomic formulas using the
logical constants $\top$ and $\bot$; the (infix) connectives $\land$,
$\lor$, and $\supset$; and existential and universal quantification.
Formulas of the last two kinds are written as
$\existsx{x:\alpha}{F}$ and $\forallx{x:\alpha}{F}$, respectively,
where $\alpha$ denotes the sort associated with the variable.
While it is essential to identify the types associated with quantified
variables, we shall often assume these can be determined from the
context, writing the quantified formulas as $\existsx{x}{F}$
and $\forallx{x}{F}$ instead, and we will also abbreviate formulas of 
the form ${\cal Q}{x_1}.\ldots.{\cal Q}{x_n}.F$ where $\cal Q$ is
$\exists$ or $\forall$ as ${\cal Q}{x_1,\ldots,x_n}.F$.
The logic treats as indistinguishable two formulas that can be made
identical by a consistent, capture-avoiding renaming of the variables
that are bound by quantifiers in them, a fact that will be used
implicitly in our discourse.

\begin{definition}[Substitutions, Unification] \label{def:substapp}
A substitution identifies a
finite sequence of terms $t_1,\ldots,t_n$ over the vocabulary
determined by a suitable language $L$ with a sequence of variables
$x_1,\ldots,x_n$ that are pairwise of the same type; such a
substitution is denoted by $\{\langle x_1,t_1 \rangle, \ldots, \langle
x_n,t_n\rangle\}$ and it has $x_1,\ldots, x_n$ as its domain and
$t_1,\ldots,t_n$ as its range.
The application of a substitution $\theta$ to a term $t$ or a formula
$F$, written, respectively, as $\appsubst{\theta}{t}$ and
$\appsubst{\theta}{F}$, corresponds to the replacement in $t$ or $F$
of the variables identified by $\theta$ by the corresponding terms,
taking care in the latter case to rename quantified variables to avoid
inadvertent capture. 
The composition of two substitutions $\theta_1$ and $\theta_2$,
denoted by $\substcomp{\theta_1}{\theta_2}$ is a substitution $\theta$ such
that $\appsubst{\theta}{e} =
\appsubst{\theta_2}{\appsubst{\theta_1}{e}}$ for any term or formula
$e$; such a substitution can always be provided explicitly.
A unification problem $\cal U$ is a collection $\{\langle e^1_1,
e^2_1\rangle,\ldots, \langle e^1_m, e^2_m \rangle \}$ of pairs of
atomic formulas or terms of identical type.
A unifier for $\cal U$ is a substitution $\theta$ such that
$\appsubst{\theta}{e^1} = \appsubst{\theta}{e^2}$ for every pair
$\langle e^1,e^2\rangle$ in $\cal U$.
$\cal U$ is said to be solvable and the pairs of terms in $\cal U$ are
said to be unifiable if there is a unifier for $\cal U$.
A most general unifier, or \mgu, for $\cal U$ is a
substitution $\theta$ from 
which any other unifier $\theta'$ for $\cal U$ can be obtained by
composition with another substitution, \ie{} there is a substitution
$\theta''$ such that $\theta' = \substcomp{\theta}{\theta''}$.
It is a known fact that, in the context of interest, $\cal U$ has a
most general unifier if it is solvable.
\end{definition}

\begin{figure}[htbp]
  \begin{center}
    \[
    \begin{array}{c c c}
      \inferrule*[right={\id, $A$ {\rm atomic}}] 
                 {\hspace*{0.1in}}
                 {\sequent{\Sigma}{\Gamma,A}{A}} &
      \inferrule*[right=\cut]
                 {\sequent{\Sigma}{\Gamma}{B} \\
                   \sequent{\Sigma}{B,\Delta}{C}}
                 {\sequent{\Sigma}{\Gamma,\Delta}{C}} &
      \inferrule*[right=\contL]
                 {\sequent{\Sigma}{\Gamma,B,B}{C}}
                 {\sequent{\Sigma}{\Gamma,B}{C}}
    \end{array}
    \]
    \[
    \begin{array}{c c}
      \inferrule*[right=\botL]
                 {\hspace*{0.1in}}
                 {\sequent{\Sigma}{\Gamma,\bot}{C}} &
      \inferrule*[right=\topR]
                 {\hspace*{0.1in}}
                 {\sequent{\Sigma}{\Gamma}{\top}} \\ \\
      \inferrule*[right=\orL]
                 {\sequent{\Sigma}{\Gamma, B}{C} \\
                   \sequent{\Sigma}{\Gamma, D}{C}}
                 {\sequent{\Sigma}{\Gamma, B \vee D}{C}} \quad & \quad
      \inferrule*[right={\orR$_i$, $i\in\setize{1,2}$}]
                 {\sequent{\Sigma}{\Gamma}{B_i}}
                 {\sequent{\Sigma}{\Gamma}{B_1 \vee B_2}} \\ \\
      \inferrule*[right={\andL$_i$, $i\in\setize{1,2}$}]
                 {\sequent{\Sigma}{\Gamma,B_i}{C}}
                 {\sequent{\Sigma}{\Gamma,B_1 \wedge B_2}{C}} &
      \inferrule*[right=\andR]
                 {\sequent{\Sigma}{\Gamma}{B} \\
                   \sequent{\Sigma}{\Gamma}{C}}
                 {\sequent{\Sigma}{\Gamma}{B \wedge C}} \\ \\
      \inferrule*[right=\impL]
                 {\sequent{\Sigma}{\Gamma}{B} \\
                   \sequent{\Sigma}{\Gamma,D}{C}}
                 {\sequent{\Sigma}{\Gamma,B \imp D}{C}} &
      \inferrule*[right=\impR]
                 {\sequent{\Sigma}{\Gamma,B}{C}}
                 {\sequent{\Sigma}{\Gamma}{B \imp C}} \\ \\
      \inferrule*[right=\forallL]
                 {\sequent{\Sigma}{\Gamma,\appsubst{\setize{\langle{}x,t\rangle}}{B}}{C}}
                 {\sequent{\Sigma}{\Gamma,\forallx{x:\alpha}{B}}{C}} &
      \inferrule*[right={\forallR, $x \notin \Sigma$}]
                 {\sequent{(\Sigma,x:\alpha)}{\Gamma}{B}}
                 {\sequent{\Sigma}{\Gamma}{\forallx{x:\alpha}{B}}} \\ \\
      \inferrule*[right={\existsL, $x \notin \Sigma$}]
                 {\sequent{(\Sigma,x:\alpha)}{\Gamma,B}{C}}
                 {\sequent{\Sigma}{\Gamma,\existsx{x:\alpha}{B}}{C}} &
      \inferrule*[right=\existsR]
                 {\sequent{\Sigma}{\Gamma}{\appsubst{\setize{\langle{}x,t\rangle}}{B}}}
                 {\sequent{\Sigma}{\Gamma}{\existsx{x:\alpha}{B}}}
    \end{array}
    \]
  \end{center}

  \caption{The rules in $\cal G$ for the logical symbols}
  \label{fig:logical-rules}
\end{figure}

\begin{figure}[htbp]
  \begin{center}
   \[
    \begin{array}{c}
      \inferrule*[right=\defR] 
                 {\sequent{\appsubst{\theta}{\Sigma}}
                          {\Gamma}
                          {\appsubst{\theta}{B}}}
                 {\sequent{\Sigma}{\Gamma}{A}} \\
       \mbox{\rm $\forallx{\overline{x}}{H \triangleq B}$ is a variant of a
             clause in $\cal D$ and $\theta$ is a substitution such
             that $A = \appsubst{\theta}{H}$}
      \\ [15pt]
      \inferrule*[right=\defL]
                 {
                   {\begin{array}{l}
                   \{\ \sequent{\appsubst{\theta}{\Sigma}}
                           {\appsubst{\theta}{\Gamma},\appsubst{\theta}{B}}
                           {\appsubst{\theta}{C}}\ \vert\ 
                   \mbox{\rm $\forallx{\overline{x}}{H \triangleq B}$ is a variant of a
                                 clause in $\cal D$} \\
                   \qquad\qquad\qquad\qquad\qquad\qquad\ 
                       \mbox{\rm named away from $\Sigma$ and $\theta$
                                 is an \mgu\ for $\{\langle A,H \rangle\}$}\ \} 
                   \end{array}}}  
                 {\sequent{\Sigma}{\Gamma,A}{C}
                }
    \end{array}
    \] 
  \end{center}
  \caption{The rules for introducing atomic formulas based on a
    definition $\cal D$}
  \label{fig:def-rules}
\end{figure}

The logic $\cal G$ is formulated in the style of a sequent calculus.
Sequents have the form $\sequent{\Sigma}{\Gamma}{F}$, where $\Gamma$
is a multiset of \emph{assumption formulas}, $F$ is
a \emph{conclusion} or \emph{goal} formula, and $\Sigma$ is a
collection of (typed) variables called the \emph{eigenvariable
  context} that represents universal quantification at the proof
level. 
For the sequent to be well-formed, the formulas in $\Gamma \cup \{
F\}$ must be well-formed relative to a given collection of constant
and predicate symbols and the variables in $\Sigma$.
We shall limit our attention to well-formed sequents;
specifically, we will consider
the proofs of only well-formed sequents and the rules of the logic are
such that they preserve this property. 
From a proof search perspective, a sequent represents a state in the
process of proving a formula $F$; the process starts out with the sequent
$\sequent{\emptyset}{\cdot}{F}$ and evolves through the use of
inference rules into a set of obligations represented by other sequents.

The core of $\cal G$ comprises a set of rules that interpret the
logical symbols.
These rules are shown in Figure~\ref{fig:logical-rules}.
The premise sequents in the rules \forallL\ and \existsR\ require the
instantiation of the quantified formulas with terms.
These terms must be of the same type as the quantified variable and
they must be constructed using only the constants of the logic and the
variables in the eigenvariables set of the sequent.
The \forallR\ and \existsL\ rules have a proviso that the quantified
variable must be fresh to the eigenvariable set of the conclusion
sequent.
This requirement can always be achieved by a renaming of the bound
variable.
The \id\ rule, which recognizes that a sequent in which
the conclusion formula is included in the assumption formulas
has a trivial proof, requires the conclusion formula to be atomic.
While this requirement is not essential, it will simplify some of the
later arguments.
Finally, we will often write the eigenvariable context of a sequent as
a listing of just its variables, assuming their types can be
determined unambiguously from their occurrences in the formulas in the
sequent. 

An aspect of $\cal G$ that sets it apart from other first-order logic
is its treatment of atomic predicates or relations through
\emph{fixed-point definitions}. 
Formally, $\cal G$ is a logic that is parameterized by a definition
$\cal D$ that comprises a set of \emph{definitional clauses} of the form
$\forallx{\bar x}{H \triangleq B}$, where $H$ is an atomic formula 
and $B$ is an arbitrary formula;\footnote{Definitions must satisfy
  certain ``stratification'' conditions for the logic to be
  consistent. We elide a discussion of this matter, noting only that
  these conditions will always be satisfied in the use we make of
  definitions to encode simple rule-based specifications of
  relations in language presentations.}
$H$ is called the head of a clause in
this form, $B$ is called its body and the sequence of variables
$\overline{x}$ is called its binder.
Two clauses that differ only in the names chosen for the variables in
their binders are considered to be equivalent and are referred to as
variants of each other.
Further, a clause is said to be named away from a sequence of
variables $\Sigma$ if the variables in its binder are distinct from
those in $\Sigma$. 
Each definitional clause is intended to be a partial specification of
the relation denoted by the predicate symbol of $H$, with all such
clauses determining its complete definition.
This interpretation is made precise by the rules in
Figure~\ref{fig:def-rules} for ``introducing'' atomic formulas on the
left and right sides of a sequent.
In these rules, we take the application of a substitution $\theta$ to an
eigenvariable context $\Sigma$ to be the removal from $\Sigma$ of the
variables in the domain of $\theta$ and the addition to $\Sigma$ of
the variables in its range.
The application of a substitution to a multiset of formulas
corresponds to its application to each member of the multiset.
The \defR\ rule encodes the idea that we may prove an instance of the
head of a clause by proving its body.
The \defL\ rule is the more novel and consequential part
of the treatment of fixed-point definitions.
Specifically, it encodes the property that an atomic formula holds
only by virtue of one of the clauses in the definition.
As such, it provides the basis for a case analysis style of reasoning
as we shall see presently.

\begin{figure}
  \begin{center}
   \[
    \begin{array}{c}
      \inferrule*[right={\id$^{**}$}] 
                 {\hspace*{0.1in}}
                 {\sequent{\Sigma}{\Gamma,A^{*^i}}{A^{*^i}}}
      \qquad
      \inferrule*[right={\id$^{@@}$}] 
                 {\hspace*{0.1in}}
                 {\sequent{\Sigma}{\Gamma,A^{@^i}}{A^{@^i}}}
      \\[5pt]
     \inferrule*[right={\id$^{*@}$}] 
                 {\hspace*{0.1in}}
                 {\sequent{\Sigma}{\Gamma,A^{*^i}}{A^{@^i}}} 
      \qquad
      \inferrule*[right={\id$^*$}] 
                 {\hspace*{0.1in}}
                 {\sequent{\Sigma}{\Gamma,A^{*^i}}{A}}
      \qquad
      \inferrule*[right={\id$^@$}] 
                 {\hspace*{0.1in}}
                 {\sequent{\Sigma}{\Gamma,A^{@^i}}{A}}
      \\[15pt]
      \inferrule*[right=\defL$^{*^i}$]
                 {
                   {\begin{array}{l}
                   \{\ \sequent{\appsubst{\theta}{\Sigma}}
                           {\appsubst{\theta}{\Gamma},(\appsubst{\theta}{B})^{*^i}}
                           {\appsubst{\theta}{C}}\ \vert\ 
                   \mbox{\rm $\forallx{\overline{x}}{H \triangleq B}$ is a variant of a
                                 clause in $\cal D$} \\
                   \qquad\qquad\qquad\qquad\qquad\qquad\qquad
                       \mbox{\rm named away from $\Sigma$ and $\theta$
                                 is an \mgu\ for $\{\langle A,H \rangle\}$}\ \} 
                   \end{array}}}  
                 {\sequent{\Sigma}{\Gamma,A^{*^i}}{C}} 
      \\[15pt]
      \inferrule*[right=\defL$^{@^i}$]
                 {
                   {\begin{array}{l}
                   \{\ \sequent{\appsubst{\theta}{\Sigma}}
                           {\appsubst{\theta}{\Gamma},(\appsubst{\theta}{B})^{*^i}}
                           {\appsubst{\theta}{C}}\ \vert\ 
                   \mbox{\rm $\forallx{\overline{x}}{H \triangleq B}$ is a variant of a
                                 clause in $\cal D$} \\
                   \qquad\qquad\qquad\qquad\qquad\qquad\qquad
                       \mbox{\rm named away from $\Sigma$ and $\theta$
                                 is an \mgu\ for $\{\langle A,H \rangle\}$}\ \} 
                   \end{array}}}  
                 {\sequent{\Sigma}{\Gamma,A^{@^i}}{C}} 
      \\[15pt] 
      \inferrule*[right={\ind{i}{m}, $A$ \mbox{\rm is atomic}}]
                 {\sequent{\Sigma}
                          {\Gamma, \forallx{\overline{x_1}}
                                           {F_1 \supset \cdots \supset
                                             \forallx{\overline{x_m}}
                                                     {A^{*^i} \supset F}}}
                          {\forallx{\overline{x_1}}
                                   {F_1 \supset \cdots \supset
                                     \forallx{\overline{x_m}}{A^{@^i} \supset F}}}}
                {\sequent{\Sigma}
                         {\Gamma}
                         {\forallx{\overline{x_1}}
                                  {F_1 \supset \cdots \supset
                                    \forallx{\overline{x_m}}{A
                                      \supset F}}}}\\[3pt]
            \mbox{\rm Annotations of the form $*^i$ and $@^i$ must not
              already appear in the conclusion sequent}
    \end{array} 
    \] 
  \end{center}
  \caption{The induction rule and associated rules for annotated formulas}
  \label{fig:ind-rules}
\end{figure}

The logic actually builds in a \emph{least fixed-point} interpretation
of definitions, thereby providing the basis for reasoning inductively
about them.
This capability is realized by associating a measure via annotations
with atomic formulas and building in the idea that this measure
decreases when an assumption formula is unfolded via a definitional
clause.\footnote{For simplicity of presentation, we limit ourselves in
  this paper to the particular realization of induction in Abella,
  rather than the more general logical treatment originating
  from~\cite{tiu04phd}.} 
The annotations are of the form $@^i$ and $*^i$ for non-zero natural
numbers $i$, where the superscript is indicates the number of
repetitions of $@$ and $*$, respectively.
These annotations work in tandem, the latter constituting a smaller
measure than the former. 
Annotations are introduced into formulas by the rule
$\ind{i}{m}$ shown in Figure~\ref{fig:ind-rules}.
This rule reduces the task of proving a formula of the
form
$\forallx{\overline{x_1}}
        {F_1 \supset \cdots \supset
          \forallx{\overline{x_m}}{A \supset F}}$,
where $A$ is an atomic formula, to showing this formula holds
when $A$ has the measure $@^i$ associated with it under the assumption
that it holds when $A$ has the smaller measure $*^i$.
The measure associated with an atomic assumption formula decreases
when it is unfolded using a definition, a fact encoded in the
\defL$^{@^i}$ variant of the \defL\ rule; the annotation notation on
a possibly non-atomic formula that is used in this rule denotes the
distribution of the annotation to its atomic constituents.
The \defL$^{*^i}$ rule reflects the fact that the lower measure is
maintained through further unfoldings.
The remaining rules adapt the \id\ rule to the situation where atomic
formulas may have annotations and the content of each of them follows
easily from the interpretation of the annotations.
In all these rules, $A$ is used as a schematic variable for an atomic
formula. 

The treatment of fixed-point definitions makes $\cal G$ a logic that
is well-suited to the task of reasoning about rule-based
specifications.
Towards illuminating this aspect, we consider a simple example built
around the definition of the {\sl append} predicate that relates three lists.
A rule-based definition of this predicate is shown below:
\begin{center}
\begin{tabular}{c}

\inferrule{\qquad}{\mbox{\sl append(nil,~L,~L)}}

\qquad\qquad
\inferrule{\mbox{\sl append(L$_1$,~L$_2$,~L$_3$)}}
          {\mbox{\sl append(cons(X,~L$_1$),~L$_2$,~cons(X,~L$_3$))}}
\end{tabular}
\end{center}
In these rules, {\sl nil} is a constructor of list type that
represents the empty list and {\sl cons} is a constructor that
represents a list obtained by adding a head element to an 
existing list. 
Moreover, tokens that begin with uppercase letters represent schematic
variables.
This definition can be rendered into a definition in $\cal G$ that
comprises the following two clauses:
\begin{tabbing}
  \qquad\=\kill
  \>$\forallx{\ell}{append(nil,\ell,\ell) \triangleq \top}$\\
  \>$\forallx{x,\ell_1,\ell_2,\ell_3}
             {append(cons(x,\ell_1),\ell_2,cons(x,\ell_3))}
                \triangleq append(\ell_1, \ell_2, \ell_3)$
\end{tabbing}
When these clauses are used to simplify the conclusion formula in a
sequent, they have the flavor of logic programming: essentially, they
figure in a backchaining-style search for a proof of the formula.
When they are used on the left of a sequent to analyze an assumption
formula, they realize a closed-world interpretation that is usually
implied in rule-based systems and that is beyond the capabilities of
logic programming.
In this particular instance, they enforce the requirement that an {\sl
  append} relation can be true only by virtue of these rules.
Moreover, the logic supports the ability to reason inductively on the
derivation of such a relation.

Towards bringing these aspects of the logic out, let us consider
proving the following formula
\[
\forallx{\ell_1,\ell_2,\ell_3,\ell_4}
        {\mbox{\sl append}(\ell_1,\ell_2,\ell_3) \imp
           \mbox{\sl append}(\ell_1,\ell_2,\ell_4) \imp
               \ell_3 = \ell_4}
\]
which asserts that once the first two arguments to {\sl append} are
fixed, the relation holds for a unique third argument.
The attempt to prove this statement starts with the following sequent:
\[ \sequent{\emptyset}{\cdot}
           {\forallx{\ell_1,\ell_2,\ell_3,\ell_4}
                    {\mbox{\sl append}(\ell_1,\ell_2,\ell_3) \imp
                          \mbox{\sl append}(\ell_1,\ell_2,\ell_4) \imp
                         \ell_3 = \ell_4}}. \]
An informal argument will proceed by induction on the derivation of
either the first or the second antecedent of the formula.
In the formal system, we may use the $\ind{1}{1}$ rule to reduce the
proof obligation to the following sequent:
\begin{tabbing}
\qquad\=\qquad\qquad\qquad\=\kill
\> $\sequent{\emptyset}
            {\forallx{\ell_1,\ell_2,\ell_3,\ell_4}
                     {\mbox{\sl append}(\ell_1,\ell_2,\ell_3)^* \imp
                           \mbox{\sl append}(\ell_1,\ell_2,\ell_4) \imp
                          \ell_3 = \ell_4}}
            {}$\\
\>\> ${\forallx{\ell_1,\ell_2,\ell_3,\ell_4}
                  {\mbox{\sl append}(\ell_1,\ell_2,\ell_3)^@ \imp
                            \mbox{\sl append}(\ell_1,\ell_2,\ell_4) \imp
                           \ell_3 = \ell_4}}$
\end{tabbing}
Using rules for logical symbols on the right of the sequent, we may
further reduce the task to proving the sequent
\begin{tabbing}
\qquad\=\qquad\qquad\=\kill
\> $\sequentctx{\ell_1,\ell_2,\ell_3,\ell_4}$\\
\>$\forallx{\ell_1,\ell_2,\ell_3,\ell_4}
                       {\mbox{\sl append}(\ell_1,\ell_2,\ell_3)^* \imp
                             \mbox{\sl append}(\ell_1,\ell_2,\ell_4) \imp
                            \ell_3 = \ell_4},$\\
\>$\sequentsansctx{\mbox{\sl append}(\ell_1,\ell_2,\ell_3)^@,
                   \mbox{\sl append}(\ell_1,\ell_2,\ell_4)}
                    {\ell_3 = \ell_4}$\\
\end{tabbing}
An informal proof would proceed at this point by unfolding the second
assumption formula, the one annotated with $@$,
something that can be achieved within $\cal G$ by
using the rule \defL$^@$.
Doing so leads to the obligation to prove the sequent
\begin{tabbing}
\qquad\=\kill
\> $\sequentctx{\ell_1,\ell_2,\ell_3,\ell_4}$\\
\>$\forallx{\ell_1,\ell_2,\ell_3,\ell_4}
         {\mbox{\sl append}(\ell_1, \ell_2, \ell_3)^* \imp
              \mbox{\sl append}(\ell_1, \ell_2, \ell_4) \imp
                            \ell_3 = \ell_4},$\\
\>$\sequentsansctx{\mbox{\sl append}(\mbox{\sl nil},\ell_2,\ell_4)}
                    {\mbox{\sl nil} = \ell_4}$\\
\end{tabbing}
and the sequent
\begin{tabbing}
\qquad\=\kill
\> $\sequentctx{\ell_1,\ell_2,\ell_3,\ell_4, x, \ell_5, \ell_6}$\\
\>$\forallx{\ell_1,\ell_2,\ell_3,\ell_4}
         {\mbox{\sl append}(\ell_1,\ell_2,\ell_3)^* \imp
             \mbox{\sl append}(\ell_1,\ell_2,\ell_4) \imp
                            \ell_3 = \ell_4},$\\
\>$\sequentsansctx{
      \mbox{\sl append}(\mbox{\sl cons}(x,\ell_5),\ell_2,\ell_4),
      \mbox{\sl append}(\ell_5,\ell_2,\ell_6)^*}
     {\mbox{\sl cons}(x,\ell_6) = \ell_4$}.
\end{tabbing}
Observe how the unfolding causes the
annotation on the assumption formula that has replaced the original
one in the second sequent to change to a $*$.
Since it is in this form, it can be used to discharge the first
antecedent in the induction hypothesis. 
The proof can now be completed by invoking case analysis, \ie{} the
\defL\ rule, with respect to the second assumption formula in the two sequents and,
in the second case, ``invoking'' the induction hypothesis through uses
of the \impL\ and relevant variants of the \id\ rule. 

We present below a substitution operation on proofs and some
metatheorems about $\cal G$ that will be useful in later
discussions.

\begin{definition}[Proof Substitutions] \label{def:proofsubst}
Let $\pi$ be a proof in $\cal G$ and let $\theta$ be a
substitution.
The application of $\theta$ to $\pi$, denoted by
$\appsubst{\theta}{\pi}$ is defined recursively on the structure of
$\pi$ as follows:
\begin{enumerate}
  \item Suppose the last rule is \defL\ or one of its variants
    and the conclusion of this rule is the sequent
    $\sequent{\Sigma}{\Gamma,A^a}{C}$, where $a$ is the empty string
    or one of $*^i$ and $@^i$. 
    Further, suppose that the premise derivations for the rules are
    $\pi_1,\ldots,\pi_n$, where $\pi_i$ derives a sequent obtained by
    considering a clause in the definition of the form
    $\forallx{\overline{x_i}}{H_i}\triangleq B_i$ and an
    \mgu\ $\theta_i$ for $\{\langle A, H_i\rangle\}$.
    For each $i$, $1 \leq i \leq n$, if
    $\{\langle \appsubst{\theta}{A},H_i \rangle \}$ 
    has a unifier, then there must be a substitution $\rho_i$ such
    that $\rho_i \circ \theta_i$ is its \mgu.
    In this situation, the rule is to be replaced by a 
    rule-like structure in which the conclusion sequent is
    $\sequent{\appsubst{\theta}{\Sigma}}
             {\appsubst{\theta}{\Gamma},\appsubst{\theta}{A^a}}
             {\appsubst{\theta}{C}}$
    and that has as premise derivations $\appsubst{\rho_i}{\pi_i}$ for
    each $i$ such that $\{\langle \appsubst{\theta}{A},H_i \rangle \}$
    has a unifier.

  \item In all other cases, if the conclusion sequent is
    $\sequent{\Sigma}{\Gamma,A}{C}$ and the premise derivations are
    $\pi_1,\ldots,\pi_n$, then this rule is replaced by a rule-like
    structure that has 
    $\sequent{\appsubst{\theta}{\Sigma}}
             {\appsubst{\theta}{\Gamma},\appsubst{\theta}{A}}{\appsubst{\theta}{C}}$
    as the conclusion sequent and
    $\appsubst{\theta}{\pi_1}, \ldots, \appsubst{\theta}{\pi_n}$ as
    premise derivations.
 \end{enumerate} 
\end{definition}

\begin{thm}\label{thm:gtheorems}
  Consider a well-formed sequent $\mathcal{S}$ of the form $\sequent{\Sigma}{\Gamma}{F}$.
  \begin{enumerate}
    \item
      If $\mathcal{S}$
      has a proof $\pi$, then any well-formed sequent
      $\sequent{\Sigma'}{\Gamma'}{F'}$
      where the formulas in $\Gamma' \cup \{F'\}$ are obtained by
      renaming the free variables in $\Gamma \cup \{F\}$ in a
      consistent and logically correct way, \ie, in a way that avoids
      inadvertent capture, has a proof whose structure is identical to
      that of $\pi$.

    \item
      If $\mathcal{S}$ has a proof of
      height $h$ then, for any substitution $\theta$, the sequent
      $\sequent{\appsubst{\theta}{\Sigma}}
              {\appsubst{\theta}{\Gamma}}
              {\appsubst{\theta}{F}}$
     has a proof of height at most $h$.
  \end{enumerate}
\end{thm}

\begin{proof}
  The first claim is proved by a straightforward induction on the
  structure of $\pi$.
  For the second claim, it suffices to show that for any proof $\pi$,
  the structure $\appsubst{\theta}{\pi}$ is also a proof.
  This can be seen again via an induction on the structure of $\pi$.
  Note that $\appsubst{\theta}{\pi}$ may lose some branches at
  \defL\ rules in comparison with $\pi$.
  While the structure of the proof may change in this manner, the
  height of the transformed proof will be at most that of the original
  one. 
\end{proof}

The first clause in the theorem assures us the particular names
chosen for the variables in the formulas in a sequent do not matter
and neither does the specific extent of the eigenvariable context so
long as it suffices to ensure the sequent is well-formed.
A consequence of the second clause is that we may adjust
the choice of \mgu\ in the use of the \defL\ rule and its variants in
a proof without increasing its height.
We will use these properties in later sections.

We have already noted how the syntactic categories in $\ntset{}$
and the constructors in $\constrset{}$ and the relations in
$\relset{}$ and $\transrelset{}$
in a language specification translate into the vocabulary of $\cal G$.
The example we have considered indicates how the the rules in
$\ruleset{}$ and $\transruleset{}$ translate into definitional
clauses.  
Once all the rules for each relation have been assembled together,
\ie, when we are reasoning about a composite language, the clauses
deriving from them can be collected into a definition in $\cal G$.
We will exploit these correspondences in our discussions to transit
seamlessly between a language specification and its formalization in
the logic.
One specific manifestation of this is that we will treat formulas of
the kind shown in Section~\ref{sec:problem}, \ie, formulas constructed
using logical symbols over relations and other vocabulary identified
in Section~\ref{sec:framework}, directly as if they are formulas in
$\cal G$ even though an actual implementation needs an
intermediate encoding step.

Following the discussion in the previous section, the metatheoretic
properties whose proofs we will consider will be represented by
formulas of the form  
$\forallx{\overline{x}}{R(\overline{t}) \imp F}$
where $R(\overline{t})$ denotes an atomic predicate.
We shall consider constructing proofs of a specific form for such
properties.
These proofs will end with a case analysis over a possibly-annotated
version of $R(\overline{t})$, followed by rules that introduce logical
symbols in the conclusion formula and some number of uses of the
induction rule; the preceding part of the proof will obviously be
distinct branches or fragments that establish each of the
sequents that arise from considering the different cases in the
definition of $R(\overline{t})$.
We shall refer to a proof that has this form as one in \emph{canonical
  form}.
In the general situation, a proof of this kind will be attempted after
some other properties have been established and can therefore be used
as lemmas by virtue of the \cut\ rule. 
If $L$ is the specification of the composite language encoded
in $\cal G$ and $P$ represents the metatheoretic property, we shall
write $\proven{L}{{\cal L}}{P}{\pi}$ to denote the fact that $\pi$ is a
complete canonical form proof of $P$ and where the
properties in the set $\cal L$ are used as lemmas.

\section{Modular Proofs for Foundational Properties}
\label{sec:foundationalProperties}

The approach to modularizing the proof of foundational properties
we had outlined in Section~\ref{sec:problem} can now be given
concreteness.
The main part of a canonical-form proof for a formula of the form
$\forallx{\overline{x}}{R(\overline{t}) \imp F}$ is the proof of a
collection of sequents, one corresponding to each of the rules
defining $R$ that is applicable to the derivation of
$R(\overline{t})$. 
Now, the rules relevant to a composite language are obtained by
combining the ones in the host language and each of the extensions.
We propose accordingly to let each language component produce a proof
of sequents corresponding to its new rules independently, with the expectation
we can combine the different fragments into a complete proof for a
composite language at the time the composition is determined.

As has been previously noted, there is an issue that needs to be addressed to make
this scheme work.
When a component attempts to construct a proof for a sequent
associated with a particular case in the definition of $R$, it does so
without knowledge of how other extensions might add to the definitions
of relations that are in the shared vocabulary, \ie, that are
identified by the host language. 
However, proofs in $\cal G$ are sensitive to the extent of
a definition.
Thus, if we do not properly restrict the manner in which proof
fragments are constructed in isolated components, we will not be able
to combine these into a valid proof for the complete language.

An observation that allows us to overcome this difficulty is that
the extensibility framework limits the way in which the definitions for
relations introduced by the host language can be enhanced: additional
rules can be added by an
extension only for those cases where the primary component for the
relation is constructed using a syntactic form introduced by that
extension.
Thus, if we limit the use of case analysis
to those situations where we know the
top-level structure of the primary component argument is a constructor
introduced by the extension or the host language, then we can be
certain the case analysis will be comprehensive even when other
components are added to the mix. 
The following definition identifying ``valid'' proof fragments is a rendition of
this idea. 
\begin{definition}[Proofs for cases for foundational properties]\label{def:proof-frag}
Let $M$ be the host language or an extension (relative to some host
language) in a library constructed within the extensibility framework
and let $L$ be the language identified by $M$. 
Further, let $P$ be a property of the form
$\forallx{\overline{x}}{R(\overline{t}) \imp F}$ in the vocabulary
determined by the host language in the library. 
Finally, let $P$ have a canonical proof relative to $L$ 
and the lemmas in $\cal L$ in which case analysis over a relation
$R'(\overline{t'})$ in the proofs of the sequents resulting from the case
analysis over $R(\overline{t})$ 
occurs only in the following situations:
\begin{enumerate}
\item the term in the primary component position for $R'$ in
  $R'(\overline{t'})$ has  a constructor as its top-level symbol, or
  
\item $M$ is an extension $E$ and $R'$ and its primary component type
  have been introduced by $E$.
\end{enumerate}
Then we denote the fact that $\frag_1,\ldots,\frag_m$ are the proofs
for the sequents arising from the case analysis on $R(\overline{t})$
using the rules introduced by $M$ in such a proof by writing 
$\hproven{L}{\lemmaset}{M}{P}{\oneTo{\frag}{m}}$. 
\end{definition}

To demonstrate the kinds of case analysis disallowed by this
definition, consider proving type preservation,
\property{exprTypePreservation},  for the sequent arising
from the \textsc{E-add} rule given by the host language.  Our initial
sequent is
\begin{multline*}
  \Gamma, e_1, e_2, \mathit{ty}, \gamma, i_1, i_2, i :
  \evalexpr{\gamma}{e_1}{\mathit{intlit}(i_1)},
  \evalexpr{\gamma}{e_2}{\mathit{intlit}(i_2)},
  \plus{i_1}{i_2}{i},
  \typeexpr{\Gamma}{\mathit{add}(e_1,e_2)}{\mathit{ty}}, \\
  \forallx{x,\mathit{ty_x},\mathit{v_x}}
          {\lookuptype{\Gamma}{x}{\mathit{ty_x}} \imp
            \lookupval{\gamma}{x}{\mathit{v_x}} \imp
            \typeexpr{\mathit{nilty}}{\mathit{v_x}}{\mathit{ty_x}}}
  \longrightarrow
  \typeexpr{\mathit{nilty}}{\mathit{intlit}(i)}{\mathit{ty}}
\end{multline*}
The case analysis that produced this sequent instantiated the
expression $e$ from the original property with $\mathit{add}(e_1,e_2)$
and introduced premises that both $e_1$ and $e_2$ evaluate to integer
literals, as well as the expression as a whole evaluates to an
integer literal of their sum.
Case analysis is \emph{not} permitted on either
$\evalexpr{\gamma}{e_1}{\mathit{intlit}(i_1)}$ or
$\evalexpr{\gamma}{e_2}{\mathit{intlit}(i_2)}$ because the evaluation
relation is introduced by the host language and the terms in their primary components,
$e_1$ and $e_2$, do not have constructors as their top-level symbols.
What expressions can instantiate $e_1$ and $e_2$ and how evaluation
may be defined on them is dependent on other extensions that may be
added in a composite language, so any case analysis carried out in the
context of the host language alone cannot predict all the relevant
rules.
Conversely, case analysis \emph{is} permitted on the typing assumption
$\typeexpr{\Gamma}{\mathit{add}(e_1,e_2)}{\mathit{ty}}$ because the
argument in its primary component position, $\mathit{add}(e_1,e_2)$,
has the $\mathit{add}$ constructor as its top-level symbol, and thus
no rule introduced by an extension can pertain to it.
This case analysis shows $\mathit{ty}$ must, in fact, be
$\mathit{int}$.  Then we can simplify the conclusion with the
\textsc{T-intlit} rule, finishing the proof.

The following lemma makes explicit the rationale for the restrictions
on proof fragments.
\begin{lemma} \label{thm:ppFragValidity}
Let $H$ be the host language and let $\{E_1,\ldots, E_n\}$ be a
collection of extensions in a library constructed in the extensibility
framework.
Let $M$ be $H$ or an $E_i$ for some $i$ such that $1\leq i
\leq n$, 
let $L$ be the language identified by $M$,
let $P$ be a  property of the form 
$\forallx{\overline{x}}{R(\overline{t}) \imp F}$
in the vocabulary determined by the host language, 
and let $\rho$ be a collection of proofs such that
$\hproven{L}{\lemmaset}{M}{P}{\rho}$ holds.
Then 
\begin{enumerate}
\item the sequents that arise from the rules for $R$ in $M$ in a case
  analysis over $R(\overline{t})$ in a canonical proof for $P$ in the
  context of the language   $\langComp{H}{\oneTo{E}{n}}$ are
  exactly the same as the ones   that are proved by the proof
  fragments in $\rho$, and  
\item
  each proof $\frag$ in $\rho$ constitutes a complete proof of the
  corresponding sequent $\cal S$ even in the context of the language
  $\langComp{H}{\oneTo{E}{n}}$.  
\end{enumerate}
\end{lemma}

\begin{proof}
  The definition of $R$ in the language
  $\langComp{H}{\oneTo{E}{n}}$ is obtained exactly by collecting
  together all the clauses for it in each of the components.
  It is obvious from this that (1) holds.

  The reason why the second clause is true can be seen by considering
  the inference steps in the logic.  The only inference rules
  that are dependent on the definition parameterizing the logic are
  \defR\ and the variants of \defL.
  In the first case, the clause that provides the basis for the
  \defR\ rule in $\frag$ will also be available relative to the larger
  language, so the proof step is still valid.
  We consider the other cases below, noting the distinction between
  the variants of \defL\ is irrelevant to the argument.

  We note first that the definition of a relation whose primary component
  is introduced by an extension cannot change with the addition of
  more extensions.
  Thus, any case analysis over such relations within an extension
  will remain unchanged when the language context is determined by the
  addition of other extensions.
  If a relation is introduced by an extension but its primary
  component type is introduced by the host language, the relation is
  defined for other extensions by instantiating the projection rule
  with each new constructor they introduce.  Thus any new rules added
  in the composition cannot apply in case analysis in the language of
  $M$ where the primary component had to be built by a constructor
  known in the language of $M$.
  Similarly, in the case a relation and its primary component type are
  introduced by the host language, case analysis must in general
  consider how the relation is modified by rules for it that are
  introduced by other extensions.
  However, the extensibility framework does not permit an extension to
  introduce rules that apply to expression forms introduced
  by other components in the library in this situation.
  For this reason, the restriction that case analysis be used in this
  case only when the argument at the location of the primary component
  has a rigid structure ensures that a case analysis that is carried out
  ignoring other extensions must remain unchanged when
  they are added to the mix.
  In summary, the restrictions on proof fragments ensure that case
  analysis carried out locally is complete even when the context is
  expanded to include other extensions. 
\end{proof}

We can now define the idea of combining proof fragments to yield a
purported proof for a foundational property relative to a composite
language. 

\begin{definition}[Proof composition]\label{def:proof-comp}
Let $H$ be the host language and let $E_1,\ldots, E_n$ be a
collection of extensions in a library constructed in the extensibility
framework.
Further, let $P$ be a property of the form  
$\forallx{\overline{x}}{R(\overline{t}) \imp F}$
in the vocabulary determined by the host language.
Finally, let $\rho_0$ be such that
$\hproven{\langComp{H}{\emptyset}}{\lemmaset}{H}{P}{\rho_0}$ holds and
for $1 \leq i \leq n$ let $\rho_i$ be such that 
$\hproven{\langComp{H}{\{E_i\}}}{\lemmaset}{E_i}{P}{\rho_i}$ holds. 
Then $\proofComp{\rho_0,\ldots,\rho_n}$ represents a proof structure
for $P$ that ends with a case analysis on $R(\overline{t})$ followed
by rules for introducing logical symbols in the conclusion formula and
some number of invocations of the induction rule, and where the proof
obligations deriving from the case analysis on $R(\overline{t})$ are
discharged by the proofs in $\rho_0\cup\rho_1\cup\ldots\cup\rho_n$.
\end{definition}

Note that clause (1) of Lemma~\ref{thm:ppFragValidity} ensures the
coherence of the above definition: there is a one-to-one
correspondence between the cases that arise for the composite language
and the proof fragments for each of the components participating in
that language.

\begin{thm}\label{thm:hostProps}
Let $H$ be the host language and let $E_1,\ldots, E_n$ be a
collection of extensions in a library constructed in the extensibility
framework.
Further, let $P$ be a property of the form  
$\forallx{\overline{x}}{R(\overline{t}) \imp F}$
in the vocabulary determined by the host language.
Finally, let $\rho_0$ be such that
$\hproven{\langComp{H}{\emptyset}}{\lemmaset}{H}{P}{\rho_0}$ holds and,
for $1 \leq i \leq n$, let $\rho_i$ be such that 
$\hproven{\langComp{H}{\{E_i\}}}{\lemmaset}{E_i}{P}{\rho_i}$ holds. 
Then $\proven{L}{\lemmaset}{P}{\proofComp{\rho_0, \ldots, \rho_n}}$ holds.
\end{thm}

\begin{proof}
As previously noted, the sequents that need to be proved in the kind
of proof of $P$ that $\proofComp{\rho_0,\ldots,\rho_n}$ is
intended to be correspond exactly to the sequents the proof
fragments in $\rho_0 \cup \ldots \cup \rho_n$ prove.
Thus, the theorem would be true if these proof fragments, which are
constructed relative to smaller languages, are also proofs of the
corresponding sequents relative to the composed language.
That this is the case is verified by Lemma~\ref{thm:ppFragValidity}.
\end{proof}

Theorem~\ref{thm:hostProps} provides the theoretical basis for our 
approach to the modular development of proofs for foundational
metatheoretic properties.
In this approach, the host language articulates the desired
metatheorem at the outset.
The designers of the host language and of each extension, who build
on the host language and hence must be cognizant of the metatheorem,
then develop proofs for the cases within the canonical proof structure
for the theorem relative to the language they each identify, taking
care to adhere to the restrictions imposed by
Definition~\ref{def:proof-frag}.
These ``proof fragments,'' which are stored by the components that
developed them, can then be automatically combined as per
Definition~\ref{def:proof-comp} to yield a proof of the property for 
the composite language at the time when the composition is
determined. 
Of course, it is not necessary actually to construct a proof
for the metatheoretic property for the composite language: by virtue
of Theorem~\ref{thm:hostProps}, the existence of proof fragments
with each component is already a guarantee such a
proof exists for the overall language.

To illustrate the above ideas, let us consider the construction of a
proof for type preservation for languages resulting from our 
example library.  The host language proves type preservation for each
evaluation rule it introduces, adhering to the restrictions given by
\refdef{proof-frag}, and thus these proofs are still valid in the
context of the composed language.
To see an example of this, look at the proof for the case of the
\textsc{E-add} rule above.  The initial sequent is the same, as it
comes from the same rule in the context of the composed language as in
the setting of the host language alone.  In the limited setting, we
analyzed the typing derivation
$\typeexpr{\Gamma}{\mathit{add}(e_1,e_2)}{\mathit{ty}}$, where the
only applicable rule was \textsc{T-add}.  In the composed language,
this is still the only applicable typing rule, so the case analysis
has the same result.  Similarly, we used the \textsc{T-intlit} rule to
simplify
$\typeexpr{\mathit{nilty}}{\mathit{intlit}(i)}{\mathit{int}}$; this
rule is also part of the composed language, so we can use it for
simplification in the expanded setting as well.  Thus our modular
proof is still applicable even though we have expanded the language in
which we are applying it.
The list extension also proves type preservation
for each rule it introduces, adhering to the restrictions so its
proofs remain valid, but with
knowledge of both itself and the host language.  The security
extension introduces no expression evaluation rules, and thus has no
cases to prove for this property.
The full set of cases we have in the composed language is the union of
those for the rules introduced by the host language and the list
extension, as proved in \reflemma{ppFragValidity}.  Since each case
also has a proof given by the host language or list extension,
depending on which one introduced the rule giving rise to the case,
we can build the full proof from their modular proofs.

A question that arises in this context is if the limitations on case
analysis in support of modular proofs severely curtail the proofs that
can be developed.
Our experimentation indicates this not to be the case, with two
apparent reasons why this is so.
First, the use of the induction hypothesis---which applies to all
expressions, including the ones constructed using the vocabulary of
(other) extensions---often obviates a ``second-level'' case analysis.
In the typical proof scenario, the purpose of the initial case
analysis is to produce assumptions with which we may use the induction
hypothesis, and thus which need not be further analyzed.
Second, in situations where such a second-level case analysis is
necessary, it can often be extracted into an auxiliary lemma, proved
separately according to our requirements, and used in the main proof.
For example, in proving decidability of equality for two expressions
$e_1$ and $e_2$, we would generally consider cases on both their
forms, but our requirements for modular proofs do not permit this.  We
can instead prove auxiliary lemmas that the \emph{form} of an
expression is decidable, then use these in the equality proof.  For
example, we can prove an expression is built by the $\mathit{add}$
constructor or not, then use this lemma for $e_2$ in the
decidability proof with a top-level case analysis on the form of $e_1$
when $e_1$ is built by $\mathit{add}$, avoiding the disallowed second
case analysis on $e_2$.

\section{Modular Proofs for Auxiliary Properties}
\label{sec:auxiliaryProperties}

We now take up the consideration of modularizing reasoning about
a metatheoretic property that is introduced by an extension. 
To recall, extensions different from the one that introduces
such a property cannot participate directly in proving it.
However, the extension that introduces the property would have
difficulty in proving it without help: it does not know the details of
the extensions with which it may interact in a composition.
To overcome this difficulty, we propose to extend the idea
introduced in Section~\ref{sec:framework} of using projection
relations to ``understand'' extensions at a distance.

The basic structure for auxiliary properties and the proofs we will
seek to construct for them will remain the same as that for 
foundational properties: these properties will still have the form 
$\forallx{\overline{x}}{R(\overline{t}) \imp F}$, and their proofs will end
with a case analysis on the definition of $R$ followed by rules that
simplify the conclusion formula and some number of uses of the
induction rule. 
The distinguishing characteristic for an auxiliary property is that
the case analysis has to be carried out completely within the
extension that introduces it.
There is no difficulty in doing this if both the relation $R$ and the
category for the primary component for $R$ are introduced by the
extension because the definition of such a relation remains unchanged
under the inclusion of other extensions.\footnote{Reasoning within
  the proofs of the sequents that result from the case analysis on
  $R(\overline{t})$ must still be properly circumscribed, but this
  matter is no different from that in the case of foundational
  properties that was discussed in the previous section.} We will
therefore not consider this situation further. 
The only other situations that are possible are those in which the
relation is introduced by either the host language or the extension
and has as its primary component a syntactic category is
introduced by the host language. 
In these cases, the approach will be to reason specifically for each
construct in the primary component category that is volunteered by the
extension or the host language and to construct a ``generic''
argument to cover the cases of constructs introduced by extensions
whose form is unknown before an actual composition.
The key to the soundness of this method for organizing the reasoning
will be to show the generic argument can, in fact, be elaborated
automatically into specific ones when all the constructs are known.

For this style of reasoning to be successful, it is important to
identify an approach that can support the construction of generic
arguments of interesting and useful properties.  
To facilitate this, we propose the notion of \emph{projection
  constraints} as a vehicle for reasoning about an extension even when
its details are unknown. 
From a methodological perspective, these constraints, which are based
on projection relations, are intended to be ones that afford 
extensions freedom in defining their semantic attributes while still
circumscribing behavior relative to the images in the host language of
the constructs they introduce.
Concretely, projection constraints take the shape of
foundational properties that can be assumed to have been established prior
to the attempt to prove an auxiliary property.
In this context, our approach to proving an auxiliary property amounts
conceptually to proving the property relative to the constructs from the host
language and those of the extension that introduces it and then to
lifting the proof to all constructs in a composition by exploiting the
projection constraints. 
This lifting, is in, fact the content of the generic argument.
The soundness of this ``lifting'' step can be established directly 
when the relation $R$ is identified by the extension: in this
situation, the projection relation figures specifically in the
definition of the relation via the projection rule for the extension and
the reasoning therefore has a straightforward inductive structure.
The matter is more complicated when the relation $R$ is introduced by
the host language.
In this case, the relation would have its own definition in other
participating extensions and it is necessary to demonstrate that
reasoning about it through the projection of the primary component is
legitimate.

We develop these ideas in the rest of the section.
We introduce the notion of projection constraints in the first
subsection below.
We then describe a method for constructing  a proof for an auxiliary
property within the logic $\cal G$ that accommodates a generic
component.
The remaining two subsections are devoted to showing the
resulting proof, which we call a \emph{proof skeleton}, can be
viewed as an actual proof. 
This is, in fact, the case without qualifications when the relation
$R$ is introduced by the extension, as we show in \subsec{elaborateExt}. 
We describe an additional condition in \subsec{elaborateHost} whose
validity ensures the soundness of proof skeletons when the relation is
introduced by the host language; to close the gap, it is necessary to
demonstrate that this condition also holds. 
The arguments in both cases are constructive: 
they identify a method for obtaining a complete proof of the
metatheoretic property for any composed 
language from the proof skeleton constructed within
the extension introducing the property.

\subsection{Projection Constraints} \label{subsec:projConstraints}

There is often a need to be able to model one extension in the context
of another in the extensibility framework.
We have seen an example of this in \subsec{viewByProj}, where 
it was necessary to have a view of the syntax
of another extension to be able to define a new relation in a manner
that covered all possible cases in a composite language.
There can also be a need to ``understand'' semantic attributes of
static and dynamic varieties for expressions from other extensions.
Consider, for instance, the definition of the visibility level of an 
expression in the security extension in the example library described
in \sec{framework}.
This definition depends on the variables appearing in an expression.
While it can be defined by a projection rule for expressions from other
extensions, its coherence depends on the variables in an expression
being preserved under projections.
As another example, consider \property{secure-correct} in
\sec{problem}, which asserts that if
a program fragment passes a security analysis then it must be the case
it will not leak private information. 
We can try to conduct the necessary reasoning about constructs from
other extensions through their projections but this will work only if
we can relate the behavior of such constructs to their images.
Thus we may need to know the evaluation of the projection of a
statement form will terminate if the evaluation of the original form
terminates and also that the states resulting from the evaluation in
the two cases will be identical.

We have already seen how projections and projection rules help address
the first of the two requirements.
Projection constraints constitute our proposal to address the
second.
These are metatheorems that are, once again, of the form
$\forallx{\overline{x}}{R(\overline{t}) \imp F}$, with the proviso
that the relation $R(\overline{t})$ is a projection relation in these
properties. 
One way to look at these properties is that they place
constraints on the behavior of extension constructs relative to 
that of the host language constructs that model them.
These constraints are foundational in nature because they are
articulated by the host language to set expectations prior to the
description of any extension. 
Moreover, these properties force the extension developer to think
carefully about how the constructs they introduce are best modelled in
the host language, \ie, about what the appropriate projections are for
these constructs from this perspective.

Some examples illustrating the structure and possible uses of
projection constraints are relevant at this point.
Relative to the extension library from \sec{sec:framework}, the
property that the set of variables in an expression are preserved
under a projection can be expressed by the following formula: 
\begin{equation} \label{eq:tc-e-eval}
  \forallx{e,e',\mathit{vs},\mathit{vs'}}
          {\projectexpr{e}{e'}  \imp  \vars{e}{\mathit{vs}}  \imp
                       \vars{e'}{\mathit{vs'}} \imp \mathit{vs} \subseteq \mathit{vs'}} 
\end{equation}
The statement of this property uses a ``subset'' relation not
introduced previously but that has an obvious definition.
The property that the evaluation of the projection of a statement must
terminate, \ie, result in producing a final state, if the evaluation
of the statement itself terminates can be expressed as
follows:
\begin{equation} \label{eq:tc-s-eval}
  \forallx{s,s',\gamma,\gamma'}{\projectstmt{s}{s'}  \imp
  \evalstmt{\gamma}{s}{\gamma'}  \imp
  \existsx{\gamma''}{\evalstmt{\gamma}{s'}{\gamma''}}}
\end{equation}
Finally, the requirement that the final state resulting from an
evaluation of a statement must be preserved under a projection is
captured by the following formula:
\begin{equation} \label{eq:tc-s-eval-results-back}
  \forallx{s,s',\gamma,\gamma_1,\gamma_2,n,v}{\projectstmt{s}{s'}  \imp
  \evalstmt{\gamma}{s}{\gamma_1}  \imp
  \evalstmt{\gamma}{s'}{\gamma_2}  \imp
  \lookupval{\gamma_2}{n}{v}  \imp
  \lookupval{\gamma_1}{n}{v}}
\end{equation}
Observe that all these properties are predicated on projection
relations for the relevant categories of expressions; as such, these
can be read genuinely as constraints on the behavior of extension
constructs based on the behavior of host language constructs that
model them.
We note also that such properties can be useful in reasoning about the
soundness of security analyses within the security extension even when
the details of the other extensions participating in a composed
language are not known, an aspect on which we elaborate in later
subsections.

Projection contraints can help in determining the purpose of a
projection relation and, hence, the form that its definition should
take. 
For example, recall that the projection of the expression
$\mathit{cons}(e_1,e_2)$ was identified as $\mathit{eq}(e_1,e_2)$ in
the list extension in \sec{framework}.
This appears strange at the outset.
However, the choice becomes a much more natural one in the context of
a projection constraint asserting that the variables appearing in
an expression are preserved under projections, a property that is
itself important to ensuring the soundness of many forms of static
analysis. 
Another thing to note is that, while the predominant use of projection
constraints is to provide a means for thinking about extension
constructs via their images, their use is not limited to this
purpose.
Thus, consider the property expressed by the following formula:
\begin{equation} \label{eq:tc-e-unique}
  \forallx{e,e_1,e_2}{\projectexpr{e}{e_1}  \imp  \projectexpr{e}{e_2}  \imp
  e_1 = e_2}
\end{equation}
This formula expresses the constraint that projections of expressions
in our example library must be unique.
Such a property can play a role in demonstrating that the security
level of an expression is unique, as we shall see later in this
section.

For projection constraints to be useful in subsequent arguments, it
is, of course, necessary to demonstrate their validity.
This can be done in a manner similar to other foundational properties.
Conceptually, this task devolves into each extension needing
to demonstrate that the constructs it introduces obey the
constraints.

\subsection{Proof Skeletons for Auxiliary Properties} \label{subsec:proveExtProp}

The main content of a canonical proof for a property
of the form $\forallx{\overline{x}}{R(\overline{t}) \imp F}$
is the consideration of the cases that arise out of the rules defining
$R$.
If $R$ is a relation whose primary component is in a category
introduced by the host language, then these cases must encompass
behavior on constructs in the category contributed by all extensions
in a composition.
However, when the proof is being constructed within a single
extension, that extension cannot have specific knowledge of what might
be introduced by other extensions.
As indicated earlier, we propose to overcome this difficulty by
reasoning about such cases
in a generic way through their projections, utilizing
projection constraints in the process. 
This subsection makes precise how this style of reasoning may be
supported.
The soundness of the method is taken up in the next two
subsections.

Support for a generic form of argument is realized in two 
steps.
First, a means is provided for representing terms whose
structure cannot be examined.
Concretely, this is done by adding a special constant
$\unknown[\tau]$ for each syntactic category $\tau$ introduced by the
host language; in what follows, knowing the specific syntactic
category will often be unimportant and we will therefore write
the constant (ambiguously) as just $\unknown$.
Second, a mechanism is provided for treating an arbitrary term via
such a constant in the reasoning process, to be interpreted through
its projections. 
Both steps are realized by identifying a ``generic'' extension
relative to a host language.

\begin{definition}[Generic Extension]\label{def:Eunknown}
Let $H = \langle \ntset{H}, \constrset{H}, \relset{H}, \ruleset{H},
\transrelset{H}, \transruleset{H}, \trset{H} \rangle$ be the host
language and $E$ be an extension in a library constructed in the
extensibility framework. 
Then $\gen{H}{E}$, a generic extension relative to $H$ and $E$, is
identified to be an extension
$\langle \emptyset, \constrset{}, \emptyset, \ruleset{}, \emptyset,
\emptyset, \emptyset \rangle$ such that 
\begin{enumerate}
  \item $\constrset{} = \{ \unknown[\tau]\ |\ \tau \in \ntset{H}\}$
    where each $\unknown[\tau]$ is a distinct constant different also
    from all the constants introduced by $H$ and $E$,
    and
    
   \item $\ruleset{}$ comprises exactly the rules of the form
\[
\inferrule{T(\unknown[\tau],y) \\
           R(x_1,\ldots,x_{i-1}, y, x_{i+1},\ldots,x_k)}
          {R(x_1,\ldots,x_{i-1},\unknown[\tau], x_{i+1},\ldots, x_k)},
          \]
for each $R \in \relset{H}$ whose $i^{th}$ argument is its primary
component, where $T \in \transrelset{H}$ is the projection relation
for $\tau$ and
$x_1,\ldots,x_{i-1},x_{i+1},\ldots,x_k$ and $y$ are distinct (schematic)
variables.
$T$ may correspond here to a projection relation with additional
parameters.
In this case, the host language is required to have specified what
these parameters should be in terms of the variables
$x_1,\ldots,x_{i-1},x_{i+1},\ldots,x_k$.
\end{enumerate}
\end{definition}

Let $H$ be a host language and $E$ be an extension and consider the
collection of rules in the language
$\langComp{H}{\setize{E,\gen{H}{E}}}$ for a  relation $R$ that is
introduced either by $H$ or by $E$ and whose primary component
category is one introduced by $H$. 
In addition to the rules explicitly provided by $H$ and $E$ this
collection will also include one of the form
\[
\inferrule{T(\unknown,y) \\
           R(t_1,\ldots,t_{i-1}, y, t_{i+1},\ldots,t_k)}
          {R(t_1,\ldots,t_{i-1},\unknown, t_{i+1},\ldots,t_k)}
          \]
where $T$ is the projection relation for the category of $\unknown$.
Moreover, there will be exactly one such rule for each relation. 
When this rule is used in a case analysis, it will lead to 
a situation where $T(\unknown,y)$ and the appropriate instance of
$R(t_1,\ldots,t_{i-1}, y, t_{i+1},\ldots,t_k)$ is added to the assumption
set of the sequent.
Thus, such a rule provides the basis for 
reasoning about a generic term representing one introduced by another
extension via a property that may be associated
with its projection. 

\begin{definition}[Proof skeleton, generic case] 
\label{def:proof-frags-e}
  Let $H$ be the host language and $E$ be an extension in a library
  constructed in the extensibility framework and let $L$ be the
  language $\langComp{H}{\setize{E,\gen{H}{E}}}$.
  Further, let $P$ be a formula of the form
  $\forallx{\overline{x}}{R(\overline{t}) \imp F}$ where $R$ is a relation
  introduced either by $H$ or by $E$ whose primary component category
  is introduced by $H$.
  By a proof skeleton for $P$ relative to $L$ and a set of lemmas
  \lemmaset{} we mean a canonical proof for $P$
  in the described context in which the clauses encoding the rules 
  from $\gen{H}{E}$ are not used in \defR\ rules in the proofs of
  sequents resulting from the case analysis over a possibly annotated
  version of $R(\overline{t})$ and the \defL\ rule and its variants
  are used relative to a relation
  $R'(\overline{t'})$ in these proofs only in the following
  situations:   
  \begin{enumerate}
  \item the primary component argument of $R'(\overline{t'})$ is a term whose
    top-level symbol is a constructor that is introduced by $H$ or
    $E$, 
  \item $R'$ and its primary component type are introduced by $E$, or
  \item the primary component argument of $R'(\overline{t'})$ is
    $\unknown$ and $R'$ is a relation introduced by $E$.
  \end{enumerate}
  The restrictions on the use of the \defR\ and (variants of the)
  \defL\ rules described above are referred to in the sequel as ``the
  constraints imposed by the definition on proof fragments''.
  If one of the sequents arising from the case analysis over
  $R(\overline{t})$ corresponds to the situation where the primary 
  component of $R$ matches \unknown, we refer to this as the
  ``generic case.''
\end{definition}

The limitation to case analysis within the proof fragments
in a proof skeleton has the purpose of ensuring such an analysis
will be complete even when other extensions participate in a
composition.
Note especially that in the last two cases the relation $R'$ is
completely determined by the extension. 
To see the effectiveness of generic reasoning that is facilitated by
Definition~\ref{def:proof-frags-e}, we consider example 
properties that might be introduced by the security extension in the 
language library described in \sec{framework}.
As the first example, consider the following formula
\begin{equation} \label{eq:level-unique}
  \forallx{\Sigma,e,sl_1,sl_2}
          {(\exprlevel{\Sigma}{e}{sl_1}) \imp
               (\exprlevel{\Sigma}{e}{sl_2})  \imp sl_1 = sl_2}
\end{equation}
asserting that the security level of an expression is unique.
This is a property whose key relation is defined by the extension
and its primary component is introduced by the host language. 
There will be a generic case for this property: cases of expressions
arising from a priori unknown extensions have also to be reasoned about.
Focusing on this case, we see that it requires us to show that
$sl_1 = sl_2$ must hold if we know $\projectexpr{\unknown}{y}$, 
$\exprlevel{\Sigma}{y}{sl_1}$, and \exprlevel{\Sigma}{\unknown}{sl_2}
hold and that the induction hypothesis can be invoked relative to
$\exprlevel{\Sigma}{y}{sl_1}$. 
Because the security level relation is introduced by the security
extension, a case analysis of the last of these relations, the second
derivation of $\mathit{level}$, is possible
and it yields that, for some $y'$, $\projectexpr{\unknown}{y'}$ and
$\exprlevel{\Sigma}{y'}{sl_2}$ must hold; intuitively the only way in
which the relation \exprlevel{\Sigma}{\unknown}{sl_2} could have been
defined is by the projection rule from the security extension, since \unknown{}
represents a term from some other extension.
We may now utilize \projConstr{tc-e-unique} to show $y = y'$
and then invoke the induction hypothesis to conclude that $sl_1 = sl_2$.

As another example, recall the auxiliary
property~\ref{eq:secure-correct} from \sec{problem} that 
essentially asserts that private information
cannot leak out from the evaluation of a secure statement.
This is a metatheoretic property posited by the security extension
but whose key relation
is introduced by the host language.
Here too there will be a generic case. 
The analysis for this case is based on the rule introduced 
by the generic extension, leading to the replacement of the premise
$\evalstmt{\gamma_A}{s}{\gamma_A'}$ by $\projectstmt{\unknown}{y}$ and
$\evalstmt{\gamma_A}{y}{\gamma_A'}$.
This process also instantiates the second premise to
$\evalstmt{\gamma_B}{\unknown}{\gamma_B'}$.
While it may seem this relation can also be analyzed by the rule
from the generic extension, such an analysis is not guaranteed to be
sound: the rule in the generic extension assumes a simulation of
behavior by the projection, which needs to be verified before it can
be used.\footnote{A similar observation applies to the case analysis
  at the outermost level, but a verification of the soundness of the
  simulation in that case will be carried out in a complete proof, as
  discussed in \subsec{elaborateHost}.}
Such a case analysis is prohibited by
Definition~\ref{def:proof-frags-e} for this reason. 
However, we can use \projConstr{tc-s-eval} to show the 
expression $y$ to which $\unknown{}$ projects must also evaluate under
$\gamma_B$ to yield a new environment $\gamma_B''$.
After analyzing the derivation of $\mathit{secure}$ and using a
projection constraint similar to \projConstr{tc-e-unique} to get a
derivation of $\mathit{secure}$ for $y$,
the inductive hypothesis tells us both $\gamma_A'$ and
$\gamma_B''$ have the same values for all public variables.
\projConstr{tc-s-eval-results-back} lets us show $\gamma_B'$ and
$\gamma_B''$ have the same values for all variables, so $\gamma_A'$
and $\gamma_B'$ have the same values for all variables.

We will need the following observation in later discussions when using
these modular proofs to build proofs for composed languages.
\begin{lemma}\label{lem:substproof}
Let $H$ be the host language and $E$ be an extension in a library
constructed in the extensibility framework.
Further, let $\cal S$ be a sequent that has a proof $\pi$ relative to
the language $\langComp{H}{\setize{E,\gen{H}{E}}}$ and the set of
lemmas $\lemmaset$ in which the constraints imposed by
Definition~\ref{def:proof-frags-e} on proof fragments are satisfied. 
Finally, let $\theta$ be a substitution determined by some language
that extends the vocabulary determined by
$\langComp{H}{\setize{E,\gen{H}{E}}}$. 
Then $\appsubst{\theta}{\pi}$ is a proof of the sequent
$\appsubst{\theta}{{\cal S}}$ based on the definitions of relations
determined by the language $\langComp{H}{\setize{E,\gen{H}{E}}}$ and
using the lemmas in $\lemmaset$ that satisfies the constraints
imposed by Definition~\ref{def:proof-frags-e} on proof fragments.
\end{lemma}
\begin{proof}
This is a refinement of clause 2 of Theorem~\ref{thm:gtheorems} that
follows from examining its proof and the definition of substitution
into derivations.
\end{proof}

\subsection{Proof Elaboration for Extension-Introduced Relations} \label{subsec:elaborateExt}

In our scheme, a proof skeleton for a metatheoretic property 
is to be taken as a complete demonstration of the validity of the
property in any well-behaved composition in the 
case where the key relation is introduced by the extension
constructing the proof skeleton.
We show this assessment to be justified in this subsection.
More specifically, we show a proof skeleton can be
elaborated into a complete proof for the property relative to any
composite language.
The main observation underlying this fact is that the proof of the
sequent arising from rules for the key relation contributed by an a
priori unknown extension can be generated via a suitable instantiation
of the proof of the generic case in a proof skeleton.
We demonstrate this in Lemma~\ref{thm:subst-unknown} below.
\begin{definition}[Term Replacement]
\label{def:term-replacement}
If $t_1$ and $t_2$ are two terms of the same type and $s$ is another
term, then $\subst{t_1}{t_2}{s}$ denotes the replacement of all
occurrences of $t_2$ in $s$ by $t_1$; this operation has an obvious
recursive definition.
The operation extends to formulas with the proviso that quantified
variables must be renamed so they do not get confused with the ones
that appear in $t_1$ or $t_2$, to sets of formulas by its application
to each member of the set, to unification problems by its distribution
to the terms in each pair in the set, and to substitutions by its application to
the terms in the range of a substitution.
The notation for terms is extended to these cases: if $F$ is
a formula, $\Gamma$ is a set of formulas, $\cal U$ is a unification
problem and $\theta$ is a substitution, then the replacement of $t_1$
by $t_2$ in each is denoted by $\subst{t_1}{t_2}{F}$,
$\subst{t_1}{t_2}{\Gamma}$, $\subst{t_1}{t_2}{{\cal U}}$, and
$\subst{t_1}{t_2}{\theta}$, respectively.
\end{definition}
  
\begin{definition}[Instance of a (Generic) Sequent] \label{def:seq-rel-e}
  A sequent ${\cal S}'$ is an instance of ${\cal S}$ determined
  by the term 
  $t$, a relationship denoted by $\seqRel{{\cal S}}{{\cal S}'}{t}$, if
  $\cal S$
  is $\sequent{\Sigma}{\Gamma}{F}$ and ${\cal S}'$ is
  $\sequent{\Sigma'}{\subst{t}{\unknown}{\Gamma}}{\subst{t}{\unknown}{F}}$;
  $\Sigma'$ must, of course, include all the variables needed to
  ensure that ${\cal S}'$ is well-formed.
\end{definition}

An example illustrating Definition~\ref{def:seq-rel-e} is in order.
Assuming $\mathit{IH}$ represents the formula 
\[  \forallx{\Sigma,e,sl_1,sl_2}
          {(\exprlevel{\Sigma}{e}{sl_1})^*  \imp
                     \exprlevel{\Sigma}{e}{sl_2}  \imp sl_1 = sl_2},
\]
let $\cal S$ be the sequent 
\[
  \sequent{\Sigma, sl_1, sl_2, x}
          {\mathit{IH},\projectexpr{\unknown}{x},
            (\exprlevel{\Sigma}{x}{sl_1})^*,
            \exprlevel{\Sigma}{\unknown}{sl_2}}
          {sl_1 = sl_2};
\]
this sequent would be the one proved in the generic case in a
proof skeleton for \property{level-unique}. 
In the proof of the property for a language determined by composition
with actual extensions, this sequent would have to be
replaced by ones resulting from it by instantiating
$\unknown{}$ with specific terms.
Thus, if the composition includes the list extension, the
$\mathit{tail}$ 
constructor in that extension will require a proof of the following
sequent, denoted by ${\cal S'}$, to be constructed:
\[
  \sequent{\Sigma, sl_1, sl_2, x, l}
          {\mathit{IH}, \projectexpr{tail(l)}{x},
            (\exprlevel{\Sigma}{x}{sl_1})^*,
            \exprlevel{\Sigma}{tail(l)}{sl_2}}
          {sl_1 = sl_2}
\]
Here, ${\cal S}'$ is an instance of $\cal S$
determined by $\mathit{tail(l)}$, \ie, the relation 
$\seqRel{{\cal S}}{{\cal S}'}{tail(l)}$ holds.

\begin{lemma} \label{thm:subst-unknown}
    Let $H$ be the host language and $E, E_1,\ldots, E_n$ be extensions
    in a library constructed in the extensibility framework.
    Further, let $\cal S$ be a sequent that has a proof relative to
    the language $\langComp{H}{\setize{E,\gen{H}{E}}}$ and the lemmas
    in $\lemmaset$ in which the constraints imposed by
    Definition~\ref{def:proof-frags-e} on proof fragments are
    satisfied. 
    Then
    \begin{enumerate}
    \item if $\unknown$ does not occur in $\cal S$, then $\cal S$ has a
      proof relative to the language
      $\langComp{H}{\setize{E,E_1,\ldots,E_n}}$ and the lemmas
      in $\lemmaset$; and

    \item if $t$ is a term in the language 
    $\langComp{H}{\setize{E,E_1,\ldots, E_n}}$ that has as its
      top-level symbol a constant introduced by $E_i$ for some i such
      that $1 \leq i \leq n$ and is such that no variables in the
    eigenvariable context of $\mathcal{S}$ occur in it, then there
    must be a proof relative to
    $\langComp{H}{\setize{E,E_1,\ldots, E_n}}$ and the lemmas in
    $\lemmaset$ for any sequent ${\cal S}'$ such that
    $\seqRel{{\cal S}}{{\cal S}'}{t}$ holds.
    \end{enumerate}
\end{lemma}
\begin{proof}
  The first part of the lemma can be proved by a simple adaptation of
  the argument for the second part.
  Alternatively, it follows from the application of the second part
  with a suitable (even if fictitious) term $t$.
  We therefore focus below only on proving the second part. 
  
  Let $\frag$ be the proof of $\cal S$ in the proof skeleton $\cal
  P$.
  The argument is by induction on the height of $\frag$ and proceeds
  by examining the last rule used. 
  The conclusion is obvious if this rule is some variant of the
  \id\ rule: for example, if $\mathcal{S}$ is
  of the form $\sequent{\Sigma}{\Gamma}{F}$ where $F$ is a member of
  $\Gamma$, then $\mathcal{S}'$ must be of the
  form
  $\sequent{\Sigma'}{\subst{t}{\unknown}{\Gamma}}{\subst{t}{\unknown}{F}}$
  and $\subst{t}{\unknown}{F}$ must be a member of $\subst{t}{\unknown}{\Gamma}$.
  For the remaining cases, it suffices to show that the same rule as
  was used to derive $\cal S$ can be used to derive ${\cal S}'$,
  possibly from premises that are $\seqRelSans{t}$ related to the
  premises in the derivation of $\cal S$.
  This is easy to see for all the rules in
  Figure~\ref{fig:logical-rules} other than \id\ and for the
  \ind{i}{m}\ rule.
  Note the use of lemmas is encompassed by the \cut\ rule. 

  The only remaining rules are those pertaining to definitions.
  Here, we must consider the fact that the set of clauses in the
  definition associated with $\langComp{H}{\setize{E,\gen{H}{E}}}$
  that can be used in $\frag$ differs from the ones associated with
  $\langComp{H}{\setize{E,E_1,\ldots, E_n}}$.  
  The clauses that can be used in $\frag$ are those
  corresponding to the rules for relations introduced by $H$ or $E$
  and the projection rules for $E$ instantiated with $\unknown$.
  For $\langComp{H}{\setize{\Eall}}$, a clause of the form
  $\forallx{\overline{x}}{H \triangleq B}$ that is derived from the
  projection rules for $E$ in the collection for
  $\langComp{H}{\setize{E,\gen{H}{E}}}$ is replaced by ones of the
  form 
  $\forallx{\overline{x},\overline{y}}
           {\subst{c(\overline{y})}{\unknown}{H} \triangleq
             \subst{c(\overline{y})}{\unknown}{B}}$
  for each constructor $c$ of the right type that is introduced by the
  extensions $E_1,\ldots,E_n$; the variables in $\overline{y}$ must be
  chosen to be distinct from those in $\overline{x}$ and we are
  assuming the precise order of the variables in a binder for a
  clause is irrelevant here.
  Additionally, the clauses for $\langComp{H}{\setize{\Eall}}$ will
  include ones deriving from the relation defining rules introduced by
  $E_1,\ldots, E_n$ and instantiations of the projection rules of
  these extensions.
  However, these clauses turn out to be irrelevant because of the
  restrictions on the use of case analysis in a proof skeleton, a fact
  that is implicit in the argument below.

  Now suppose that $\frag$ ends with the \defR\ rule.
  In this case, $\cal S$ must have the form
  $\sequent{\Sigma}{\Gamma}{A}$ for some atomic formula $A$ and
  it must be derived from a sequent of the form
  $\sequent{\Sigma'}{\Gamma}{\appsubst{\theta}{B}}$ using a clause
  $\forallx{\overline{x}}{H \triangleq B}$ from the available
  collection relative to $\langComp{H}{\setize{E,\gen{H}{E}}}$, where
  $\theta$ is such that $A = \appsubst{\theta}{H}$.
  Using Theorem~\ref{thm:gtheorems}, clause (1), we may assume the
  domain of $\theta$ to be disjoint from the set of variables
  appearing in $t$. 
  Using Theorem~\ref{thm:subst-distr}
  (Appendix~\ref{app:unification}), we then see that 
  $\subst{t}{\unknown}{A} = 
    \appsubst{\subst{t}{\unknown}{\theta}}{\subst{t}{\unknown}{H}}$.
  Now, corresponding to the clause
  $\forallx{\overline{x}}{H \triangleq B}$, there is one in the
  collection available relative to
  $\langComp{H}{\setize{E,E_1,\ldots, E_n}}$ that has the clause
  $\forallx{\overline{x}}
           {\subst{t}{\unknown}{H} \triangleq
               \subst{t}{\unknown}{B}}$
  as an instance; this is trivially the case for the clauses
  corresponding to the rules introduced by $H$ and $E$ since
  $\unknown$ does not appear in such clauses, and it can be arranged
  via suitable substitutions for the ones derived from the
  projection rules for $E$.
  But then, using the substitution $\subst{t}{\unknown}{\theta}$, we
  see that the sequent ${\cal S}'$, 
  which is, in fact, of the form
  $\sequent{\Sigma'}{\subst{t}{\unknown}{\Gamma}}{\subst{t}{\unknown}{A}}$,
  may be derived by a \defR\ rule from a sequent of the form
  $\sequent{\Sigma''}{\subst{t}{\unknown}{\Gamma}}
           {\appsubst{\subst{t}{\unknown}{\theta}}
                     {\subst{t}{\unknown}{B}}}$ relative to the
  language $\langComp{H}{\setize{E,E_1,\ldots, E_n}}$.
 By Theorem~\ref{thm:subst-distr}, 
 $\appsubst{\subst{t}{\unknown}{\theta}}
             {\subst{t}{\unknown}{B}} =
    \subst{t}{\unknown}{\appsubst{\theta}{B}}$.
  Thus, ${\cal S}'$ may be derived using a \defR\ rule from a premise
  sequent that is related by $\seqRelSans{t}$ to the one used
  in $\frag$, as desired.
  
  To complete the proof, we must consider the case where $\frag$ ends
  with some version of the \defL\ rule.
  The versions differ only in that some formulas may be annotated and
  that annotations may be affected by the rule.
  However, annotations are not relevant to the objective at hand and
  we will ignore them in our argument. 
  Now, in this case, $\cal S$ must be of the form
  $\sequent{\Sigma}{\Gamma,R'(\overline{t'})}{C}$ with the 
  \defL\ rule in $\frag$ pertaining to $R'(\overline{t'})$.
  Then ${\cal S}'$ must be of the form 
  $\sequent{\Sigma'}{\subst{t}{\unknown}{\Gamma},
                     \subst{t}{\unknown}{R'(\overline{t'})}}
           {\subst{t}{\unknown}{C}}$
  and we will want to show that it can be derived by a \defL\ rule
  from premises that are related by $\seqRelSans{t}$ to the premises
  for the \defL\ rule in $\frag$.
  The argument proceeds by considering each of the cases for \defL\ in
  $\frag$ that are permitted by Definition~\ref{def:proof-frags-e}.

  \smallskip
  \noindent {\it The primary component argument of $R'(\overline{t'})$
    has a constructor introduced by $H$ or $E$ as its top-level symbol}.
  Clearly, the primary component argument of
  $\subst{t}{\unknown}{R'(\overline{t'})}$ will continue to have the
  same constructor as its top-level symbol.
  Now, every rule given by an extension for a relation identified by the
  host language must have a term in the primary component location
  that has a constructor the extension introduces as its
  top-level symbol. 
  From this and from the manner in which projection rules are
  transformed into actual rules in a composition, it follows
  that the only clauses relative to
  $\langComp{H}{\setize{E,E_1,\ldots, E_n}}$
  that could be relevant to a case analysis on
  $\subst{t}{\unknown}{R'(\overline{t'})}$ are the ones arising
  from the rules in $H$ and $E$, \ie, the same clauses
  considered in the \defL\ rule in $\frag$.
  Let $\forallx{\overline{x}}{H \triangleq B}$ be one of these
  clauses.
  Since $\unknown$ does not appear in $H$, $H =
  \subst{t}{\unknown}{H}$. 
  By Theorem~\ref{thm:substunif} (Appendix~\ref{app:unification}),
  $\{\langle \subst{t}{\unknown}{R'(\overline{t'})},
                  \subst{t}{\unknown}{H} \rangle \}$
  has a unifier only if $\{\langle R'(\overline{t'}), H \rangle \}$
  has one; in applying the theorem, we note the top-level symbol
  for $t$ cannot appear in $R'(\overline{t'})$ because $\cal S$ is
  well-formed relative to the language
  $\langComp{H}{\setize{E,\gen{H}{E}}}$. 
  Thus, there is a premise arising from this clause in the \defL\ rule
  we are considering for deriving ${\cal S}'$ only if there is one in
  the \defL\ rule in $\frag$.
  Suppose that there is such a premise in the \defL\ rule in
  $\frag$.
  Then this premise would have the form
  $\sequent{\appsubst{\theta}{\Sigma}}
           {\appsubst{\theta}{\Gamma}, \appsubst{\theta}{B}}
           {\appsubst{\theta}{C}}$,
  for some \mgu\ $\theta$ for
  $\{\langle R'(\overline{t'}), H \rangle \}$. 
  Using Theorem~\ref{thm:gtheorems}, Theorem~\ref{thm:unifsubst}, and
  Lemma~\ref{lem:substproof}, we may assume $\theta$ is such
  that $\subst{t}{\unknown}{\theta}$ is an \mgu\ for
  $\{\langle \subst{t}{\unknown}{R'(\overline{t'})},
             \subst{t}{\unknown}{H} \rangle \}$; 
  the variables in $\overline{x}$ would need to be disjoint from those
  appearing in $\overline{t}$, but this is easily arranged.
  Note also that the domain of $\theta$ can be limited to the
  variables in $\Sigma$ and in $\overline{x}$, and hence to be
  disjoint from the variables appearing in $\overline{t}$.
  Now, we may pick
  $\sequent{\appsubst{\subst{t}{\unknown}{\theta}}{\Sigma'}}
           {\appsubst{\subst{t}{\unknown}{\theta}}
                     {\subst{t}{\unknown}{\Gamma}},
              \appsubst{\subst{t}{\unknown}{\theta}}
                       {\subst{t}{\unknown}{B}}}
           {\appsubst{\subst{t}{\unknown}{\theta}}
                     {\subst{t}{\unknown}{C}}}$
  as the premise based on this clause in the derivation we are wanting
  to construct for ${\cal S}'$.
  Using Theorem~\ref{thm:subst-distr}, we see that this sequent
  is the same as
  $\sequent{\appsubst{\subst{t}{\unknown}{\theta}}{\Sigma'}}
           {\subst{t}{\unknown}{\appsubst{\theta}{\Gamma}}
             \subst{t}{\unknown}{\appsubst{\theta}{B}}}
           {\subst{t}{\unknown}{\appsubst{\theta}{C}}}$,
  which is obviously related by $\seqRelSans{t}$ to the corresponding
  premise in $\frag$.
  Since this argument is independent of the particular clause
  selected, the desired result follows.

  \smallskip
  \noindent {\it $R'$ and its primary component type are introduced by
    $E$}.
  In this case, again, the clauses that are available in the case
  analysis of $\subst{t}{\unknown}{R'(\overline{t'})}$ are identical
  to those available for case analysis of $R'(\overline{t'})$ and
  $\unknown$ does not appear in these clauses.
  An argument identical to the previous situation suffices to show
  what is needed.

  \smallskip
  \noindent {\it $R'$ is a relation introduced by $E$ and the primary
    component argument of $R'(\overline{t'})$ is $\unknown$}.
  Clauses of two varieties can figure in the \defL\ rule in $\frag$
  in this situation.
  First, the clause may encode a rule contributed directly by $E$.
  Such a clause will persist unchanged in the definition corresponding
  to the language $\langComp{H}{\setize{E,E_1,\ldots, E_n}}$ and
  $\unknown$ will not occur in it.
  We can argue as before in this case that if the clause gives
  rise to a premise for the \defL\ rule that we want to use for
  deriving ${\cal S}'$ then there must be a premise for the
  \defL\ rule in $\frag$ to which it is related by $\seqRelSans{t}$.
  Second, the clause may have arisen from a projection rule in $E$.
  Here we observe the following easily established fact: if a clause
  of the form  $\forallx{\overline{x}}{H \triangleq B}$ arising from
  such a projection rule is relevant to the analysis of
  $R'(\overline{t'})$ 
  relative to $\langComp{H}{\setize{E,\gen{H,E}}}$, then it is only a
  clause of the form
  $\forallx{\overline{x}}{\subst{t}{\unknown}{H} \triangleq
                          \subst{t}{\unknown}{B}}$
  that will be relevant to the analysis of
  $\subst{t}{\unknown}{R'(\overline{t'})}$ relative to
  $\langComp{H}{\setize{E,E_1,\ldots, E_n}}$.
  Although the details in this situation differ from that in the
  earlier cases---\unknown\ appears in the clause
  $\forallx{\overline{x}}{H \triangleq B}$ and it is a clause after a
  replacement of $\unknown$ that must be considered relative to 
  $\langComp{H}{\setize{E,E_1,\ldots, E_n}}$---that argument can now be 
  easily adapted to show that if the latter clause results in a
  premise sequent in a derivation of ${\cal S}'$ by a \defL\ rule,
  then there must be a premise for the \defL\ rule in 
  $\frag$ that is related by $\seqRelSans{t}$ to it.
\end{proof}

We can now state and prove the main result of this subsection. 

\begin{thm} \label{thm:full-proof-ext}
  Let $H$ be the host language and \Eall{} be extensions in a 
  library constructed in the extensibility framework, and let $L =
  \langComp{H}{\setize{\Eall}}$. 
  Further, let $P$ be a formula of the form
  $\forallx{\overline{x}}{R(\overline{t}) \imp F}$ in the vocabulary of
  $\langComp{H}{\setize{E}}$ with $R$ being a relation introduced by $E$.
  Finally, let $\cal P$ be a proof skeleton for $P$ relative to
  $\langComp{H}{\setize{E,\gen{H}{E}}}$ and the lemmas
  in $\cal L$.
  Then there is a proof of $P$ relative to $L$ and the lemmas in
  $\cal L$. 
\end{thm}
\begin{proof}
  As in $\cal P$, the proof of $P$ relative to $L$ may end with
  rules for introducing the top-level logical symbols in $P$ and some
  number of uses of the induction rule.
  This would leave a obligation to provide a proof for a sequent of the
  form $\sequent{\Sigma}{\Gamma,R(\overline{t})}{F}$ or
  $\sequent{\Sigma}{\Gamma,(R(\overline{t}))^{@^i}}{F}$ that is 
  identical to the one in $\cal P$. 
  We shall assume the sequent to have the former form below; the
  argument in the other case is similar.
  Now, in $\cal P$, this sequent is proved by the use of the
  \defL\ rule relative to $R(\overline{t})$.
  We may consider using the \defL\ rule in the proof relative to
  $L$ as well.
  Noting $R$ is introduced by $E$, we see there are two
  sources for the premises for the \defL\ rule in the two cases: 
  they either derive from clauses that encode the rules for $R$ in 
  $E$ or the rules generated from the projection rule for $E$ through
  instantiation relative to $\langComp{H}{\setize{E,\gen{H}{E}}}$
  and $L$, respectively.
  The premise sequents resulting from the former must be identical
  and they must also not contain $\unknown$.
  Since these sequents have a proof relative to the language
  $\langComp{H}{\setize{E,\gen{H}{E}}}$, Lemma~\ref{thm:subst-unknown}
    assures us they must also have a proof relative to the
    language $\langComp{H}{\setize{E,E_1,\ldots,E_n}}$. 
  
  Thus it only remains to be shown the premises in the proposed use
  of \defL\ relative to $L$ that derive from the clauses that encode
  the rules for $R$ generated from the projection rule in $E$ have
  derivations.
  There is a relationship between the clauses that come from the
  projection rule relative to the language
  $\langComp{H}{\setize{E,\gen{H}{E}}}$ and $L$:
  a clause of the form $\forallx{\overline{x}}{H \triangleq B}$ in the former
  case will be replaced by a family of clauses of the form
  $\forallx{\overline{x},\overline{y}}
           {\subst{c(\overline{y})}{\unknown}{H} \triangleq
               \subst{c(\overline{y})}{\unknown}{B}}$
  in the latter, with one such clause for each constant $c$ of the
  appropriate type introduced by the extensions in $E_1,\ldots,E_n$.
  Moreover, $H$ should have $R$ as its top-level predicate symbol and
  $\unknown$ should be the argument in the primary component location
  for $R$.
  Since neither $\unknown$ nor $c$ can appear in $R(\overline{t})$,
  the heads of these clauses can unify with $R(\overline{t})$  only if
  the argument in the primary component location for $R$ is a variable.
  Using arguments similar to those seen in the proof of
  Lemma~\ref{thm:subst-unknown}, it can be established that the clauses
  deriving from the projection rule will yield premise sequents
  relative to $L$ only if there is a premise sequent corresponding to it
  relative to $\langComp{H}{\setize{E,\gen{H}{E}}}$.
  Moreover, it can be arranged such that if the premise sequent is
  $\cal S$ in the latter case and they are
  ${\cal S}_1,\ldots,{\cal S}_n$ in the former case, then
  $\seqRel{{\cal S}_i}{{\cal S}}{c_i(\overline{y_i})}$ holds with $c_i$
  and the variables in $\overline{y_i}$ not appearing in ${\cal S}$.
  The existence of the proof skeleton $\cal P$ assures us that $\cal
  S$ has a proof.
  Using Lemma~\ref{thm:subst-unknown}, we may conclude that, for $1
  \leq i \leq n$, ${\cal S}_i$ must have a proof. 
\end{proof}

Substantiating earlier comments, we note that the proofs of
Lemma~\ref{thm:subst-unknown} and Theorem~\ref{thm:full-proof-ext}
provide the basis for automatically 
constructing a proof for the property relative to the full language
from a proof skeleton once the components comprising the language are
known.

\subsection{Proof Elaboration for Host-Introduced
  Relations} \label{subsec:elaborateHost}

We now consider the soundness of treating a proof skeleton for a
metatheoretic property whose key relation is introduced by the host
language as a demonstration of the validity of the property in any
language composition.
In the construction of the proof skeleton in this case, the behavior
of the key relation in an arbitrary undetermined extension is modelled
by the rule for it that is provided by the generic extension.
Two assumptions underlie the treatment through this rule.
First, it is assumed that the behavior of the relation is equivalent
to that when the term that is its primary component is replaced by a
term to which it projects.
Second, it is assumed that the definition of the relation when such a
replacement is done is smaller in an inductive ordering.
Clearly, for the proof skeleton to constitute an adequate
demonstration of the property, it will be necessary to show these
assumptions are justified.

We describe an approach to meeting the above requirement in this
subsection. 
The starting point for our approach is to associate with the key
relation a new relation that builds in the described assumptions; we
refer to the latter as the \emph{projection version} of the original
relation.
One obligation in the overall scheme is to show the equivalence of the
two relations.
This obligation has the structure of a foundational property that we
expect to be validated in a modular manner along the lines discussed
in Section~\ref{sec:foundationalProperties}.
Assuming the equivalence of the relation and its projection version,
we then show a proof skeleton for the property can
be used to generate a complete proof for it relative to a
language that includes other well-behaved extensions.

The rest of this subsection develops these ideas.
We first define the notion of a projection version of a relation. 
We then show how a proof skeleton can be elaborated into a complete
proof when the two relations are equivalent.
We conclude the subsection by discussing what is involved in showing
the equivalence of a relation and its projection version and
reflecting on what the reliance on such a property entails for the
generality of the approach we propose.

\subsubsection{The Projection Version of a Relation}

The validity of the reasoning embodied in a proof skeleton is
dependent on viewing the key relation through the projection of its
primary component.
To capture this requirement, we build it into the definition of a new
relation we will use as a proxy in the reasoning process.

\begin{definition}\label{def:projectionVersion}
Let $H$ be a host language and $E_1,\ldots, E_n$ be a set of extensions
in a library constructed in the extensibility framework.
The projection version of the relation $R$ introduced by the host
language ($R\in\relset{H}$) is written as $\transRel{R}$.
Further, the definition of this relation is derived from that of $R$
as follows.
Assuming that the $i^{th}$ argument of $R$ is its primary component,
let 
\[
\inferrule{\overline{R(\overline{s})} \\ \overline{U}}
          {R(t_1, \ldots, t_{i-1}, t_i, t_{i+1}, \ldots, t_m)}
\]
be a rule for it, where $\overline{R(\overline{s})}$ denotes a set of
premises in which $R$ appears and $\overline{U}$ denotes the remaining
premises. 
If this is a rule in the collection corresponding to the host
language, then the definition of $\transRel{R}$ includes the rule 
\[
\inferrule{\overline{\transRel{R}(\overline{s})} \\ \overline{U}}
          {\transRel{R}(t_1, \ldots, t_{i-1}, t_i, t_{i+1}, \ldots, t_m)};
\]
in other words, the definition of $\transRel{R}$ includes an identical
rule, except that $R$ is replaced in it by $\transRel{R}$.
If this rule is in the collection corresponding to one of the
extensions $E_1,\ldots, E_n$ on the other hand, then the counterpart
rule for $\transRel{R}$ has the form
\[
\inferrule{\overline{\transRel{R}(\overline{s})} \\ \overline{U} \\
           T(t_i,x) \\
           \transRel{R}(t_1, \ldots, t_{i-1}, x, t_{i+1}, \ldots, t_m)}
          {\transRel{R}(t_1, \ldots, t_{i-1}, t_i, t_{i+1}, \ldots, t_m)},
\]
where $T$ represents the projection relation for the category of the
primary component of $R$ and $x$ is a variable that is fresh to the
rule. 
If $T$ corresponds here to a projection relation with additional
parameters, these parameters must be filled in adherence to the 
dependencies on the arguments of $R$ specified by the host language
and that are utilized also in constructing the generic extension. 
Observe the additional premises in this case encode the
requirement that the behavior of $\transRel{R}$ must remain the same
as that under a projection of its primary component and that the
definition of the relation for the projection must be smaller in the
inductive ordering. 
\end{definition}

\begin{figure}
  \[
  \inferrule*[Right=X-seq]
             {\evalstmt{\gamma}{s_1}{\gamma'} \\
               \evalstmt{\gamma'}{s_2}{\gamma''}} 
             {\evalstmt{\gamma}{\mathit{seq(s_1, s_2)}}{\gamma''}}
  \]
  \[
  \inferrule*[Right=\transRel{\mbox{X-seq}}]
             {\evalstmtT{\gamma}{s_1}{\gamma'} \\
               \evalstmtT{\gamma'}{s_2}{\gamma''}} 
             {\evalstmtT{\gamma}{\mathit{seq(s_1, s_2)}}{\gamma''}}
  \]

  \[
  \inferrule*[Right=X-splitlist]
            {\evalexpr{\gamma}{e}{\mathit{cons}(v_1, v_2)} \\
              n_{hd} \neq n_{tl}}
            {\evalstmt{\gamma}{\mathit{splitlist}(n_{hd}, n_{tl}, e)}
              {\consval{n_{hd}}{v_1}
                {\consval{n_{tl}}{v_2}
                  {\consval{n_{hd}}{\mathit{cons}(v_1, v_2)}{\gamma}}}}}
  \]
  \[
  \inferrule*[Right=\transRel{\mbox{X-splitlist}}]
            {\evalexpr{\gamma}{e}{\mathit{cons}(v_1, v_2)} \\
              n_{hd} \neq n_{tl} \\
              \projectstmt{\mathit{splitlist}(n_{hd}, n_{tl}, e)}{x_T} \\\\
            \evalstmtT{\gamma}{x_T}
              {\consval{n_{hd}}{v_1}
                {\consval{n_{tl}}{v_2}
                  {\consval{n_{hd}}{\mathit{cons}(v_1, v_2)}{\gamma}}}}
            }
            {\evalstmtT{\gamma}{\mathit{splitlist}(n_{hd}, n_{tl}, e)}
              {\consval{n_{hd}}{v_1}
                {\consval{n_{tl}}{v_2}
                  {\consval{n_{hd}}{\mathit{cons}(v_1, v_2)}{\gamma}}}}}
  \]

  \[
  \inferrule*[Right=X-secdecl]
             {\evalexpr{\gamma}{e}{v}}
             {\evalstmt{\gamma}{\mathit{secdecl(n, ty, sl, e)}}
               {\consval{n}{v}{\gamma}}}
  \]
  \[
  \inferrule*[Right=\transRel{\mbox{X-secdecl}}]
             {\evalexpr{\gamma}{e}{v} \\
               \projectstmt{\mathit{secdecl}(n, ty, sl, e)}{x_T} \\
               \evalstmtT{\gamma}{x_T}{\consval{n}{v}{\gamma}}}
             {\evalstmtT{\gamma}{\mathit{secdecl(n, ty, sl, e)}}
               {\consval{n}{v}{\gamma}}}
  \]
\caption{Some evaluation rules for the example language library and their
  projection versions} \label{fig:R_T-rules}
\end{figure}

\fig{R_T-rules} illustrates the above definition by showing some of
the rules for statement evaluation (denoted by $\evalstmtRel$) for our
example language library and their correlates in the definition of the
projection version of this relation (denoted by $\transRel{\evalstmtRel}$).
Corresponding to the \textsc{X-seq} rule, which is introduced by the
host language, is the $\transRel{\textsc{X-seq}}$ rule with the
same structure except that it defines the projection version of the
relation.
Corresponding to the two extension-introduced rules 
\textsc{X-splitlist} (introduced by the list extension) and
\textsc{X-secdecl} (introduced by the security extension), we have the
rules $\transRel{\textsc{X-splitlist}}$ and
$\transRel{\textsc{X-secdecl}}$ that include two additional premises
identifying the behavior of the projection version of the relation
with that under projections.

We desire to use $\transRel{R}$ in the reasoning process as a proxy
for $R$.
Our ability to do so will depend on showing the two relations to be
equivalent.
One part of the equivalence, which we call the \dropT{R}
property, is given by the following formula:
\begin{equation*}
  \forallx{\overline{x}}{\transRel{R}(\overline{x}) \imp R(\overline{x})}
\end{equation*}
This property must obviously be true---the premises for $\transRel{R}$
include all that is required of $R$ and possibly more---and a modular
proof for it can be constructed in a mechanical fashion.
We shall assume this has been done and use \dropT{R} freely as a
lemma in what follows. 
The other part of the equivalence, the \addT{R}
property, is expressed by the formula
\begin{equation*}
  \forallx{\overline{x}}{R(\overline{x})\imp\transRel{R}(\overline{x})}
\end{equation*}
This property is more substantive and must be proved explicitly.
We shall assume it has been proved in the next subsection and
will return later to the question of how this might be done. 

\subsubsection{Converting a Proof Skeleton into a Complete Proof}
\label{subsec:proveProp}

While our ultimate goal is to prove a property of the form
$\forallx{\overline{x}}{R(\overline{t})\imp F}$, we shall focus here on
transforming a proof skeleton into a proof of the property
$\forallx{\overline{x}}{\transRel{R}(\overline{t}) \imp F}$ instead.
Once we have a proof of the latter as well as of the \addT{R} property,
we can easily obtain one of the property of real
interest.

The proofs we want to construct will end as usual with a case
analysis on a possibly-annotated version of the key relation, which
now will be \transRel{R}, followed by rules that introduce logical
symbols in the conclusion formula, and some number of uses of the
induction rule. 
As before, the proof skeleton provides the information from which
the proof is to be constructed automatically.
However, additional care is needed now in elaborating this process.
One reason for such care is that the key relation for the property
verified by the proof skeleton is replaced in the context of
the composite language by its proxy.
This change can be accommodated, as we shall see, by invoking the
\dropT{R} and \addT{R} lemmas at relevant places.
The more complex issue is that of constructing a proof fragment for the
cases arising from the rules for \transRel{R} contributed by
the extensions in the composition that are different from the ones
based on which the proof skeleton was developed.
While similar in spirit to the ideas developed in Section~\ref{subsec:elaborateExt},
there are differing details to be considered. 
The rule that gives rise to the generic proof fragment in the proof
skeleton is one that has the form
\[
\inferrule{T(\unknown,y) \\
           R(x_1,\ldots,x_{i-1}, y, x_{i+1},\ldots,x_k)}
          {R(x_1,\ldots,x_{i-1},\unknown, x_{i+1},\ldots,x_k)},
          \]
where the $x_i$'s and $y$ are distinct (schematic) variables.
In the actual language obtained by composition with extensions
$E_1,\ldots,E_n$, this rule is replaced by ones of the form
\[
\inferrule{\overline{\transRel{R}(\overline{s})} \\ \overline{U} \\
           T(t_i,y) \\
           \transRel{R}(t_1, \ldots, t_{i-1}, y, t_{i+1}, \ldots, t_m)}
          {\transRel{R}(t_1, \ldots, t_{i-1}, t_i, t_{i+1}, \ldots, t_m)},
\]
where, once again, $y$ is a variable that is fresh to the rule.
Ignoring the distinction between $R$ and $\transRel{R}$, we see
the conclusion of the latter rule is obtained from the former not just
by replacing $\unknown$ by $t_i$ but also by substituting the terms
$t_1,\ldots,t_{i-1},t_{i+1},\ldots,t_m$, respectively, for the
variables $x_1,\ldots,x_{i-1},x_{i+1},\ldots,x_m$.
Thus the generic proof fragment would have proved a more general
sequent than would arise in the case of the actual language
identified by the composition.

We define a relation between sequents that helps us bridge the first
of these differences.

\begin{definition}[Projection version of a sequent]
\label{def:seq-rel-h}
  Let ${\cal S} = \sequent{\Sigma}{\Gamma}{F}$ and  ${\cal S}' =
  \sequent{\Sigma'}{\Gamma'}{F'}$ be two sequents that are
  well-formed with respect to some given vocabulary.
  We say that $S'$ is an $R$-projection version of $S$, a relationship
  denoted by $\seqTRelnew{R}{S}{S'}$, if there is a subset
  $\widehat{\Gamma'}$ of $\Gamma'$ such that $\widehat{\Gamma'}$ and $F'$ result
  from $\Gamma$ and $F$ as follows: for each $i > 0$ either every
  formula of the form $(R(\overline{t}))^{*^i}$ and
  $(R(\overline{t}))^{@^i}$ in them has been replaced respectively by  
  $(\transRel{R}(\overline{t}))^{*^i}$ and
  $(\transRel{R}(\overline{t}))^{@^i}$ or none has been so replaced,
  and some occurrences of formulas of the form $R(\overline{t})$
  have been replaced by $\transRel{R}(\overline{t})$. 
\end{definition}

We illustrate this definition using sequents that might arise in the
context of trying to prove \property{secure-correct} and its
projection version.
Let $\mathit{IH}$ represent the formula
\begin{multline*}
  \forallx{s,\Sigma,sl,\gamma_A,\gamma_B,\Sigma',\gamma_A',\gamma_B'}
  {(\evalstmt{\gamma_A}{s}{\gamma_A'})^*  \imp
  \evalstmt{\gamma_B}{s}{\gamma_B'}  \imp
  \secure{\Sigma}{sl}{s}{\Sigma'}  \imp} \\
  \eqpublicvals{\Sigma}{\gamma_A}{\gamma_B}  \imp
  \eqpublicvals{\Sigma'}{\gamma_A'}{\gamma_B'}.
\end{multline*}
Then the sequent that arises in case analysis when the statement
evaluation relation is considered to be derived using the rule for the
$\mathit{seq}$ construct introduced by the host language (rule \textsc{X-seq}
in \fig{R_T-rules}) is the following: 
\begin{multline}
  \Sigma, sl, \gamma_A, \gamma_B,
  \Sigma', \gamma_A', \gamma_B', s_1, s_2, \gamma_A'' : \\
        \mathit{IH}, (\evalstmt{\gamma_A}{s_1}{\gamma_A''})^*,
        (\evalstmt{\gamma_A''}{s_2}{\gamma_A'})^*,
        \evalstmt{\gamma_B}{\mathit{seq}(s_1, s_2)}{\gamma_B'},
        \secure{\Sigma}{sl}{\mathit{seq}(s_1, s_2)}{\Sigma'}, \\
        \eqpublicvals{\Sigma}{\gamma_A}{\gamma_B}
  \longrightarrow
  \eqpublicvals{\Sigma'}{\gamma_A'}{\gamma_B'}
  \label{eq:seqcase}
\end{multline}
There would be a corresponding sequent if we were to try to construct
a proof for the projection version of \property{secure-correct} in the
context of the composed language.
If $\mathit{IH'}$ is the formula
\begin{multline*}
  \forallx{s,\Sigma,sl,\gamma_A,\gamma_B,\Sigma',\gamma_A',\gamma_B'}
  {(\evalstmtT{\gamma_A}{s}{\gamma_A'})^*  \imp
  \evalstmt{\gamma_B}{s}{\gamma_B'}  \imp
  \secure{\Sigma}{sl}{s}{\Sigma'}  \imp} \\
  \eqpublicvals{\Sigma}{\gamma_A}{\gamma_B}  \imp
  \eqpublicvals{\Sigma'}{\gamma_A'}{\gamma_B'},
\end{multline*}
then the sequent that arises during case analysis from considering 
$\transRel{\textsc{X-seq}}$ in \fig{R_T-rules}, \ie, the rule for the
projection version of the relation $\evalstmtRel$, as the basis for
evaluation is the following:
\begin{multline}
  \Sigma, sl, \gamma_A, \gamma_B,
  \Sigma', \gamma_A', \gamma_B', s_1, s_2, \gamma_A'' : \\
        \mathit{IH'}, (\evalstmtT{\gamma_A}{s_1}{\gamma_A''})^*,
        (\evalstmtT{\gamma_A''}{s_2}{\gamma_A'})^*,
        \evalstmt{\gamma_B}{\mathit{seq}(s_1, s_2)}{\gamma_B'},
        \secure{\Sigma}{sl}{\mathit{seq}(s_1, s_2)}{\Sigma'}, \\
        \eqpublicvals{\Sigma}{\gamma_A}{\gamma_B}
  \longrightarrow
  \eqpublicvals{\Sigma'}{\gamma_A'}{\gamma_B'}
  \label{eq:seqprojcase}
\end{multline}
If $\cal S$ is the sequent in the display numbered \ref{eq:seqcase}
and ${\cal S}'$ is the sequent in the display numbered
\ref{eq:seqprojcase}, then $\seqTRelnew{eval}{S}{S'}$ holds. 

\begin{lemma} \label{thm:seq-to-transRel}
    Let $H$ be the host language and $E, E_1,\ldots, E_n$ be extensions
    in a library constructed in the extensibility framework.
    If there is a proof for a sequent $\cal S$ relative to
    $\langComp{H}{\setize{E,E_1,\ldots, E_n}}$ and a set of lemmas
    that includes $\dropT{R}$ and $\addT{R}$, then there must be a
    proof relative to the same language and set of lemmas for any
    sequent ${\cal S}'$ such that
    $\seqTRelnew{R}{\mathcal{S}}{\mathcal{S}'}$ holds.  
\end{lemma}
\begin{proof}
  Let $\frag$ be the derivation for $\cal S$.
  We prove the lemma by induction on the height of $\frag$. 
  The argument proceeds by considering the cases for the
  last inference rule in $\frag$.
  
  We first deal collectively with the situations where the last
  rule is not \defR\ or some variant of \defL\ or \id.
  In all these cases, it is easy to see there is a counterpart in
  $\mathcal{S}'$ to the formula in $\mathcal{S}$ to which the rule
  pertains that enables it to be used in proving $\mathcal{S}'$ from
  premises related via $\seqTRelnewsans{R}$ to the premises
  for $\mathcal{S}$ in $\frag$.
  Obvious applications of the induction hypothesis now complete the
  proof.
  
  Suppose the last rule in $\frag$ is some variant of \defL. 
  If this formula is not of the form $(R(\overline{t}))^{*^i}$,
  $(R(\overline{t}))^{@^i}$, or $R(\overline{t})$ that has been
  replaced by $(\transRel{R}(\overline{t}))^{*^i}$,
  $(\transRel{R}(\overline{t}))^{@^i}$, or
  $\transRel{R}(\overline{t})$  in ${\cal S}'$, then the same
  rule can be used to prove $\mathcal{S}'$ from premises
  that must, again, have proofs by virtue of the induction hypothesis.  
  If the assumption formula is of the form $(R(\overline{t}))^{*^i}$
  or $(R(\overline{t}))^{@^i}$ and has been replaced by
  $(\transRel{R}(\overline{t}))^{*^i}$ or
  $(\transRel{R}(\overline{t}))^{@^i}$ in ${\cal S}'$, then, noting
  the relationship between the rules for $R$ and  $\transRel{R}$, it
  can be seen that a  \defL$^{*^i}$ or a \defL$^{@^i}$ rule can be
  used to derive ${\cal S}'$ from premise sequents related
  via $\seqTRelnewsans{R}$ to those for  $\mathcal{S}$ in $\frag$. 
  Invocations to the induction hypothesis help complete the proof in
  this case.
  A virtually identical argument yields the desired conclusion when
  the formula in $\cal S$ is $R(\overline{t})$ and has been replaced
  by $\transRel{R}(\overline{t})$ in ${\cal S}'$. 
 
  Suppose the last inference rule in $\frag$ is a \defR\ rule.
  Then the conclusion formula for $\cal S$ must not be of the form 
  $(R(\overline{t}))^{*^i}$ or $(R(\overline{t}))^{@^i}$.
  If the same atomic formula is the conclusion formula of
  $\mathcal{S}'$, an identical \defR\ rule would be applicable
  the sequent, leading to a premise sequent that is related via
  $\seqTRelnewsans{R}$ to that for $\mathcal{S}$ in $\frag$.
  The argument can, once again, be completed by invoking the induction
  hypothesis.
  The only remaining possibility is that the conclusion formula is of
  the form $R(\overline{t})$ in $\cal S$ and of the form
  $\transRel{R}(\overline{t})$ in ${\cal S}'$.
  Here we use the argument just outlined to construct a proof for a
  sequent like ${\cal S}'$ except the conclusion formula
  has been replaced by $R(\overline{t})$ and then use \addT{R} as a
  lemma to extend this into a proof for ${\cal S}'$. 

  Finally, suppose that $\frag$ ends with some variant of the
  \id\ rule.
  If the conclusion formula in $\mathcal{S}$ is not of the form
  $R(\overline{t})$, it is easy to see that $\mathcal{S}'$ must be
  derivable by the same variant of the \id\ rule. 
  If the conclusion formula in $\mathcal{S}$ is of the form
  $R(\overline{t})$, it could be matched with an identical assumption
  formula or with one of the form $(R(\overline{t}))^{*^i}$ or
  $(R(\overline{t}))^{@^i}$. 
  In the first case, $R(\overline{t})$ in the assumption set or in
  the conclusion may have been replaced by
  $\transRel{R}(\overline{t})$ in ${\cal S}'$.
  Taking recourse to the lemmas \addT{R} and \dropT{R} we may 
  reduce the provability of ${\cal S}'$ in either situation to that of 
  a sequent in which $R(\overline{t})$ appears in both the assumption
  set and as the conclusion formula.
  In the remaining two cases, the argument is obvious if
  $(R(\overline{t}))^{*^i}$ or $(R(\overline{t}))^{@^i}$ persist in
  ${\cal S}'$. 
  Otherwise, either $(\transRel{R}(\overline{t}))^{*^i}$ or
  $(\transRel{R}(\overline{t}))^{@^i}$ must appear in the assumption
  set in $\mathcal{S}'$ and the conclusion formula may
  be either $\transRel{R}(\overline{t})$ or $R(\overline{t})$.
  Thus ${\cal S}'$ is already in a form to which the \id$^*$ or
  \id$^{@^i}$ rule
  applies or the provability of ${\cal S}'$ can be reduced to that of
  a sequent to which the \id\ rule applies by using an \id$^*$ or an
  \id$^@$ rule and the $\dropT{R}$ lemma.
\end{proof}

We can now state and prove the main result of this subsection.

\begin{thm} \label{thm:full-proof-R_T}
  Let $H$ be a host language and let $E, E_1,\ldots, E_n$ be extensions
  in a library constructed in the extensibility framework.
  Further, let $P$ be a property of the form
  $\forallx{\overline{x}}{R(\overline{t})\imp F}$ in the vocabulary of
  the language $\langComp{H}{\setize{E}}$ with $R$ being a relation introduced
  by $H$.
  Finally let $\cal P$ be a proof skeleton for $P$ relative to the
  language $\langComp{H}{\setize{E,\gen{H}{E}}}$ and the set of lemmas 
  $\lemmaset$ that includes \dropT{R} and \addT{R}.
  Then there must be a proof 
  for
  $\forallx{\overline{x}}
           {\transRel{R}(\overline{t})\imp{}F}$
  relative to the language $\langComp{H}{\setize{\Eall}}$ and $\lemmaset$. 
\end{thm}

\begin{proof}
  The proof skeleton $\cal P$ ends with rules for introducing the
  top-level logical symbols in $P$ and some number of uses of the
  induction rule. 
  The proof of $P$ relative to $\langComp{H}{\setize{\Eall}}$ and
  $\lemmaset$ may end similarly. 
  The preceding part of $\cal P$ is a proof of a sequent $\cal S$ that
  is of the form
  $\sequent{\Sigma}{\Gamma, (R(\overline{t}))^{@^i}}{F}$ or
  $\sequent{\Sigma}{\Gamma, R(\overline{t})}{F}$.
  In the proof we are wanting to construct for $P$ relative to
  $\langComp{H}{\setize{\Eall}}$ and $\lemmaset$, it can easily be
  ascertained that we will be left with an obligation to 
  construct a proof for a sequent ${\cal S}'$ that is of the form
  $\sequent{\Sigma'}{\Gamma',(\transRel{R}(\overline{t}))^{@^i}}{F'}$
  or $\sequent{\Sigma'}{\Gamma',\transRel{R}(\overline{t})}{F'}$
  where $\Gamma'$ and $F'$ are identical to $\Gamma$ and $F$ except
  that, for some $i$, all occurrences of formulas of the form
  $(R(\overline{s}))^{*^i}$ and $(R(\overline{s}))^{@^i}$ 
  in them have been replaced by $(\transRel{R}(\overline{s}))^{*^i}$
  and $(\transRel{R}(\overline{s}))^{@^i}$ respectively, and some
  occurrences of formulas of the form $R(\overline{s})$ have been
  replaced by $\transRel{R}(\overline{s})$.
  We note here that neither $\cal S$ nor ${\cal S}'$ can have
  occurrences of $\unknown$ or any of the constants introduced by the
  extensions $E_1,\ldots, E_n$.
  Now, the last step in the proof of $\cal S$ in $\cal P$ is a
  \defL$^{@^i}$ or a \defL\ rule based on the atomic formula
  $(R(\overline{t}))^{@^i}$ or $R(\overline{t})$.
  The premises in this rule are sequents resulting from considering
  clauses for $R$ that come from two sources: the clause could be one 
  that is contributed directly by $H$ and $E$, or it could be the
  ``pseudo'' projection rule for $R$ contributed by $\gen{H}{E}$.
  Using the fact that all these sequents have derivations, we will 
  show that we can construct a proof for ${\cal S}'$ by using a
  \defL$^{@^i}$ or \defL\ rule oriented around the formula 
  $(\transRel{R}(\overline{t}))^{@^i}$ or
  $\transRel{R}(\overline{t})$, thereby verifying the theorem. 

  Similar to the \defL$^{@^i}$ or \defL\ rule that concludes the proof
  of $\cal S$, the premises for the proposed \defL$^{@^i}$ or
  \defL\ rule for deriving ${\cal S}'$ 
  arise from considering clauses from two sources.  These could be
  ones contributed by $H$ and $E$ or they could be ones contributed by
  one of the extensions $E_1,\ldots,E_n$.
  We have to show each of these premises has a proof.
  Given the relationship between the definition of $\transRel{R}$ and
  $R$ (in the complete language), it is easy to see that if ${\cal
    S}'_1$ is a premise that results from considering a clause for
  $\transRel{R}$ from $H$ or $E$, then there must be a premise
  ${\cal S}_1$ for the rule that derives $\cal S$ in $\cal P$ that is
  such that $\seqTRelnew{R}{{\cal S}_1}{{\cal S}'_1}$ holds.
  Moreover, $\unknown$ cannot appear in either of these sequents.
  Since ${\cal S}_1$ has a proof relative to
  $\langComp{H}{\setize{E,\gen{H}{E}}}$ in which the constraints
  imposed by Definition~\ref{def:proof-frags-e} on proof fragments are
  satisfied, it follows from Lemma~\ref{thm:subst-unknown} that
  ${\cal S}_1$ must have a proof relative to
  $\langComp{H}{\setize{E_1,\ldots,E_n}}$. 
    But, then, by Lemma~\ref{thm:seq-to-transRel}, it must be the case
    that ${\cal S}'_1$ has a proof relative to the same language.

    It only remains to be shown that any premise for the \defL$^{@^i}$
    or \defL\ rule in the proof we are wanting to construct for ${\cal
      S}'$ that derives from a clause for $\transRel{R}$ from one of
    the extensions in $E_1,\ldots,E_n$ has a proof.
    We will do this assuming the rule is \defL$^{@^i}$; the
    argument if it is \defL\ is similar.
    Let ${\cal S}'_2$ be any one of these premises.
    If there is such a premise, we claim there must be a premise
    sequent ${\cal S}_2$ for the \defL$^{@^i}$ rule in $\cal P$ that
    arises from the clause corresponding to the rule for $R$ from
    $\gen{H}{E}$ 
    and that ${\cal S}_2$ has associated with it a sequent ${\cal S}''_2$, 
    a substitution $\theta$, and a term $t$ with a constant
    introduced by one of $E_1,\ldots,E_n$ as its top-level symbol and
    all of whose variables are distinct from those in the
    eigenvariable context for ${\cal S}_2$, that together are such that
    $\seqRel{\appsubst{\theta}{{\cal S}_2}}{{\cal S}''_2}{t}$ and
    $\seqTRelnew{R}{{\cal S}''_2}{{\cal S}'_2}$ hold. 
    If this claim is true, then using Lemmas~\ref{lem:substproof},
    \ref{thm:subst-unknown}, and \ref{thm:seq-to-transRel} and the
    fact that ${\cal S}_2$ has a proof, it follows that there must be
    one for ${\cal S}'_2$ as well.

    To verify the claim, we examine the clause encoding the rule for
    $R$ from $\gen{H}{E}$ and, correspondingly, the clauses encoding
    the rules for $\transRel{R}$ from $E_1,\ldots, E_n$.
    If the former has the form
    $\forallx{\overline{x}}{H \triangleq B}$, then the latter have the
    form
    $\forallx{\overline{x'}}
             {\subst{c(\overline{y})}{\unknown}
                    {\appsubst{\rho_1}{\hat{H}}}
               \triangleq
               \subst{c(\overline{y})}{\unknown}
                     {\appsubst{\rho_1}{(\hat{B} \land B')}}}$,
    where $c$ is a constant introduced by one of the extensions
    $E_1,\ldots,E_n$, $H$ and $\hat{H}$ and, similarly, $B$ and
    $\hat{B}$ are identical formulas except that occurrences of $R$ in
    the first formulas in the pairs have been replaced by
    $\transRel{R}$ in the second formulas in the pairs, and
    $\overline{y}$ is a sequence of variables fresh to
    $\appsubst{\rho_1}{H}$ (and hence also to
    $\appsubst{\rho_1}{\hat{H}}$).
    We may also assume that the variables in $\overline{x}$ and
    $\overline{x'}$ have been named away from $\Sigma$ and $\Sigma'$,
    the eigenvariable contexts of the two sequents
    ${\cal S}$ and ${\cal S}'$. 
    Now, a premise of the kind ${\cal S}'_2$ must result from the use of a
    clause of the second form relative to the sequent ${\cal S}'$, \ie,
    $\sequent{\Sigma'}{\Gamma',(\transRel{R}(\overline{t}))^{@^i}}{F'}$.
    In this case, there must be an \mgu\ for
    $\{\langle \subst{c(\overline{y})}{\unknown}
                     {\appsubst{\rho_1}{\hat{H}}},
               \transRel{R}(\overline{t}) \rangle \}$.
    Using Theorems~\ref{thm:substunif} and \ref{thm:unifsubst}
    (Appendix~\ref{app:unification}), we see 
    we can write such an \mgu\ as
    $\subst{c(\overline{y})}{\unknown}{\rho_2}$, 
    where $\rho_2$ is an \mgu\ for
    $\{\langle \appsubst{\rho_1}{\hat{H}},
               \transRel{R}(\overline{t}) \rangle \}$; we make use of
    the fact that $\unknown$ does not occur in
    $\transRel{R}(\overline{t})$ and, hence,
    $\transRel{R}(\overline{t}) =
    \subst{c(\overline{y})}{\unknown}{\transRel{R}(\overline{t})}$
    as also the observation that $\rho_2$, being an \mgu\ for 
    $\{\langle \appsubst{\rho_1}{\hat{H}},
               \transRel{R}(\overline{t}) \rangle \}$,
    must not include variables not appearing in the unification
    problem in its domain.
    Using such an \mgu, we may determine ${\cal S}'_2$ to be a sequent of the form
    \[\sequent{\Sigma_2'}
             {\appsubst{\subst{c(\overline{y})}{\unknown}{\rho_2}}
                       {\Gamma'},
              (\appsubst{\subst{c(\overline{y})}{\unknown}{\rho_2}}
                       {\subst{c(\overline{y})}{\unknown}
                              {\appsubst{\rho_1}{B'}}})^{*^i},
              (\appsubst{\subst{c(\overline{y})}{\unknown}{\rho_2}}
                       {\subst{c(\overline{y})}{\unknown}
                              {\appsubst{\rho_1}{\hat{B}}}})^{*^i}}
            {\appsubst{\subst{c(\overline{y})}{\unknown}{\rho_2}}{F'}}.\]
   Since there are no occurrences of $\unknown$ or the variables in
   the domain of $\rho_1$ in $\Gamma'$ and $F'$, we may write these
   respectively as
   $\subst{c(\overline{y})}{\unknown}{\appsubst{\rho_1}{\Gamma}}$
   and
   $\subst{c(\overline{y})}{\unknown}{\appsubst{\rho_1}{F}}$.
   But then, using Theorem~\ref{thm:subst-distr}, ${\cal S}'_2$ may be
   rewritten as
   \[
    \sequent{\Sigma_2'}
              {\subst{c(\overline{y})}
                     {\unknown}
                     {\appsubst{\rho_2 \circ \rho_1}{\Gamma'}},
               (\subst{c(\overline{y})}{\unknown}
                     {\appsubst{\rho_2 \circ \rho_1}{B'}})^{*^i},
               (\subst{c(\overline{y})}{\unknown}
                     {\appsubst{\rho_2 \circ \rho_1}{\hat{B}}})^{*^i}}
              {\subst{c(\overline{y})}{\unknown}
                     {\appsubst{\rho_2 \circ \rho_1}{F'}}}.
   \]
   Now, since $\transRel{R}(\overline{t})$ is unaffected by $\rho_1$
   and $\rho_2$ unifies
   $\{\langle \appsubst{\rho_1}{\hat{H}},
               \transRel{R}(\overline{t}) \rangle \}$,
   it follows that $\rho_2 \circ \rho_1$ is a unifier for
   $\{\langle \hat{H},\transRel{R}(\overline{t}) \rangle \}$.               
   Hence, there must be an \mgu\ for
   $\{\langle \hat{H},\transRel{R}(\overline{t}) \rangle \}$ or,
   identically for $\{\langle H, R(\overline{t}) \rangle \}$ and
   there must therefore be a premise of the form ${\cal S}_2$ claimed
   to exist for the \defL$^{@^i}$ rule in $\cal P$.
   In more detail, ${\cal S}_2$ must be a sequent of the form
   $\sequent{\Sigma_2}
            {\appsubst{\delta}{\Gamma}, (\appsubst{\delta}{B})^{*^i}}
            {\appsubst{\delta}{F}}$
   where $\delta$ is some \mgu\ for $\{\langle H, R(\overline{t})
   \rangle \}$.
   Since $\rho_1 \circ \rho_2$ must be a unifier for
   $\{\langle H, R(\overline{t}) \rangle \}$,
   there must be a substitution $\delta'$ such that
   $\rho_2 \circ \rho_1 = \delta' \circ \delta$.
   But then, if we pick $t$ to be the term $c(\overline{y})$,
   $\theta$ to be the substitution $\delta'$ and ${\cal S}''_2$ to be
   the sequent
   $\subst{c(\overline{y})}
             {\unknown}
             {\appsubst{\delta'}{{\cal S}_2}}$,
   it can be verified that
   $\seqRel{\appsubst{\theta}{{\cal S}_2}}{{\cal S}''_2}{t}$ and
   $\seqTRelnew{R}{{\cal S}''_2}{{\cal S}'_2}$ hold. 
\end{proof}

The proofs of Theorem~\ref{thm:full-proof-R_T} and the
lemmas on which it depends are clearly constructive, implying thereby that
there is a procedure for extracting a proof of the formula
$\forallx{\overline{x}}{\transRel{R}(\overline{t})\imp{}F}$ relative to a
composite language once we have a proof skeleton for
the formula. 
We can, in turn, use this and \addT{R} to build a proof of the
original property.
\begin{thm} \label{thm:full-proof-host}
  Let $H$ be a host language and let $E, E_1,\ldots, E_n$ be extensions
  in a library constructed in the extensibility framework.
  Further, let $P$ be a property of the form
  $\forallx{\overline{x}}{R(\overline{t})\imp F}$ where
  $R$ is a relation introduced by $H$ and let $\cal P$ be a proof
  skeleton constructed for $P$ by $E$ using a set of lemmas
  $\lemmaset$ that includes \dropT{R} and \addT{R}.
  Then there must be a proof of $P$ relative to the language
  $\langComp{H}{\setize{\Eall}}$ using the lemmas in $\lemmaset$.
\end{thm}
\begin{proof}
  Using $\cal P$ and \refthm{full-proof-R_T}, we can construct a
  proof for $P'=\forallx{\overline{x}}{\transRel{R}(\overline{t})\imp
    F}$.
  If we are given $R(\overline{t})$, \addT{R} allows us to conclude
  that $\transRel{R}(\overline{t})$ holds.
  We can now use this and $P'$ to derive $F$, thereby concluding that
  $\forallx{\overline{x}}{R(\overline{t})\imp F}$ must hold.
\end{proof}

\subsubsection{Showing that Behavior is Preserved Through a Projection}
\label{subsec:addT}

The scheme we have identified for an extension to prove a
metatheoretic property whose key relation, $R$, is introduced by the
host language relies crucially on showing this relation is
preserved under projections and that the definition of the relation 
is inductively smaller when viewed through projections; this is, in
fact, the real content of the \addT{R} property. 
The requirement that $R$ satisfies these conditions is obviously a
foundational one and, in particular, has the character of a projection
constraint. 
As such, any mechanism for proving such properties modularly may be
used to establish \addT{R}.
However, there are aspects of this property that merit special
treatment for which we identify the structure described below. 

The only systematic way to prove a property of the form
$\forallx{\overline{x}}{R(\overline{x})\imp\transRel{R}(\overline{x})}$  would be to use
an induction on the definition of the relation $R(\overline{x})$.
A difficulty with doing this directly is that the definition of
$R(\overline{x})$ does not give us a handle on the projections that are
possible on the primary component of the relation, something
playing a critical role in the definition of the relation
$\transRel{R}(\overline{x})$.
To deal with this situation, we introduce an intermediate relation
that additionally records the number of possible projection steps.
\begin{definition}[Extension Size of a Relation] \label{def:extSize}
Let $H$ be a host language and $E_1,\ldots, E_n$ be a set of extensions
in a library constructed in the extensibility framework.
The extension size version of a relation $R$ introduced by the host
language ($R\in\relset{H}$), written $\extSize{R}$, is a relation
that adds a natural number argument to those of $R$ and is
defined by rules derived from those for $R$ as follows.\footnote{We
  assume the usual inductive encoding of natural numbers using a
  constant $z$ to denote $0$ and the unary constructor $s$ to denote
  the successor operation. Moreover, addition and ordering relations
  on terms denoting natural numbers are defined in the obvious
  way. The equality relation and the summation operation on natural
  numbers are shorthands for a more elaborate presentation using the
  addition relation.}
Assuming the $i^{th}$ argument of $R$ is its primary component, let 
\[
\inferrule{\overline{R(\overline{s})} \\ \overline{U}}
          {R(\overline{t})}
          \]
be a rule for it, where $\overline{R(\overline{s})}$ denotes a set of
premises in which $R$ appears and $\overline{U}$ denotes the remaining
premises. If this is a rule in the collection corresponding to the
host language, then the definition of $\extSize{R}$ includes the rule
\[
\inferrule{\overline{\extSize{R}(\overline{s}, n_i)} \\ \overline{U} \\
           n = \sum{n_i}}
          {\extSize{R}(\overline{t}, n)}
\]
If the rule is in the collection corresponding to one of the
extensions $E_1,\ldots, E_n$, on the other hand, then the counterpart
rule for $\extSize{R}$ has the form 
\[
\inferrule{\overline{\extSize{R}(\overline{s}, n_i)} \\ \overline{U} \\
           n = 1 + \sum{n_i}}
          {\extSize{R}(\overline{t}, n)}
\]
Observe the definition of $\extSize{R}$ builds in a count of the
number of projections that are possible in the definition of
$R(\overline{t})$. 
\end{definition}

The proof of \addT{R} can now be split into showing the following two
properties: 
\begin{eqnarray*} 
  \forallx{\overline{x}}
          {R(\overline{x}) \imp \existsx{n}{\extSize{R}(\overline{x},n)}}
  \label{eq:toR_ES}\\
  \forallx{\overline{x},n}{\extSize{R}(\overline{x}, n) \imp
  \transRel{R}(\overline{x})} \label{eq:toR_T}
\end{eqnarray*}
The first of these has an obvious inductive proof and, in fact, one
that can be automatically generated based on the relationship between
the definitions of $R$ and $\extSize{R}$.
The second property has a structure similar to \addT{R}, except that 
we now have the capability of basing an inductive proof not just on
the derivation of the relation but also on the natural number argument
that counts the number of possible projections.

In a modular proof, arguments for different cases for the definition
of $\extSize{R}$ will have to be provided by the host language and
each extension, as for any other foundational property.
The treatment in the case of the host language can be based on an
induction on the definition of $\extSize{R}$ and has a
straightforward, formulaic structure.
The treatment in the case of an extension is more complex and this is
actually where the substance of the argument lies.
In this case, the projection described by the extension would have to
be analyzed explicitly and shown to have a behavior identical to that
described for $R$ directly in the extension.
In establishing this, the induction on the natural number argument of
$\extSize{R}$ will allow for the assumption of the property on any
embedded projections.

\subsubsection{Assessing the Proposed Proof Construction Approach}
\label{sssec:generic-assessment}

The approach we have described to establishing a metatheoretic
property introduced by an extension and whose key relation is
introduced by the host language is useful only when two requirements
are met.  
First, there should be relations basic to the language
structure that the host language may describe and whose behavior on
extension-introduced constructs it can tightly constrain.
Second, extensions should be able to describe interesting
metatheoretic properties oriented around such relations.
There are paradigmatic situations in which the first of these
requirements is readily obtained.
One example of this is that where extensions preserve the basic
structure of the host language but provide both convenience in syntax
and new abstraction mechanisms.
Semantic attributes that adhere to the syntactic structure of
expressions and identify static characteristics in such cases 
may not have the desired property.
However, ones describing dynamic behavior, such as evaluation
relations, often do.
We posit that metatheoretic properties with which an extension might want to
associate a conglomeration of a priori unknown language
components will usually be based not on the specific syntactic
characteristics of expressions but rather on their dynamic semantics.

We see this in our example language library with statement evaluation.
Statement evaluation can be viewed as simply updating the evaluation
context and branching, in which case it is clear that any behavior an
extension might introduce for a new statement form should be
achievable by some combination of existing statement forms.
It is also natural for extensions to describe interesting properties
oriented around evaluation, as our security extension does.
That extension is
able to introduce a property oriented around statement evaluation,
demonstrating its new static analysis accurately reflect the dynamic
behavior of a program as far as information flow from private to
public variables is concerned.

\section{A Practical Realization of Modular Reasoning}
\label{sec:implementation}

The results of this paper are meant to provide the basis for an
actual system for the metatheoretic analysis of extensible languages
and we have in fact begun an implemention towards this end. 
We sketch this work below to provide a sense of how we expect our
theoretical results to be utlized; a comprehensive description is
orthogonal to the focus of this paper.

Our implementation comprises two systems called {\it
  Sterling} and 
{\it Extensibella}~\cite{mmel.website}.
Sterling can be viewed as providing a front-end to the verification system:  this system
provides facilities for developers to identify the host language and
extensions for particular language libraries and further to specify
each of these components in the style described in
Section~\ref{sec:framework}. 
More specifically, for each component, the developer
identifies its syntactic categories and their associated constructors,
the semantic relations with their types and primary components, and
the rules defining these relations.
Sterling checks these specifications are well-formed and
fit together in the sense explained in
\sec{framework-composing-languages}.
If a component passes this test, it is converted into a form
that can provide the basis for reasoning within the logic $\cal G$
described in Section~\ref{sec:abella}.

Extensibella is the system that supports the actual development of
proofs. 
In substance, Extensibella supports the interactive proof
development style intrinsic to {\it
  Abella}~\cite{baelde14jfr}, the proof assistant for the logic $\cal
G$.
Abella is a system that supports a tactics-style
  development of a proof: a state in the proof assistant is
  characterized by an ordered collection of sequents to be proved, 
  and a tactic command is invoked to replace the first of these
  with a sequence of new sequents representing proof obligations
  relative to an inference rule.
  Extensibella retains this structure.
  In fact, Extensibella implements its functionality by delegating the
  realization of proof steps to Abella.
However, the development of proofs in the extensibility context must
adhere to constraints not imposed in Abella.
One such constraint is that proofs must adhere to a canonical
structure.
Moreover, the same canonical form must be used for proofs constructed
in the different components for foundational properties.
Finally, uses of case analysis must be restricted as per
Definition~\ref{def:proof-frag} when a proof is being constructed for
a foundational property and as per Definition~\ref{def:proof-frags-e}
when a proof is being constructed for an auxiliary property.
Extensibella checks these constraints are satisfied before
utilizing Abella to realize the proof step.

There is additional structure relevant to the development of
proofs for auxiliary properties.
The proofs of these properties are the responsibility of only the
extensions that introduce them.
However, within that context, it is necessary to include a proof of
the generic case if this is relevant for the formula to be proved.
Further, when an extension introduces a property where the key
relation $R$ is introduced by the host language, a proof must be generated
for the \dropT{R} property and it must be ascertained that the \addT{R}
property is available as a property introduced by the host language.
These tasks are also the responsibility of Extensibella.

Much of the discussion up to this point has been oriented around the
proof of a single property, with others being available as lemmas.
However, the reality is that it is a \emph{collection} of properties associated
with languages.
Moreover, there must be a notion of ordering associated with this
collection that allows some properties to be used as lemmas in the
proofs of others.
This ordering is realized as follows.
First, the host language identifies the order for the foundational
properties it introduces.
Then each extension uses this order and intersperses the properties it
introduces between host language properties.
Of course, an extension property that needs to use a host language
property in its proof must appear later in the sequence; this must be
the relationship, for example, between \addT{R} and an auxiliary
property whose key relation is $R$.
Once the ordering is determined and checked for the properties, the
developer of each component can construct their proofs in the manner
discussed in this paper, using properties earlier in the ordering as
lemmas.
If each extension's ordering is consistent with that inherited from the host
language, we are guaranteed there will be an
overall ordering for the properties that will work for any
composition.

Our implementation does not currently span all the aspects of what is
needed in a completely configured system for the modular development
of metatheory for extensible languages.
At the moment, it provides the functionality needed to develop proof
fragments for foundational properties and proof skeletons for
auxiliary properties.
The mechanism for composing proof fragments for foundational
properties and for elaborating proof skeletons using the constructive
content of Theorems~\ref{thm:hostProps}, \ref{thm:full-proof-ext}, and
\ref{thm:full-proof-host} is under development.
We would also like to build an interface for automatically generating
language specifications for Extensibella from those provided using
formalisms such as attribute grammars and that are used in systems
such as Silver~\cite{vanwyk10scp} to support concrete realizations of extensible 
languages in the style considered in this paper.

\section{Related Work}
\label{sec:relatedWork}

The work described in this paper builds around a particular vision for
programming languages, that where extensibility is realized by
providing targeted features by picking particular extensions from a
library built around a basic core.
Within this context, we have also taken it as a given that the
compatibility of an extension in the library with the core, represented
by a host language, and with other extensions should be determinable
at the time of its conception rather than when a composition is
attempted.
These perspectives have three important consequences with regard to our
approach to metatheory.
First, it should be possible to think of metatheoretic properties
pertinent to features introduced by an extension in addition
to those that apply to the language as a whole.
Second, it should be possible to modularize and thereby to distribute
the work associated with verifying properties to each component in
the library so it can be carried out as the component is being
elaborated. 
Finally, such independently-performed verification work must still
provide a guarantee of the relevant properties for any composite
language, \ie, the soundness of proof composition and elaboration is
essential to our work.

While there has been other work related to metatheoretic analysis for
extensible languages, our effort is distinctive in what it takes as
fundamental requirements.
The particular focus and the results obtained are therefore also
different from that in most other work in this realm.
We bring this observation out by discussing such work specifically
below.
These other efforts may be characterized broadly based on the view
they take of language extensibility, whether this is realized by
combining parts with equal status at a linguistic level or by
building around a core language as in this paper.
We break down our discussion accordingly.

\subsection{Extensibility via the Combination of Complementary Components}
In languages built by complementary components, no portion of the
language is more essential than any other, so there is no requirement
for an identified host
language.
The language components build on a shared set of
declarations of syntactic categories and semantic analyses.
Each component declares a portion of the language, adding constructors for
the syntactic categories and rules for the semantic analyses for those
constructors.
For example, one component might introduce Boolean
expressions and declare rules for typing and evaluating them, while
another might introduce arithmetic expressions with their typing and
evaluation rules.
Because the shared base limits the semantics the components can
describe, there isn't a possibility for them to add new analyses.
Correspondingly, in this setting, the idea of associating properties
with such analyses---the notion of auxiliary properties in this
paper---is not meaningful.

Work has nevertheless been done on developing proofs for
metatheoretic properties for extensible language that are obtained
through this kind of composition.
Like their syntactic categories and
semantic analyses, properties for such languages are declared ahead
of the components being written, with the proof work distributed
across the components.
This makes them similar to our
foundational properties, those introduced by a host language.
One early work, Proof Weaving~\cite{mulhern06icfp}, has each component
provide a full proof of each property in Coq, then pulls these proofs
apart to try to rebuild them into a proof for a composed language.
However, there is no notion corresponding to the soundness of proof
composition in this paper: 
the process described
can fill out
the structure of the full proof and the portions known when each
component proof was written but can, and often does, leave 
holes the language composer would need to fill to complete the
proof.

Another work, Meta-Theory a la
Carte~\cite{delaware13thesis,delaware13popl}, also requires each
language component to provide proofs for their properties in Coq.
However, the encoding used for the language description and proofs
prevents examining the shape of sub-terms and how values are computed.
This is similar to our restrictions on case analysis, and allows the
proofs to be composed automatically to form a working proof.  There is
one wrinkle in this work, however, as the authors note canonical forms lemmas (\eg, a
value of an integer type is an integer) cannot be stated to be shared
by all components, as their statements rely on the specific
vocabularies introduced by certain components;
thus the proofs must be produced in an ad hoc manner at composition
time.
Similar work was done in Agda at the same time~\cite{schwaab13plpv},
but the authors
report they did not succeed in getting Agda to accept the
composed proofs.

\subsection{Extensibility by Building around a Core Language}
One example of this approach is the
\textsc{fpop}~\cite{jin23pldi} system that encodes a version of
extensible languages
in Coq using the ideas of inheritance and family polymorphism,
translating this down into basic Coq.  Extensions inherit from a base
language, adding to its inductive types for syntax and adding new cases
to functions defining semantics, all without needing to modify the
inherited definitions.  Properties are proven in the context of the
base language and each extension.  An extension may either reuse an
inherited proof or override it.
Unlike our approach, \textsc{fpop} does not have automatic composition
of language semantics nor a guarantee of properties introduced by an
extension holding for a composition.  Compositions of both semantics
and proofs rely on glue code for new constructors.  Because new
semantic definitions and proof
cases must be written for constructors from other extensions, it is
sometimes the case that a property given by one extension will not
hold for a constructor from another extension.
Thus \textsc{fpop} is most useful for adding features to a language
one at a time, with each building on the last, as opposed to our view
where features are added independently of each other and able to be
added to a composition or left out as desired.

Another exemplar of this approach to extensibility, with an
extensibility model much more similar to ours than \textsc{fpop}'s
model, is the SoundX~\cite{lorenzen16popl}
system.
This system allows extensions to introduce
new syntax and typing rules for that syntax, which then desugar to the
host language.
The focus in this work is to provide an automatic method for checking
that a well-typed term
in the extended language desugars into a well-typed term in the host
language.
This property permits the host language designer to prove properties
about typing, such as type soundness, for the host language alone.
Programs written using extension-introduced syntax then have these
properties when their desugared versions are run without proving the
properties for the extension-introduced syntax because the desugared
versions are well-typed.
We can prove properties similarly in our framework;
however, our framework goes beyond this model in allowing evaluation
to be defined 
on the extended syntax, with the properties about typing proven for
the extended syntax directly.
Note also SoundX is limited to just this one property.
Moreover, because the approach is based on desugaring, extensions
cannot introduce new analyses and properties about them, as is done in
the security extension example.

Finally, another framework for proving properties~\cite{kaminski2017sle} works
for extensible languages written in attribute grammars specified in Silver.  It allows
the host language or any extension to introduce properties and prove them only for
constructs from the host language.  The property then transfers to
other constructs across projection, called forwarding~\cite{vanwyk2002cc} in the setting
of attribute grammars, by requiring attribute values to
be essentially equal across projection and proving each property holds
when relevant values are essentially equal.
We share some of the same broad outlines of ideas, with properties
carrying across projection, but our system is less restrictive.
Rather than requiring values to be essentially equal, we allow the
language designer to decide how values should be related---this idea
is embedded in the notion of projection constraints.  This may lead
to more complex proof work as a trade-off, since the extension
designer introducing a property must show \emph{why} it holds across
translation rather than relying on values being basically the same.

\section{Conclusion and Future Work}
\label{sec:conclusion}

We have presented a modular approach to proving properties for
programming languages constructed from a host language
through the addition of a collection of independently-developed
extensions.
This approach allows both the host language and each extension to
introduce properties and to reason about them in the context of
the host language alone or the host language augmented by the extension,
with a guarantee the reasoning will ensure the properties
will hold for any composition of extensions with the host language
without any additional work required of the person determining
the composition.
More specifically, we have shown how the modular reasoning
can be used automatically to extract proofs of the relevant
properties for any composed language.

While the approach we have described is quite broad, its actual
realization in this paper is limited.
We intend to address some of these limitations in future work.
One limitation we would like to ease arises from the restricted
form we permit for projection rules.
We currently require the rules to have exactly two premises, one that
determines a projection and another that ensures the relation of
interest holds for the projection, with all other arguments being
identical.
A particularly limiting form of this requirement is seen in the pseudo
projection rules used in the generic extension to reason
about the behavior in undetermined extensions of relations
introduced by the host language.
As we have seen in Section~\ref{subsec:elaborateHost} and discussed
more specifically in Section~\ref{sssec:generic-assessment}, this
requires the behavior of the relation to be identical to that of
the projection.
However, in many reasoning examples, we can do with weaker
requirements.
An example of this is provided by \property{tc-s-eval-results-back},
where the requirement is only that the evaluation of a statement and
its projection should produce states that bind variables to the same
values.
Easing the restriction on projection rules should allow languages more
freedom in their definitions and should also expand the reasoning
ability.

Another current limitation is our restriction to a host language and
extensions that build on only it, where each extension is
independent of every other extension.
Systems implementing extensible
languages, such as the Silver attribute grammar
system~\cite{vanwyk10scp} or the Sterling system we use for language
specifications for our implementation, permit extensions to
build on other extensions in addition to the host language.
A use of this capability can be 
seen in the \textsc{ableC}~\cite{kaminski2017oopsla} extensible
specification of C written in Silver, where some extensions build on
other extensions that provide intermediate levels of abstraction
between them and the lower-level C host language.
Our framework will be applicable to more language specifications if
we can handle more general extension schemes.

In another direction, we intend to continue developing examples of the
use of our framework for proving properties.
These examples will allow us to explore the trade-off between the strength of
projection constraints and the freedom extensions have for defining their
semantics.
Strong projection constraints make the semantics of extensions easier
to model even without detailed knowledge of them, and thus make it
easier for other extensions to introduce and prove properties.
For example, we might have a constraint for expression evaluation
requiring evaluation results to be closely related across projections.
This would support properties about the values resulting from
evaluation, something \projConstr{eq:tc-e-eval}, that preserves only
variables under a projection, would not support.
However, strong projection constraints will also restrict the semantic
definitions extension writers can create, and may make it impossible
to construct some extensions.
We believe the example language we have used in
this paper leans toward loose projection constraints, but further
investigation of the trade-offs is needed.

\bibliographystyle{acm}
\bibliography{bibs}

\begin{thebibliography}{10}

\bibitem{baelde14jfr}
{\sc Baelde, D., Chaudhuri, K., Gacek, A., Miller, D., Nadathur, G., Tiu, A.,
  and Wang, Y.}
\newblock Abella: A system for reasoning about relational specifications.
\newblock {\em Journal of Formalized Reasoning 7}, 2 (December 2014).

\bibitem{delaware13thesis}
{\sc Delaware, B.}
\newblock {\em Feature Modularity in Mechanized Reasoning}.
\newblock PhD thesis, University of Texas at Austin, Austin, Texas, USA, 2013.

\bibitem{delaware13popl}
{\sc Delaware, B., d.~S.~Oliveira, B.~C., and Schrijvers, T.}
\newblock {M}eta-theory \`{a} la carte.
\newblock In {\em Proceedings of the 40th Annual ACM SIGPLAN-SIGACT Symposium
  on Principles of Programming Languages\/} (New York, NY, USA, 2013), POPL
  '13, ACM, pp.~207--218.

\bibitem{ekman2007oopsla}
{\sc Ekman, T., and Hedin, G.}
\newblock The {JastAdd} extensible {Java} compiler.
\newblock In {\em Proceedings of the Conference on Object Oriented Programming,
  Systems, Languages, and Systems (OOPSLA)\/} (2007), ACM, pp.~1--18.

\bibitem{erdweg11oopsla}
{\sc Erdweg, S., Rendel, T., Kastner, C., and Ostermann, K.}
\newblock {SugarJ}: Library-based syntactic language extensibility.
\newblock In {\em Proceedings of the Conference on Object Oriented Programming,
  Systems, Languages, and Systems (OOPSLA)\/} (2011), ACM, pp.~391--406.

\bibitem{gacek11ic}
{\sc Gacek, A., Miller, D., and Nadathur, G.}
\newblock Nominal abstraction.
\newblock {\em Information and Computation 209}, 1 (2011), 48--73.

\bibitem{jin23pldi}
{\sc Jin, E., Amin, N., and Zhang, Y.}
\newblock Extensible metatheory mechanization via family polymorphism.
\newblock {\em Proc. ACM Program. Lang. 7}, PLDI (jun 2023).

\bibitem{kaminski17phd}
{\sc Kaminski, T.}
\newblock {\em Reliably Composable Language Extensions}.
\newblock PhD thesis, University of Minnesota, Minneapolis, Minnesota, USA,
  2017.

\bibitem{kaminski2017oopsla}
{\sc Kaminski, T., Kramer, L., Carlson, T., and Van~Wyk, E.}
\newblock Reliable and automatic composition of language extensions to {C}: The
  {ableC} extensible language framework.
\newblock {\em Proceedings of the ACM on Programming Languages 1}, OOPSLA (Oct.
  2017), 98:1--98:29.

\bibitem{kaminski2012sle}
{\sc Kaminski, T., and Van~Wyk, E.}
\newblock Modular well-definedness analysis for attribute grammars.
\newblock In {\em Proceedings of the 5th International Conference on Software
  Language Engineering (SLE)\/} (Sept. 2012), vol.~7745 of {\em Lecture Notes
  in Computer Science}, Springer-Verlag, pp.~352--371.

\bibitem{kaminski2017sle}
{\sc Kaminski, T., and Van~Wyk, E.}
\newblock Ensuring non-interference of composable language extensions.
\newblock In {\em Proceedings of the 10th ACM SIGPLAN International Conference
  on Software Language Engineering (SLE)\/} (October 2017), ACM, pp.~163--174.

\bibitem{lorenzen16popl}
{\sc Lorenzen, F., and Erdweg, S.}
\newblock Sound type-dependent syntactic language extension.
\newblock {\em SIGPLAN Not. 51}, 1 (Jan. 2016), 204--216.

\bibitem{martelli82}
{\sc Martelli, A., and Montanari, U.}
\newblock An efficient unification algorithm.
\newblock {\em ACM Transactions on Programming Languages and Systems 4}, 2
  (Apr. 1982), 258--282.

\bibitem{mmel.website}
{\sc Michaelson, D., Nadathur, G., and Van~Wyk, E.}
\newblock Modular metatheory for extensible languages webpage, Dec. 2023.
\newblock http://mmel.cs.umn.edu.

\bibitem{mulhern06icfp}
{\sc Mulhern, A.}
\newblock Proof {W}eaving.
\newblock In {\em Proceedings of the First Informal {ACM} {SIGPLAN} Workshop on
  Mechanizing Metatheory\/} (Portland, Oregon, September 2006), The Eleventh
  ACM SIGPLAN International Conference on Functional Programming.

\bibitem{schwaab13plpv}
{\sc Schwaab, C., and Siek, J.~G.}
\newblock Modular {T}ype-{S}afety {P}roofs in {A}gda.
\newblock In {\em Proceedings of the 7th Workshop on Programming Languages
  Meets Program Verification\/} (New York, NY, USA, 2013), PLPV '13, ACM,
  pp.~3--12.

\bibitem{schwerdfeger2009pldi}
{\sc Schwerdfeger, A., and Van~Wyk, E.}
\newblock Verifiable composition of deterministic grammars.
\newblock In {\em Proceedings of the ACM SIGPLAN Conference on Programming
  Language Design and Implementation (PLDI)\/} (2009), ACM, pp.~199--210.

\bibitem{tiu04phd}
{\sc Tiu, A.}
\newblock {\em A Logical Framework for Reasoning about Logical Specifications}.
\newblock PhD thesis, Pennsylvania State University, May 2004.

\bibitem{vanwyk10scp}
{\sc Van~Wyk, E., Bodin, D., Gao, J., and Krishnan, L.}
\newblock Silver: an extensible attribute grammar system.
\newblock {\em Science of Computer Programming 75}, 1--2 (January 2010),
  39--54.

\bibitem{vanwyk2002cc}
{\sc Van~Wyk, E., de~Moor, O., Backhouse, K., and Kwiatkowski, P.}
\newblock Forwarding in attribute grammars for modular language design.
\newblock In {\em Proceedings of the 11th Conference on Compiler Construction
  (CC)\/} (2002), vol.~2304 of {\em Lecture Notes in Computer Science},
  Springer-Verlag, pp.~128--142.

\end{thebibliography}

\newpage

\appendix

\section{Full Language Library} \label{app:language}

We have here the details of the host language and each extension
included in our language library.

\newcommand{\grmmrsep}{~~$|$~~}
\newcommand{\oneColWid}{\textwidth}
\newcommand{\twoColWidL}{0.5\textwidth}
\newcommand{\twoColWidR}{0.5\textwidth}
\newcommand{\threeColWid}{0.33\textwidth}
\newcommand{\shortenSpace}{\vspace*{-0.2in}}
\newcommand{\tableTogetherSpace}{\vspace{-0.3in}}
\newcommand{\preTableSpace}{\vspace*{-0.1in}}

\subsection{Host Language $H$}
\subsubsection{Syntax}
$\ntset{H} = \{ s, e, n, i, ty, \Gamma, \gamma \}$

\bigskip\noindent
$\constrset{H}$:

\begin{tabular}{lclp{1in}lcl}
$s$ & ::= & $\mathit{skip}$
      \grmmrsep $\mathit{decl}(n, ty, e)$
      \grmmrsep $\mathit{assign}(n, e)$
      \grmmrsep $\mathit{seq}(s, s)$
      \grmmrsep $\mathit{ifte}(e, s, s)$
      \grmmrsep $\mathit{while}(e, s)$ \\
$e$ & ::= & $\mathit{var}(n)$
  \grmmrsep $\mathit{intlit}(i)$
  \grmmrsep $\mathit{true}$
  \grmmrsep $\mathit{false}$
  \grmmrsep $\mathit{add}(e, e)$
  \grmmrsep $\mathit{eq}(e, e)$
  \grmmrsep $\mathit{gt}(e, e)$
  \grmmrsep $\mathit{not}(e)$ \\
$\mathit{ty}$ & ::= &  $\mathit{int}$
      \grmmrsep $\mathit{bool}$ \\
$\Gamma$ & ::= & $\mathit{nilty}$
       \grmmrsep $\mathit{consty}(n, ty, \Gamma)$ \\
$\gamma$ & ::= & $\mathit{nilval}$
       \grmmrsep $\consval{n}{e}{\gamma}$
\end{tabular}

\subsubsection{Relations}
\begin{tabular}{ll}
  \hspace*{-0.15in}
  \relset{H} = & \hspace*{-0.15in}
                 \{\lookuptype{\Gamma^*}{n}{ty}, \ \
                 \notpresenttype{\Gamma^*}{n}, \ \
                 \lookupval{\gamma^*}{n}{e}, \ \
                 \val{e^*}, \ \
                 \vars{e^*}{2^{n}}, \\
               & \typeexpr{\Gamma}{e^*}{ty}, \ \
                 \typestmt{\Gamma}{s^*}{\Gamma}, \ \
                 \updateCtx{\gamma^*}{n}{e}{\gamma}, \ \
                 \removeCtx{\gamma^*}{n}{\gamma}, \ \
                 \evalexpr{\gamma}{e^*}{e}, \ \
                 \evalstmt{\gamma}{s^*}{\gamma}\}
\end{tabular}

\bigskip
\noindent $\ruleset{H}$ contains the following rules:

\newcommand{\spaceholder}{\hspace*{0.1in}}

\begin{center}
\framebox{$\lookuptype{\Gamma^*}{n}{ty}$}
\end{center}
\preTableSpace
\begin{longtable}{b{\twoColWidL}b{\twoColWidR}}
\shortenSpace
\[
  \inferrule*[right=LT-Cons-Head]
     { \spaceholder }
     {\lookuptype{\mathit{consty(n,ty,\Gamma)}}{n}{ty}}
\]
&
\shortenSpace
\[
  \inferrule*[right=LT-Cons-Tail]
     {n \not = n' \\
      \lookuptype{\Gamma}{n}{ty}}
     {\lookuptype{\mathit{consty(n',ty',\Gamma)}}{n}{ty}}
\]
\\
\end{longtable}

\begin{center}
\framebox{$\notpresenttype{\Gamma^*}{n}$}
\end{center}
\preTableSpace
\begin{longtable}{b{\twoColWidL}b{\twoColWidR}}
\shortenSpace
\[
  \inferrule*[right=NBT-Nil]
    { \spaceholder }
    {\notpresenttype{\mathit{nilty}}{n}}
\]
&
\shortenSpace
\[
  \inferrule*[right=NBT-Cons]
     {n \not = n' \\
      \notpresenttype{\Gamma}{n}}
     {\notpresenttype{\mathit{consty(n',ty',\Gamma)}}{n}}
\]
\\
\end{longtable}

\begin{center}
\framebox{$\lookupval{\gamma^*}{n}{e}$}
\end{center}
\preTableSpace
\begin{longtable}{b{\twoColWidL}b{\twoColWidR}}
\shortenSpace
\[
  \inferrule*[right=LV-Cons-Head]
     { \spaceholder }
     {\lookupval{(\consval{n}{v}{\gamma})}{n}{v}}
\]
&
\shortenSpace
\[
  \inferrule*[right=LV-Cons-Tail]
     {n \not = n' \\ 
      \lookupval{\gamma}{n}{v}}
     {\lookupval{(\consval{n'}{v'}{\gamma})}{n}{v}}
\]
\end{longtable}

\newpage
\begin{center}
\framebox{$\updateCtx{\gamma^*}{n}{v}{\gamma}$}
\end{center}
\preTableSpace
\begin{longtable}{b{\oneColWid}}
\shortenSpace
\[
  \inferrule*[right=Update]
     { \removeCtx{\gamma}{n}{\gamma'} }
     { \updateCtx{\gamma}{n}{v}{(\consval{n}{v}{\gamma'})} }
\]
\\
\end{longtable}

\begin{center}
\framebox{$\removeCtx{\gamma^*}{n}{\gamma}$}
\end{center}
\preTableSpace
\begin{longtable}{b{\twoColWidL}b{\twoColWidR}}
\shortenSpace \shortenSpace
\[
  \inferrule*[right=R-Cons-Head]
     { \spaceholder }
     {\removeCtx{(\consval{n}{v}{\gamma})}{n}{\gamma}}
\]
&
\shortenSpace \shortenSpace
\[
  \inferrule*[right=R-Cons-Tail]
     {\removeCtx{\gamma}{n}{\gamma'}}
     {\removeCtx{(\consval{n'}{v'}{\gamma})}{n}{(\consval{n'}{v'}{\gamma'})}}
\]
\\
\end{longtable}

\begin{center}
\framebox{$\val{e^*}$}
\end{center}
\preTableSpace
\begin{longtable}{b{\threeColWid}b{\threeColWid}b{\threeColWid}}
\shortenSpace
\[
  \inferrule*[right=V-Int]{$ $}{\val{intlit(i)}}
\]
&
\shortenSpace
\[
  \inferrule*[right=V-True]{$ $}{\val{\mathit{true}}}
\]
&
\shortenSpace
\[
  \inferrule*[right=V-False]{$ $}{\val{\mathit{false}}}
\]
\\
\end{longtable}

\begin{center}
\framebox{$\vars{e^*}{ns}$}
\end{center}
\preTableSpace
\begin{longtable}{b{\twoColWidL}b{\twoColWidR}}
\shortenSpace
\[
  \inferrule*[right=VR-var]
    { }
    {\vars{\mathit{var(n)}}{\{n\}}}
\]
&
\shortenSpace
\[
  \inferrule*[right=VR-true]
    { }
    {\vars{\mathit{true}}{\emptyset}}
\]
\\
\shortenSpace
\[
  \inferrule*[right=VR-intlit]
    { }
    {\vars{\mathit{intlit(i)}}{\emptyset}}
\]
&
\shortenSpace
\[
  \inferrule*[right=VR-false]
    { }
    {\vars{\mathit{false}}{\emptyset}}
\]
\\
\shortenSpace
\[
  \inferrule*[right=VR-add]
    {\vars{e_1}{vr_1} \\
     \vars{e_2}{vr_2}}
    {\vars{\mathit{add(e_1, e_2)}}{(vr_1 \cup vr_2)}}
\]
&
\shortenSpace
\[
  \inferrule*[right=VR-gt]
    {\vars{e_1}{vr_1} \\
     \vars{e_2}{vr_2}}
    {\vars{\mathit{gt(e_1, e2)}}{(vr_1 \cup vr_2)}}
\]
\\
\shortenSpace
\[
  \inferrule*[right=VR-eq]
    {\vars{e_1}{vr_1} \\
     \vars{e_2}{vr_2}}
    {\vars{\mathit{eq(e_1, e_2)}}{(vr_1 \cup vr_2)}}
\]
&
\shortenSpace
\[
  \inferrule*[right=VR-not]
    {\vars{e}{vr}}
    {\vars{\mathit{not(e)}}{vr}}
\]
\\
\end{longtable}

\begin{center}
\framebox{$\typeexpr{\Gamma}{e^*}{ty}$}
\end{center}
\preTableSpace
\begin{longtable}{b{\twoColWidL}b{\twoColWidR}}
\shortenSpace
\[
  \inferrule*[right=T-var]
    {\lookuptype{\Gamma}{n}{ty}}
    {\typeexpr{\Gamma}{var(n)}{ty}}
\]
&
\shortenSpace
\[
  \inferrule*[right=T-true]
    { \spaceholder }
    {\typeexpr{\Gamma}{\mathit{true}}{\mathit{bool}}}
\]
\\
\shortenSpace \shortenSpace
\[
  \inferrule*[right=T-intlit]
    { \spaceholder }
    {\typeexpr{\Gamma}{\mathit{intlit(i)}}{\mathit{int}}}
\]
&
\shortenSpace \shortenSpace
\[
  \inferrule*[right=T-false]
    { \spaceholder }
    {\typeexpr{\Gamma}{\mathit{false}}{\mathit{bool}}}
\]
\\
\shortenSpace
\[
  \inferrule*[right=T-add]
    {\typeexpr{\Gamma}{e_1}{\mathit{int}} \\ 
     \typeexpr{\Gamma}{e_2}{\mathit{int}}}
    {\typeexpr{\Gamma}{\mathit{add(e_1, e_2)}}{\mathit{int}}}
\]
&
\shortenSpace
\[
  \inferrule*[right=T-gt]
    {\typeexpr{\Gamma}{e_1}{int} \\ 
     \typeexpr{\Gamma}{e_2}{int}}
    {\typeexpr{\Gamma}{\mathit{gt(e_1, e_2)}}{\mathit{bool}}}
\]
\\
\shortenSpace
\[
  \inferrule*[right=T-eq]
    {\typeexpr{\Gamma}{e_1}{int} \\ 
     \typeexpr{\Gamma}{e_2}{int}}
    {\typeexpr{\Gamma}{\mathit{eq(e_1, e_2)}}{\mathit{bool}}}
\]
&
\shortenSpace
\[
  \inferrule*[right=T-not]
    {\typeexpr{\Gamma}{e}{bool}}
    {\typeexpr{\Gamma}{\mathit{not(e)}}{\mathit{bool}}}
\]
\\
\end{longtable}

\begin{center}
\framebox{$\typestmt{\Gamma}{s^*}{\Gamma}$}
\end{center}
\preTableSpace
\begin{longtable}{b{\twoColWidL}b{\twoColWidR}}
\shortenSpace
\[
  \inferrule*[right=TS-skip]
    { \spaceholder }
    {\typestmt{\Gamma}{\mathit{skip}}{\Gamma}}
\]
&
\shortenSpace
\[
  \inferrule*[right=TS-assign]
    {\typeexpr{\Gamma}{e}{ty} \\
     \lookuptype{\Gamma}{n}{ty}}
    {\typestmt{\Gamma}{\mathit{assign(n, e)}}{\Gamma}}
\]
\\
\shortenSpace
\[
  \inferrule*[right=TS-seq]
    {\typestmt{\Gamma}{s_1}{\Gamma'} \\
     \typestmt{\Gamma'}{s_2}{\Gamma''}} 
    {\typestmt{\Gamma}{\mathit{seq(s_1, s_2)}}{\Gamma''}}
\]
&
\shortenSpace
\[
  \inferrule*[right=TS-ifte]
    {\typeexpr{\Gamma}{e}{\mathit{bool}} \\
     \typestmt{\Gamma}{s_1}{\Gamma'} \\
     \typestmt{\Gamma}{s_1}{\Gamma''}} 
    {\typestmt{\Gamma}{\mathit{ifte(e, s_1, s_2)}}{\Gamma}}
\]
\\
\shortenSpace
\[
  \inferrule*[right=TS-decl]
    {\typeexpr{\Gamma}{e}{ty} \\
     \notpresenttype{\Gamma}{n}}
    {\typestmt{\Gamma}{\mathit{decl(n, ty, e)}}{\mathit{consty(n, ty, \Gamma)}}}
\]
&
\shortenSpace
\[
  \inferrule*[right=TS-while]
    {\typeexpr{\Gamma}{e}{\mathit{bool}} \\
     \typestmt{\Gamma}{s}{\Gamma'}}
    {\typestmt{\Gamma}{\mathit{while(e, s)}}{\Gamma}}
\]
\\
\end{longtable}

\begin{center}
\framebox{$\evalexpr{\gamma}{e^*}{e}$}
\end{center}
\preTableSpace
\begin{longtable}{b{\twoColWidL}b{\twoColWidR}}
\shortenSpace
\[
  \inferrule*[right=E-var]
    {\lookupval{\gamma}{n}{v}}
    {\evalexpr{\gamma}{\mathit{var(n)}}{v}}
\]
& \shortenSpace
\[
  \inferrule*[right=E-true]
    { \spaceholder }
    {\evalexpr{\gamma}{\mathit{true}}{\mathit{true}}}
\]
\\
\shortenSpace
\[
  \inferrule*[right=E-intlit]
    { \spaceholder }
    {\evalexpr{\gamma}{\mathit{intlit(i)}}{\mathit{intlit(i)}}}
\]
& \shortenSpace
\[
  \inferrule*[right=E-false]
    { \spaceholder }
    {\evalexpr{\gamma}{\mathit{false}}{\mathit{false}}}
\]
\\
\shortenSpace
\[
  \inferrule*[right=E-eq-True]
     {\evalexpr{\gamma}{e_1}{v_1} \\
      \evalexpr{\gamma}{e_2}{v_2} \\
      v_1 = v_2}
     {\evalexpr{\gamma}{\mathit{eq(e_1, e_2)}}{\mathit{true}}}
\]
& \shortenSpace
\[
  \inferrule*[right=E-eq-False]
     {\evalexpr{\gamma}{e_1}{v_1} \\ 
      \evalexpr{\gamma}{e_2}{v_2} \\
      v_1 \not = v_2}
     {\evalexpr{\gamma}{\mathit{eq(e_1, e_2)}}{\mathit{false}}}
\]
\\
\shortenSpace
\[
  \inferrule*[right=E-gt-True]
     {\evalexpr{\gamma}{e_1}{\mathit{intlit}(i_1)} \\
      \evalexpr{\gamma}{e_2}{\mathit{intlit}(i_2)} \\
      i_1 > i_2}
     {\evalexpr{\gamma}{\mathit{gt(e_1, e_2)}}{\mathit{true}}}
\]
& \shortenSpace
\[
  \inferrule*[right=E-gt-False]
     {\evalexpr{\gamma}{e_1}{\mathit{intlit}(i_1)} \\ 
      \evalexpr{\gamma}{e_2}{\mathit{intlit}(i_2)} \\
      i_1 \leq i_2}
     {\evalexpr{\gamma}{\mathit{gt(e_1, e_2)}}{\mathit{false}}}
\]
\\
\shortenSpace
\[
  \inferrule*[right=E-not-True]
     {\evalexpr{\gamma}{e}{\mathit{false}}}
     {\evalexpr{\gamma}{\mathit{not(e)}}{\mathit{true}}}
\]
& \shortenSpace
\[
  \inferrule*[right=E-not-False]
     {\evalexpr{\gamma}{e}{\mathit{true}}}
     {\evalexpr{\gamma}{\mathit{not(e)}}{\mathit{false}}}
\]
\\
\end{longtable}
\tableTogetherSpace
\begin{longtable}{b{\oneColWid}}
\shortenSpace
\[
  \inferrule*[right=E-add]
     {\evalexpr{\gamma}{e_1}{\mathit{intlit}(i_1)} \\
      \evalexpr{\gamma}{e_2}{\mathit{intlit}(i_2)} \\
      \plus{i_1}{i_2}{i}}
     {\evalexpr{\gamma}{\mathit{add(e_1, e_2)}}{\mathit{intlit}(i)}}
\]
\end{longtable}

\newpage
\begin{center}
\framebox{$\evalstmt{\gamma}{s^*}{\gamma}$}
\end{center}
\preTableSpace
\begin{longtable}{b{\twoColWidL}b{\twoColWidR}}
\shortenSpace
\[
  \inferrule*[right=X-skip]
    { \spaceholder }
    {\evalstmt{\gamma}{\mathit{skip}}{\gamma}}
\]
&
\shortenSpace
\[
  \inferrule*[right=X-seq]
    {\evalstmt{\gamma}{s_1}{\gamma'} \\
     \evalstmt{\gamma'}{s_2}{\gamma''}} 
    {\evalstmt{\gamma}{\mathit{seq(s_1, s_2)}}{\gamma''}}
\]
\\
\shortenSpace
\[
  \inferrule*[right=X-ifte-True]
    {\evalexpr{\gamma}{e}{\mathit{true}} \\
     \evalstmt{\gamma}{s_1}{\gamma'}} 
    {\evalstmt{\gamma}{\mathit{ifte(e, s_1, s_2)}}{\gamma'}}
\]
&
\shortenSpace
\[
  \inferrule*[right=X-ifte-False]
    {\evalexpr{\gamma}{e}{\mathit{false}} \\
     \evalstmt{\gamma}{s_2}{\gamma'}} 
    {\evalstmt{\gamma}{\mathit{ifte(e, s_1, s_2)}}{\gamma'}}
\]
\\
\shortenSpace
\[
  \inferrule*[right=X-decl]
    {\evalexpr{\gamma}{e}{v}}
    {\evalstmt{\gamma}{\mathit{decl(n, ty, e)}}{\consval{n}{v}{\gamma}}}
\]
&
\shortenSpace
\[
  \inferrule*[right=X-while-True]
    {\evalexpr{\gamma}{e}{\mathit{true}} \\
     \evalstmt{\gamma}{s}{\gamma'} \\
     \evalstmt{\gamma'}{\mathit{while(e, s)}}{\gamma''}}
    {\evalstmt{\gamma}{\mathit{while(e, s)}}{\gamma''}}
\]
\\
\shortenSpace
\[
  \inferrule*[right=X-assign]
    {\evalexpr{\gamma}{e}{v} \\
      \updateCtx{\gamma}{n}{v}{\gamma'}}
    {\evalstmt{\gamma}{\mathit{assign(n, e)}}{\gamma'}}
\]
&
\shortenSpace
\[
  \inferrule*[right=X-while-False]
    {\evalexpr{\gamma}{e}{\mathit{false}}}
    {\evalstmt{\gamma}{\mathit{while(e, s)}}{\gamma}}
\]
\\
\end{longtable}

\subsubsection{Projections}
$\transrelset{H} = \{ \proj{e} : e,\ \proj{s} : s,\ \proj{ty} : ty \}$

\bigskip
\noindent $\transruleset{H} = \emptyset$

\bigskip
\noindent $\trset{H} = \emptyset$

\subsection{List Extension $L$}
\subsubsection{Syntax}
$\ntset{L} = \emptyset$

\bigskip\noindent
$\constrset{L}$:

\begin{tabular}{lclp{1in}lcl}
$s$ & ::= & $\mathit{splitlist}(n, n, e)$ \\
$e$ & ::= & $\mathit{nil}$
  \grmmrsep $\mathit{cons}(e, e)$
  \grmmrsep $\mathit{null}(e)$
  \grmmrsep $\mathit{head}(e)$
  \grmmrsep $\mathit{tail}(e)$ \\
$\mathit{ty}$ & ::= & $\mathit{list}(ty)$ \\
\end{tabular}

\newpage
\subsubsection{Relations}
$\relset{L} = \emptyset$

\bigskip
\noindent $\ruleset{L}$ contains the following rules:

\begin{center}
\framebox{$\typeexpr{\Gamma}{e^*}{ty}$}
\end{center}
\preTableSpace
\begin{longtable}{b{\twoColWidL}b{\twoColWidR}}
\shortenSpace
\[
  \inferrule*[right=T-nil]
    { \spaceholder }
    {\typeexpr{\Gamma}{\mathit{nil}}{\mathit{list(ty)}}}
\]
&
\shortenSpace
\[
  \inferrule*[right=T-cons]
    {\typeexpr{\Gamma}{e_1}{ty} \\ 
     \typeexpr{\Gamma}{e_2}{\mathit{list(ty)}}}
    {\typeexpr{\Gamma}{\mathit{cons(e_1, e_2)}}{\mathit{list(ty)}}}
\]
\\
\shortenSpace
\[
  \inferrule*[right=T-head]
    {\typeexpr{\Gamma}{e}{\mathit{list(ty)}}}
    {\typeexpr{\Gamma}{\mathit{head(e)}}{ty}}
\]
&
\shortenSpace
\[
  \inferrule*[right=T-tail]
    {\typeexpr{\Gamma}{e}{\mathit{list(ty)}}}
    {\typeexpr{\Gamma}{\mathit{tail(e)}}{\mathit{list(ty)}}}
\]
\\
\end{longtable}
\tableTogetherSpace
\begin{longtable}{b{\oneColWid}}
\shortenSpace
\[
  \inferrule*[right=T-null]
    {\typeexpr{\Gamma}{e}{\mathit{list(ty)}}}
    {\typeexpr{\Gamma}{\mathit{null(e)}}{\mathit{bool}}}
\]
\end{longtable}

\begin{center}
\framebox{$\val{e^*}$}
\end{center}
\preTableSpace
\begin{longtable}{b{\twoColWidL}b{\twoColWidR}}
\shortenSpace
\[
  \inferrule*[right=V-nil]
    { \spaceholder }
    {\val{\mathit{nil}}}
\]
&
\shortenSpace
\[
  \inferrule*[right=V-cons]
    {\val{e_1} \\ 
     \val{e_2}}
    {\val{\mathit{cons(e_1, e_2)}}}
\]
\end{longtable}

\begin{center}
\framebox{$\vars{e^*}{2^{n}}$}
\end{center}
\preTableSpace
\begin{longtable}{b{\twoColWidL}b{\twoColWidR}}
\shortenSpace
\[
  \inferrule*[right=VR-nil]
    { \spaceholder }
    {\vars{\mathit{nil}}{\emptyset}}
\]
&
\shortenSpace
\[
  \inferrule*[right=VR-head]
    {\vars{e}{vr}}
    {\vars{\mathit{head(e)}}{vr}}
\]
\\
\shortenSpace
\[
  \inferrule*[right=VR-cons]
    {\vars{e_1}{vr_1} \\ 
     \vars{e_2}{vr_2}}
    {\vars{\mathit{cons(e_1, e_2)}}{(vr_1 \cup vr_2)}}
\]
&
\shortenSpace
\[
  \inferrule*[right=VR-tail]
    {\vars{e}{vr}}
    {\vars{\mathit{tail(e)}}{vr}}
\]
\\
\end{longtable}
\tableTogetherSpace
\begin{longtable}{b{\oneColWid}}
\shortenSpace
\[
  \inferrule*[right=VR-null]
    {\vars{e}{vr}}
    {\vars{\mathit{null(e)}}{vr}}
\]
\\
\end{longtable}

\begin{center}
\framebox{$\evalexpr{\gamma}{e^*}{e}$}
\end{center}
\preTableSpace
\begin{longtable}{b{\twoColWidL}b{\twoColWidR}}
\shortenSpace
\[
  \inferrule*[right=E-nil]
    { \spaceholder }
    {\evalexpr{\gamma}{\mathit{nil}}{\mathit{nil}}}
\]
&
\shortenSpace
\[
  \inferrule*[right=E-head]
    {\evalexpr{\gamma}{e}{\mathit{cons(v_1, v_2)}}}
    {\evalexpr{\gamma}{\mathit{head(e)}}{v_1}}
\]
\\
\shortenSpace
\[
  \inferrule*[right=E-cons]
    {\evalexpr{\gamma}{e_1}{v_1} \\ 
     \evalexpr{\gamma}{e_2}{v_2}}
    {\evalexpr{\gamma}{\mathit{cons(e_1, e_2)}}
                      {\mathit{cons(v_1, v_2)}}}
\]
&
\shortenSpace
\[
  \inferrule*[right=E-tail]
    {\evalexpr{\gamma}{e}{\mathit{cons(v_1, v_2)}}}
    {\evalexpr{\gamma}{\mathit{tail(e)}}{v_2}}
\]
\\
\shortenSpace
\[
  \inferrule*[right=E-null-true]
    {\evalexpr{\gamma}{e}{\mathit{nil}}}
    {\evalexpr{\gamma}{\mathit{null(e)}}{\mathit{true}}}
\]
&
\shortenSpace
\[
  \inferrule*[right=E-null-false]
    {\evalexpr{\gamma}{e}{\mathit{cons(v_1, v_2)}}}
    {\evalexpr{\gamma}{\mathit{null(e)}}{\mathit{false}}}
\]
\\
\end{longtable}

\begin{center}
\framebox{$\typestmt{\Gamma}{s^*}{\Gamma}$}
\end{center}
\preTableSpace
\begin{longtable}{b{\oneColWid}}
\shortenSpace
\[
  \inferrule*[right=TS-splitlist]
             {\typeexpr{\Gamma}{e}{\mathit{list(ty)}} \\
               \lookuptype{\Gamma}{n_{hd}}{ty} \\
               \lookuptype{\Gamma}{n_{tl}}{\mathit{list(ty)}}}
             {\typestmt{\Gamma}{\mathit{splitlist}(n_{hd}, n_{tl}, e)}{\Gamma}}
\]
\\
\end{longtable}

\begin{center}
\framebox{$\evalstmt{\gamma}{s^*}{\gamma}$}
\end{center}
\preTableSpace
\begin{longtable}{b{\oneColWid}}
\shortenSpace
\[
  \inferrule*[right=X-splitlist]
             {\evalexpr{\gamma}{e}{\mathit{cons}(v_1, v_2)} \\
               n_{hd} \neq n_{tl}}
             {\evalstmt{\gamma}{\mathit{splitlist}(n_{hd}, n_{tl}, e)}
              {\consval{n_{hd}}{v_1}
                {\consval{n_{tl}}{v_2}{\gamma}}}}
\]
\\
\end{longtable}

\subsubsection{Projections}
$\transrelset{L} = \emptyset$

\bigskip
\noindent $\transruleset{L}$ contains the following rules: \\
\begin{longtable}{b{\twoColWidL}b{\twoColWidR}}
\shortenSpace
\[
  \inferrule*[right=P-null]
    { }
    {\projectexpr{\mathit{null(e)}}{e}}
\]
&
\shortenSpace
\[
  \inferrule*[right=P-nil]
    { }
    {\projectexpr{\mathit{nil}}{\mathit{true}}}
\]
\\
\shortenSpace
\[
  \inferrule*[right=P-head]
    { }
    {\projectexpr{\mathit{head(e)}}{e}}
\]
&
\shortenSpace
\[
  \inferrule*[right=P-cons]
    { }
    {\projectexpr{\mathit{cons}(e_1, e_2)}{\mathit{eq}(e_1, e_2)}}
\]
\\
\shortenSpace
\[
  \inferrule*[right=P-tail]
   { }
   {\projectexpr{\mathit{tail(e)}}{e}}
\]
&
\shortenSpace
\[
  \inferrule*[right=P-list]
    { }
    {\projectty{\mathit{list}(ty)}{ty}}
\]
\\
\end{longtable}
\tableTogetherSpace
\begin{longtable}{b{\oneColWid}}
\shortenSpace
\[
  \inferrule*[right=P-splitlist]
    {n_{hd} \neq n_{tl}}
    {\proj{s}(\mathit{splitlist}(n_{hd}, n_{tl}, e), \\\\
        seq(seq(assign(n_{hd}, e), assign(n_{tl},
           tail(var(n_{hd})))), assign(n_{hd}, head(var(n_{hd})))))}
\]
\\
\end{longtable}

\noindent $\trset{L} = \emptyset$

\subsection{Security Extension $S$}
\subsubsection{Syntax}
$\ntset{S} = \{ \mathit{sl}, \Sigma \}$

\bigskip\noindent
$\constrset{S}$:

\begin{tabular}{lclp{1in}lcl}
$s$ & ::= & $\mathit{secdecl}(n, ty, \mathit{sl}, e)$ \\
$\mathit{sl}$ & ::=  & $\mathit{public}$
             \grmmrsep $\mathit{private}$ \\
$\Sigma$ & ::= & $\mathit{nilsec}$
       \grmmrsep $\mathit{conssec}(n, \mathit{sl}, \Sigma)$ \\
\end{tabular}

\subsubsection{Relations}
\begin{tabular}{ll}
  \hspace*{-0.15in}
  \relset{S} = & \hspace*{-0.15in}
               \{\lookupsec{\Sigma^*}{n}{\mathit{sl}}, \ \
                 \notpresentsec{\Sigma^*}{n}, \ \
                 \joinlevel{\mathit{sl^*}}{\mathit{sl}}{\mathit{sl}}, \ \
                 \exprlevel{\Sigma}{e^*}{\mathit{sl}}, \\
               & \secure{\Sigma}{\mathit{sl}}{s^*}{\Sigma}\}
\end{tabular}

\bigskip
\noindent $\ruleset{S}$ contains the following rules:

\begin{center}
\framebox{$\typestmt{\Gamma}{s^*}{\Gamma}$}
\end{center}
\preTableSpace
\begin{longtable}{b{\oneColWid}}
\shortenSpace
\[
  \inferrule*[right=TS-secdecl]
    {\typeexpr{\gamma}{e}{ty} \\
     \notpresenttype{\Gamma}{n}}
    {\typestmt{\Gamma}{\mathit{secdecl(n, ty, sl, e)}}{\mathit{consty(n, ty, \Gamma)}}}
\]
\\
\end{longtable}

\begin{center}
\framebox{$\evalstmt{\gamma}{s^*}{\gamma}$}
\end{center}
\preTableSpace
\begin{longtable}{b{\oneColWid}}
\shortenSpace
\[
  \inferrule*[right=X-secdecl]
    {\evalexpr{\gamma}{e}{v}}
    {\evalstmt{\gamma}{\mathit{secdecl(n, ty, sl, e)}}{\consval{n}{v}{\gamma}}}
\]
\\
\end{longtable}

\begin{center}
\framebox{$\lookupsec{\Sigma^*}{n}{\mathit{sl}}$}
\end{center}
\preTableSpace
\begin{longtable}{b{\twoColWidL}b{\twoColWidR}}
\shortenSpace
\[
  \inferrule*[right=LS-Cons-Head]
     { \spaceholder }
     {\lookupsec{\mathit{conssec(n,sl,\Sigma)}}{n}{sl}}
\]
&
\shortenSpace
\[
  \inferrule*[right=LS-Cons-Tail]
     {n \not = n' \\
      \lookupsec{\Sigma}{n}{sl}}
     {\lookupsec{\mathit{conssec(n',sl',\Sigma)}}{n}{sl}}
\]
\\
\end{longtable}

\begin{center}
\framebox{$\notpresentsec{\Sigma^*}{n}$}
\end{center}
\preTableSpace
\begin{longtable}{b{\twoColWidL}b{\twoColWidR}}
\shortenSpace
\[
  \inferrule*[right=NBS-Nil]
    { \spaceholder }
    {\notpresentsec{\mathit{nilsec}}{n}}
\]
&
\shortenSpace
\[
  \inferrule*[right=NBS-Cons]
     {n \not = n' \\
      \notpresentsec{\Sigma}{n}}
     {\notpresentsec{\mathit{conssec(n',ty',\Sigma)}}{n}}
\]
\\
\end{longtable}

\newpage
\begin{center}
\framebox{$\joinlevel{\mathit{sl}^*}{\mathit{sl}}{\mathit{sl}}$}
\end{center}
\preTableSpace
\begin{longtable}{b{\oneColWid}}
\shortenSpace
\[
  \inferrule*[right=J-public]
    { }
    {\joinlevel{\mathit{public}}{\mathit{public}}{\mathit{public}}}
\]
\\
\end{longtable}
\tableTogetherSpace
\begin{longtable}{b{\twoColWidL}b{\twoColWidR}}
\shortenSpace
\[
  \inferrule*[right=J-private-l]
    { }
    {\joinlevel{\mathit{private}}{\ell}{\mathit{private}}}
\]
&
\shortenSpace
\[
  \inferrule*[right=J-private-r]
    { }
    {\joinlevel{\ell}{\mathit{private}}{\mathit{private}}}
\]
\\
\end{longtable}

\begin{center}
\framebox{$\exprlevel{\Sigma}{e^*}{\mathit{sl}}$}
\end{center}
\preTableSpace
\begin{longtable}{b{\twoColWidL}b{\twoColWidR}}
\shortenSpace
\[
  \inferrule*[right=L-var]
    {\lookupsec{\Sigma}{n}{\ell}}
    {\exprlevel{\Sigma}{\mathit{var(n)}}{\ell}}
\]
&
\shortenSpace
\[
  \inferrule*[right=L-true]
    { \spaceholder }
    {\exprlevel{\Sigma}{\mathit{true}}{\mathit{public}}}
\]
\\
\shortenSpace
\[
  \inferrule*[right=L-int]
    { \spaceholder }
    {\exprlevel{\Sigma}{\mathit{intlit(i)}}{\mathit{public}}}
\]
&
\shortenSpace
\[
  \inferrule*[right=L-false]
    { \spaceholder }
    {\exprlevel{\Sigma}{\mathit{false}}{\mathit{public}}}
\]
\\
\shortenSpace
\[
  \inferrule*[right=L-add]
    {\exprlevel{\Sigma}{e_1}{\ell_1} \\
     \exprlevel{\Sigma}{e_2}{\ell_2} \\
     \joinlevel{\ell_1}{\ell_2}{\ell}}
    {\exprlevel{\Sigma}{add(e_1, e_2)}{\ell}}
\]
&
\shortenSpace
\[
  \inferrule*[right=L-gt]
    {\exprlevel{\Sigma}{e_1}{\ell_1} \\
     \exprlevel{\Sigma}{e_2}{\ell_2} \\
     \joinlevel{\ell_1}{\ell_2}{\ell}}
    {\exprlevel{\Sigma}{gt(e_1, e_2)}{\ell}}
\]
\\
\shortenSpace
\[
  \inferrule*[right=L-eq]
    {\exprlevel{\Sigma}{e_1}{\ell_1} \\
     \exprlevel{\Sigma}{e_2}{\ell_2} \\
     \joinlevel{\ell_1}{\ell_2}{\ell}}
    {\exprlevel{\Sigma}{eq(e_1, e_2)}{\ell}}
\]
&
\shortenSpace
\[
  \inferrule*[right=L-not]
    {\exprlevel{\Sigma}{e}{\ell}}
    {\exprlevel{\Sigma}{not(e)}{\ell}}
\]
\\
\end{longtable}

\begin{center}
\framebox{$\secure{\Sigma}{\mathit{sl}}{s^*}{\Sigma}$}
\end{center}
\preTableSpace
\begin{longtable}{b{\oneColWid}}
\shortenSpace
\[
  \inferrule*[right=S-skip]
    { \spaceholder }
    {\secure{\Sigma}{\ell}{\mathit{skip}}{\Sigma}}
\]
\\
\shortenSpace
\[
  \inferrule*[right=S-seq]
    {\secure{\Sigma}{\ell}{s_1}{\Sigma'} \\
     \secure{\Sigma'}{\ell}{s_2}{\Sigma''} }
    {\secure{\Sigma}{\ell}{\mathit{seq}(s_1, s_2)}{\Sigma''}}
\]
\\
\shortenSpace
\[
  \inferrule*[right=S-decl]
    {\exprlevel{\Sigma}{e}{\mathit{public}} \\
     \notpresentsec{\Sigma}{n}}
    {\secure{\Sigma}{\mathit{public}}
      {\mathit{decl(n, ty, e)}}
      {\mathit{conssec(n, public, \Sigma)}}}
\]
\\
\shortenSpace
\[
  \inferrule*[right=S-assign-private]
    {\exprlevel{\Sigma}{e}{\ell} \\
     \lookupsec{\Sigma}{n}{\mathit{private}}}
    {\secure{\Sigma}{\ell'}
      {\mathit{assign(n, e)}}{\Sigma}}
\]
\\
\shortenSpace
\[
  \inferrule*[right=S-assign-public]
    {\exprlevel{\Sigma}{e}{\mathit{public}} \\
     \lookupsec{\Sigma}{n}{\mathit{public}}}
    {\secure{\Sigma}{\mathit{public}}
      {\mathit{assign(n, e)}}{\Sigma}}
\]
\\
\shortenSpace
\[
  \inferrule*[right=S-ifte]
    {\exprlevel{\Sigma}{e}{\ell} \\
     \joinlevel{\ell'}{\ell}{\ell''} \\
     \secure{\Sigma}{\ell''}{s_1}{\Sigma_1} \\
     \secure{\Sigma}{\ell''}{s_2}{\Sigma_2}}
    {\secure{\Sigma}{\ell'}{\mathit{ifte(e, s_1, s_2)}}{\Sigma}}
\]
\\
\shortenSpace
\[
  \inferrule*[right=S-while]
    {\exprlevel{\Sigma}{e}{\ell} \\
     \joinlevel{\ell'}{\ell}{\ell''} \\
     \secure{\Sigma}{\ell''}{s}{\Sigma'}}
    {\secure{\Sigma}{\ell'}{\mathit{while(e, s)}}{\Sigma}}
\]
\\
\shortenSpace
\[
  \inferrule*[right=S-secdecl-private]
    {\exprlevel{\Sigma}{e}{\ell} \\
     \notpresentsec{\Sigma}{n}}
    {\secure{\Sigma}{\ell'}
      {\mathit{secdecl(n, ty, private, e)}}
      {\mathit{conssec(n, private, \Sigma)}}}
\]
\\
\shortenSpace
\[
  \inferrule*[right=S-secdecl-public]
    {\exprlevel{\Sigma}{e}{\mathit{public}} \\
     \notpresentsec{\Sigma}{n}}
    {\secure{\Sigma}{\mathit{public}}
      {\mathit{secdecl(n, ty, public, e)}}
      {\mathit{conssec(n, public, \Sigma)}}}
\]
\\
\end{longtable}

\subsubsection{Projections}
$\transrelset{S} = \emptyset$

\bigskip
\noindent $\transruleset{S}$ contains the following rule:
\begin{longtable}{b{\oneColWid}}
\shortenSpace
\[
  \inferrule*[right=P-secdecl]{ }{\projectstmt{\mathit{secdecl}(n, ty,
      \ell, e)}{\mathit{decl}(n, ty, e)}}
\]
\\
\end{longtable}

\bigskip
\noindent $\trset{S}$ contains the following rules:
\shortenSpace
\begin{longtable}{b{\twoColWidL}b{\twoColWidR}}
\shortenSpace
\[
  \inferrule*[right=P-level]
    {\projectexpr{e}{e'} \\
     \exprlevel{\Sigma}{e'}{\ell}}
    {\exprlevel{\Sigma}{e}{\ell}}
\]
&
\shortenSpace
\[
  \inferrule*[right=P-secure]
    {\projectstmt{s}{s'} \\
     \secure{\Sigma}{\ell}{s'}{\Sigma'} }
    {\secure{\Sigma}{\ell}{s}{\Sigma'}}
\]
\\
\end{longtable}

\section{Substitution, Unification and Term Replacement} \label{app:unification}

This appendix contains observations used in the proof of
Lemma~\ref{thm:subst-unknown} in Section~\ref{subsec:elaborateExt}.

We first show a distributivity property for term replacement.

\begin{thm}\label{thm:subst-distr}
  Let $\theta$ be a substitution whose domain does not contain any of
  the variables that appear in the sequence of terms $\overline{t}$.
  Then
  \begin{enumerate}
    \item for any term $s$,
      $\appsubst{\subst{c(\overline{t})}{\unknown}{\theta}}
                {\subst{c(\overline{t})}{\unknown}{s}} =
                \subst{c(\overline{t})}{\unknown}{\appsubst{\theta}{s}}$.

    \item for any formula $F$,
     $\appsubst{\subst{c(\overline{t})}{\unknown}{\theta}}
                {\subst{c(\overline{t})}{\unknown}{F}} =
                \subst{c(\overline{t})}{\unknown}{\appsubst{\theta}{F}}$.
                
    \item for any unification problem $\cal U$,
      $\appsubst{\subst{c(\overline{t})}{\unknown}{\theta}}
                {\subst{c(\overline{t})}{\unknown}{{\cal U}}} =
                \subst{c(\overline{t})}{\unknown}{\appsubst{\theta}{{\cal U}}}$.
  \end{enumerate}
\end{thm} 

\begin{proof}
  The second and third clauses follow easily from the first. 
  We prove the first clause by induction on the structure of $s$.

  If $s$ is a variable, the argument is straightforward: essentially,
  both sides of the equation reduce to $x$ if $x$ is not in the domain
  of $\theta$ and to $\subst{c(\overline{t})}{\unknown}{t'}$ if
  $\langle x, t'\rangle \in \theta$.
  If $s$ is $\unknown$, then both sides reduce to $c(\overline{t})$;
    we make use of the fact that no variables in $t$ appear in the
    domain of $\theta$ to ensure that the lefthand side of the
    equation reduces to this term.
  If $s$ is some other constant $c'$, then both sides reduce to $c'$.
  Finally, suppose that $s$ is a compound term of the form
  $c'(r_1,\dots,r_n)$.
  In this case, we make use of the fact that substitution and term
  replacement distribute to the arguments of the term and the
  induction hypothesis to conclude that the equality holds; to allow
  for the application of the induction hypothesis, we note that, for
  $1 \leq i \leq n$, the variables that appear in $r_i$ are included
  in the collection of variables that appear in $s$.
\end{proof}

We next show that the unifiability of
  $\subst{c(\overline{t})}{\unknown}{{\cal U}}$ implies the
  unifiability of   $\cal U$ in the case the constructor $c$ and
  the variables in $\overline{t}$ do not appear.
  In the proof, we will find a ``generalized inversion
of replacement'' operation on terms, useful.
\begin{definition}[Inversion of Term Substitution] \label{def:inversion}
The ``inversion'' of a term $s$ is an operation that replaces all
subterms of $t$ of the form $c(\overline{t'})$ for any
sequence of terms $\overline{t'}$, \ie, any term in which $c$ appears
as the top-level symbol, with $\unknown$.
We will write the result of this operation applied to $s$ as
$\inv{s}$.
This operation is extended to substitutions, unification problems,
etc., by essentially distributing it to the terms in them.
The notation is also lifted accordingly; \eg, we write
$\inv{\theta}$ to denote the inversion applied to $\theta$.
\end{definition}
\begin{lemma}\label{lem:appsubst-inv}
  Let $s$ be a term in which $c$ and the variables in the sequence of
  terms $\overline{t}$ do not appear. Then, for any substitution
  $\theta$, it is the case that
  $\inv{(\appsubst{\theta}{\subst{c(\overline{t})}{\unknown}{s}})} =
  \appsubst{\inv{\theta}}{s}$.
\end{lemma}

\begin{proof}
  By induction on the structure of $s$.
  We consider by cases this structure.

  Suppose $s$ is a variable $x$.
  In this case, $\subst{c(\overline{t})}{\unknown}{s} = x$.
  Now the analysis breaks up into two subcases, depending on whether
  or not $x$ is in the domain of $\theta$.
  If it is not in this domain, then it will also not be in the domain
  of $\inv{\theta}$ and both
  $\inv{(\appsubst{\theta}{\subst{c(\overline{t})}{\unknown}{s}})}$ and
  $\appsubst{\inv{\theta}}{s}$ will be $x$.
  Alternatively, let $\langle x, t\rangle \in \theta$.
  Then $\langle x, \inv{t} \rangle \in \inv{\theta}$
  The desired conclusion now follows by observing that
  $\inv{(\appsubst{\theta}{\subst{c(\overline{t})}{\unknown}{s}})}
  = \inv{(\appsubst{\theta}{x})} = \appsubst{\inv{\theta}}{s}$.

  Suppose $s$ is a constant, which includes the case where $s$ is
  $\unknown$.
  In this case, we claim that both
  $\inv{(\appsubst{\theta}{\subst{c(\overline{t})}{\unknown}{s}})}$
  and $\appsubst{\inv{\theta}}{s}$ are identical to $s$.
  This is obvious when $s$ is a constant different from $\unknown$.
  When $s$ is $\unknown$, the claim follows easily from observing
  $\subst{c(\overline{t})}{\unknown}{s} = c(\overline{t})$,
  substitution into $c(\overline{t})$ preserves the top-level symbol,
  and that $\inv{(\cdot)}$ only looks at this symbol, reducing
  the term to $\unknown$ if it is $c$.

  The last case to consider is that when $s$ is of the form
  $c'(r_1,\ldots,r_n)$.
  Here we use the induction hypothesis to conclude that, for $1 \leq i
  \leq n$,
  $\inv{(\appsubst{\theta}{\subst{c(\overline{t})}{\unknown}{r_i}})} =
  \appsubst{\inv{\theta}}{r_i}$. 
  Noting substitution and the $\inv{(\cdot)}$ operation
  preserve the top-level symbol in a compound term when this symbol is
  different from $c$ and that they distribute to the arguments now
  suffices to complete the proof. 
\end{proof}

\begin{thm}\label{thm:substunif}
If $\cal U$ is a unification problem in which $c$ and the variables in
the sequence of terms $\overline{t}$ do not appear and $\theta$
unifies $\subst{c(\overline{t})}{\unknown}{{\cal U}}$, then
$\inv{\theta}$ unifies $\cal U$.
\end{thm}

\begin{proof}
We associate a measure given by a pair of natural numbers $\langle
n_1,n_2\rangle$ with a unification problem when $n_1$ sums up the
sizes of the terms in the problem and $n_2$ counts the number of pairs
of terms. 
The proof is by lexicographic induction on the measure associated with
$\cal U$. 
The argument is based on picking a pair $\langle s_1,s_2 \rangle$ in
$\cal U$ and reducing the truth of the lemma for $\cal U$ to its truth
on a smaller unification problem.

Suppose that one of $s_1$ and $s_2$ is a variable $x$. Without loss of
generality, let this be $s_1$.
Then we have the following cases for $s_2$:
\begin{enumerate}
  \item $s_2$ is $\unknown$. In this case,
    $\subst{c(\overline{t})}{\unknown}{\langle s_1,s_2 \rangle}$ will
    be a pair of the form $\langle x, c(\overline{t'})\rangle$.
    Since $\theta$ makes these two terms equal, it must have a
    substitution of the form $\langle x, c(\overline{t''}) \rangle$ in
    it for $x$.
    But, then, $\inv{\theta}$ must have the pair $\langle x,
    \unknown\rangle$ in it.

  \item $s_2$ is a variable $y$.
    In this case, $\langle s_1, s_2 \rangle$ is also a member of
    $\subst{c(\overline{t})}{\unknown}{{\cal U}}$ and, since $\theta$
    is a unifier for this problem, it must be the case that 
    $\appsubst{s_1}{\theta} = \appsubst{s_2}{\theta}$.
    From this it follows easily that
    $\appsubst{s_1}{\inv{\theta}} = \appsubst{s_2}{\inv{\theta}}$.

  \item $s_2$ is a term different from a variable and $\unknown$.
    In this case, $\theta$ must have a pair of the form
    $\langle x,
    \appsubst{\subst{c(\overline{t})}{\unknown}{s_2}}{\theta} \rangle$ in
      it.
      But, then, $\inv{\theta}$ will have the pair
      $\langle x,
      \inv{(\appsubst{\subst{c(\overline{t})}{\unknown}{s_2}}{\theta})} \rangle$
      in it.
      By Lemma~\ref{lem:appsubst-inv}, this pair is the same as
      $\langle x, \appsubst{s_2}{\inv{\theta}} \rangle$,
      \ie, $\inv{\theta}$ must unify the two terms in this pair.
   \end{enumerate}
   Thus, in all these cases, the truth of the lemma is dependent on
   its truth for the unification problem that results from $\cal U$ by
   leaving out the pair $\langle s_1, s_2 \rangle$.
   We may therefore invoke the induction hypothesis to complete the proof.

   The cases that remain are those where $s_1$ and $s_2$ both have
   a constant or a constructor as their top-level symbol.
   If either of them is a constant, then the other must be an
   identical constant else $\theta$ cannot unify
   $\subst{c(\overline{t})}{\unknown}{s_1}$ and
   $\subst{c(\overline{t})}{\unknown}{s_2}$; if the constant 
   is $\unknown$, we need to observe that the top-level symbol will
   become $c$ under the replacement and the only way the other term
   will have the same top-level symbol is if it too is $\unknown$.
   Thus, in this situation, $\inv{\theta}$ must unify this pair and we
   are left to show the lemma holds for a smaller unification 
   problem, which must be the case by the induction hypothesis.

   To conclude the proof, we have to consider the case where $s_1$ and
   $s_2$ are both compound terms. 
   Since the replacement of $\unknown$ by $c(\overline{t})$ must
   preserve the top-level function symbol and $\theta$ unifies $\cal
   U$, this symbol in $s_1$ and $s_2$ must be identical.
   Thus, $s_1$ and $s_2$ must be terms of the form
   $c'(r^1_1,\ldots,r^1_n)$ and $c'(r^2_1,\ldots,r^2_n)$,
   respectively.
   Since $\theta$ unifies $\subst{c(\overline{t})}{\unknown}{s_1}$
   and $\subst{c(\overline{t})}{\unknown}{s_2}$, it easily follows
   that it must unify
   $ \{ \langle \subst{c(\overline{t})}{\unknown}{r^1_1},
                \subst{c(\overline{t})}{\unknown}{r^2_1} \rangle,
        \ldots,
        \langle \subst{c(\overline{t})}{\unknown}{r^1_n},
                \subst{c(\overline{t})}{\unknown}{r^2_n}      
        \rangle \} $.
   Letting ${\cal U}'$ represent the unification problem obtained from
   $\cal U$ by removing $\langle s_1, s_2 \rangle$, it then follows that
   $\theta$ must unify
   $\subst{c(\overline{t})}
          {\unknown}
          {( \{\langle r^1_1,r^2_1\rangle, \ldots, \langle
                      r^1_n,r^2_n\rangle \} \cup
            {\cal U}')}$.
     Now, the sum of the sizes of the terms in the unification problem
     $ \{\langle r^1_1,r^2_1\rangle, \ldots, \langle
                      r^1_n,r^2_n\rangle \} \cup
            {\cal U}'$     
     is smaller than that of $\cal U$, and hence the induction
     hypothesis can be applied to it.
     Doing so implies that $\inv{\theta}$ is a unifier for this
     problem, from which it easily follows that $\inv{\theta}$ is
     a unifier for $\cal U$.
\end{proof}

Finally we show there is an \mgu\ for $\cal U$ from which we can
get an \mgu\ for $\subst{c(\overline{t})}{\unknown}{{\cal U}}$ under
suitable circumstances. 
\begin{lemma}\label{lem:non-occurrence}
  Let $x$ be a variable that does not appear in the terms in the
  sequence $\overline{t}$. Then $x$ occurs in a term $s$ if and only
  if it occurs in $\subst{c(\overline{t})}{\unknown}{s}$.
\end{lemma}

\begin{proof}
  By an easy induction on the structure of $s$.
\end{proof}

\begin{thm}\label{thm:unifsubst}
  Let $\cal U$ be a unification problem in which
  the variables in the sequence of terms $\overline{t}$ do not appear.
  If $\cal U$ is unifiable, then there is an \mgu\ $\theta$ for $\cal
  U$ whose domain does not contain variables appearing in $\overline{t}$
  and is such that $\subst{c(\overline{t})}{\unknown}{\theta}$ is
  an \mgu\ for  $\subst{c(\overline{t})}{\unknown}{{\cal U}}$; note
  the domain of this \mgu\ for
  $\subst{c(\overline{t})}{\unknown}{{\cal U}}$ must also be disjoint
  from the collection of variables appearing in $t$.
\end{thm}

\begin{proof}
  The theorem is vacuously true if $\cal U$ is not unifiable.
  We therefore assume it is.
  We then use the property proved in \cite{martelli82} that an
  \mgu\ for it can be obtained by applying the following
  transformations in any order: 
  \begin{description}
    \item[Reorder] If there is a pair of terms $\langle s, x\rangle$
      where $s$ is not a variable and $x$ is one, replace it with
      $\langle x, s\rangle$.

    \item[Drop Trivial] If there is a pair of the form
      $\langle x, x\rangle$ where $x$ is a variable, drop the pair.

    \item[Variable Elimination] If there is a pair of the form
      $\langle x, s\rangle$ where $x$ is a variable, $s$ is not $x$
      and $x$ does not occur in $s$ but it does occur in a term in
      some other pair in  the current 
      unification problem, transform the problem by applying
      the substitution $\{\langle x, s\rangle\}$ to it. 
      Note that $x$ must not appear in $s$ for this step to be
      applicable but this is guaranteed to be the case because the
      original unification problem is solvable.

    \item[Term Reduction] If there is a pair of the form
      $\langle c'(r^1_1,\ldots,r^1_n), c'(r^2_1,\dots,r^2_n) \rangle$
      for some constructor $c'$, then transform the unification
      problem by replacing this pair with the set of pairs
      $\{ \langle r^1_1,r^2_1 \rangle,\ldots,\langle r^1_n, r^2_n\rangle \}$;
      in the degenerate case, the two terms may be
      constants, in which case the pair gets dropped.
      Once again, this must be the only case for a pair of terms with
      a constructor as the top-level symbol because the original
      problem is unifiable.
  \end{description}
  The application of these steps is guaranteed to terminate.
  When they do, what will be left behind will be a collection of pairs
  of terms whose first component is a variable that does not appear
  anywhere else in the collection.
  A unification problem in this form is called a \emph{solved form} and
  if the transformation steps yield a unification problem in such a
  form, it is an \mgu\ for the original problem. 
  Since none of the steps introduce any new variables, the domain of
  the \mgu\ that is produced in this way for $\cal U$ must be limited
  to the variables appearing in $\cal U$ and hence must be disjoint
  from the collection of variables that occur in $\overline{t}$. 

  We will show below that if ${\cal U}'$ is a unification problem in
  which neither the constructor $c$ nor the variables in
  $\overline{t}$ appear and it can be transformed into the unification
  problem ${\cal U}''$ by the application of one of the steps
  described, then $\subst{c(\overline{t})}{\unknown}{{\cal U}'}$ can
  be transformed into $\subst{c(\overline{t})}{\unknown}{{\cal U}''}$
  by the application of a finite sequence of these steps.
  Using Lemma~\ref{lem:non-occurrence}, it is easy to see that if
  $\cal U$ is in solved form then
  $\subst{c(\overline{t})}{\unknown}{{\cal U}'}$ must also be in
  solved form.
  The theorem follows from these facts.

  It only remains to be shown that a transformation step applied to
  ${\cal U}'$ can be mimicked with respect to
  $\subst{c(\overline{t})}{\unknown}{{\cal U}'}$.
  Since the term replacement operation leaves variables unchanged,
  this is obvious for the {\it reorder}  and {\it drop trivial} steps.
  For {\it variable elimination}, we observe first that if a pair of
  the form $\langle x, s\rangle$ appears in ${\cal U}'$, then the pair
  $\langle x, \subst{c(\overline{t})}{\unknown}{s} \rangle$ must appear in 
  $\subst{c(\overline{t})}{\unknown}{{\cal U}'}$. Further, using
  Lemma~\ref{lem:non-occurrence}, we see that the conditions for the
  application of the same transformation step to
  $\subst{c(\overline{t})}{\unknown}{{\cal U}'}$ must be satisfied.
  Applying it will yield the unification problem
  $\appsubst{\{x,\subst{c(\overline{t})}{\unknown}{s}\rangle}
            {\subst{c(\overline{t})}{\unknown}{{\cal U}'}}$ which, by
  Theorem~\ref{thm:subst-distr}, must be the same as
  $\subst{c(\overline{t})}{\unknown}
         {\appsubst{\{\langle x, s\rangle\}}{{\cal U}'}}$.
  But this is, in fact, nothing other than
  $\subst{c(\overline{t})}{\unknown}{{\cal U}''}$.
  Finally, for {\it term reduction} we use the fact that term
  replacement leaves constructors other than $\unknown$ 
  unchanged and distribute to the arguments.
  In the only case that escapes this consideration, \ie, when the pair
  is $\langle \unknown,\unknown \rangle$, the observation is that the
  corresponding pair in
  $\subst{c(\overline{t})}{\unknown}{{\cal U}'}$ will be
    $\langle c(\overline{t}), c(\overline{t})\rangle$. 
  However, this pair can be eliminated from the unification problem by
  repeated applications of {\it term reduction} and {\it drop
    trivial}.
\end{proof}

\end{document}